\documentclass{aa}

\newcommand{\HII}{H\,{\scriptsize II}}

\newcommand{\CII}{[C\,{\scriptsize II}]}

\newcommand{\OI}{[O\,{\scriptsize I}]}

\newcommand{\NII}{[N\,{\scriptsize II}]}

\newcommand{\CI}{[C\,{\scriptsize I}]}

\usepackage{graphicx}
\usepackage{hyperref}

\usepackage{txfonts}
\begin{document}

\title{Globules and pillars  in Cygnus X}
\subtitle{III. {\sl Herschel} and upGREAT/SOFIA far-infrared spectroscopy of the globule IRAS 20319+3958 in
  Cygnus X\thanks{The \CII\ data shown in Figs. 4 and 5 are available in fits format form the CDS via anonymous ftp to cdsarc.u-strasbg.fr 
(130.79.128.5) or via http://cdsweb.u-strasbg.fr/cgi-bin/qcat?J/A+A/.}}
  \author{N. Schneider\inst{1} 
  \and M. R\"ollig \inst{1}
  \and E.T. Polehampton \inst{2}
  \and F. Comer\'on \inst{3}
  \and A.A. Djupvik \inst{4,5}
  \and Z. Makai \inst{1,6}
  \and C. Buchbender \inst{1}
  \and R. Simon \inst{1}
  \and S. Bontemps \inst{7}  
  \and R. G\"usten \inst{8}   
  \and G. White \inst{2,9}
  \and Y. Okada \inst{1}
  \and A. Parikka \inst{10}   
  \and N. Rothbart \inst{11}   
  }
 \institute{
  I. Physik. Institut, University of Cologne, Z\"ulpicher Str. 77, 50937 Cologne, Germany\\
  \email{nschneid@ph1.uni-koeln.de}
  \and
  RAL Space, STFC Rutherford Appleton Laboratory, Chilton, Didcot, Oxfordshire, OX11 0QX, UK 
  \and 
  ESO, Karl Schwarzschild Str. 2, 85748, Garching, Germany
  \and 
  Nordic Optical Telescope, Rambla Jos\'{e} Ana Fern\'{a}ndez P\'{e}rez 7, 38711 Bre\~{n}a Baja, Spain
  \and
  Department of Physics and Astronomy, Aarhus University, Ny Munkegade 120, 8000 Aarhus C, Denmark
  \and
  Department of Physics and Astronomy, West Virginia University, Morgantown, WV 26506, USA
  \and
  Laboratoire d’Astrophysique de Bordeaux, Universit\'e de Bordeaux, CNRS, B18N, all\'ee G. Saint-Hilaire, 33615 Pessac,
  France
  \and
   Max-Planck Institut f\"ur Radioastronomie, Auf dem H\"ugel 69, 53121 Bonn, Germany 
  \and
  Department of Physics and Astronomy, The Open University, Walton Hall, Milton Keynes, MK7 6AA, UK
  \and
  SOFIA-USRA, NASA Ames Research Center, MS 232-12, Moffett Field, CA 94035, USA
  \and 
  DLR, Rutherfordstraße 2, 12489 Berlin-Adlershof, Germany
  }   
%

\titlerunning{Globules and pillars in Cygnus X III. PDR}
\authorrunning{N. Schneider}
\date{\today}

\abstract {IRAS 20319+3958 in Cygnus X South is a rare example of a
  free-floating globule (mass $\sim$240 M$_\odot$, length $\sim$1.5
  pc) with an internal \HII\ region created by the stellar feedback of
  embedded intermediate-mass stars, in particular, one Herbig Be star.
  In \citet{schneider2012} and \citet{djupvik2017}, we proposed that
  the emission of the far-infrared (FIR) lines of \CII\ at 158 $\mu$m
  and \OI\ at 145 $\mu$m in the globule head are mostly due to an
  internal photodissociation region (PDR). Here, we present a {\sl
    Herschel}/HIFI \CII\ 158 $\mu$m map of the whole globule and a
  large set of other FIR lines (mid-to high-J CO lines observed with
  {\sl Herschel}/PACS and SPIRE, the \OI\ 63 $\mu$m line and the
  $^{12}$CO 16$\to$15 line observed with upGREAT on SOFIA), covering
  the globule head and partly a position in the tail.  The \CII\ map
  revealed that the whole globule is probably rotating.  Highly
  collimated, high-velocity \CII\ emission is detected close to the
  Herbig Be star. We performed a PDR analysis using the KOSMA-$\tau$
  PDR code for one position in the head and one in the tail.  The
  observed FIR lines in the head can be reproduced with a
  two-component model: an extended, non-clumpy outer PDR shell and a
  clumpy, dense, and thin inner PDR layer, representing the interface
  between the \HII\ region cavity and the external PDR. The modelled
  internal UV field of $\sim$2500 G$_\circ$ is similar to what we
  obtained from the {\sl Herschel} FIR fluxes, but lower than what we
  estimated from the census of the embedded stars. External
  illumination from the $\sim$30 pc distant Cyg OB2 cluster, producing
  an UV field of $\sim$150-600 G$_\circ$ as an upper limit, is
  responsible for most of the \CII\ emission. For the tail, we modelled
  the emission with a non-clumpy component, exposed to a UV-field of
  around 140 G$_\circ$.}

\keywords{interstellar medium: clouds
          -- individual objects: Cygnus X   
          -- molecules
          -- kinematics and dynamics
          -- Radio lines: ISM}
\maketitle


\section{Introduction} \label{intro} 

\begin{figure}[htb]
\begin{center} 
\includegraphics [width=9cm, angle={0}]{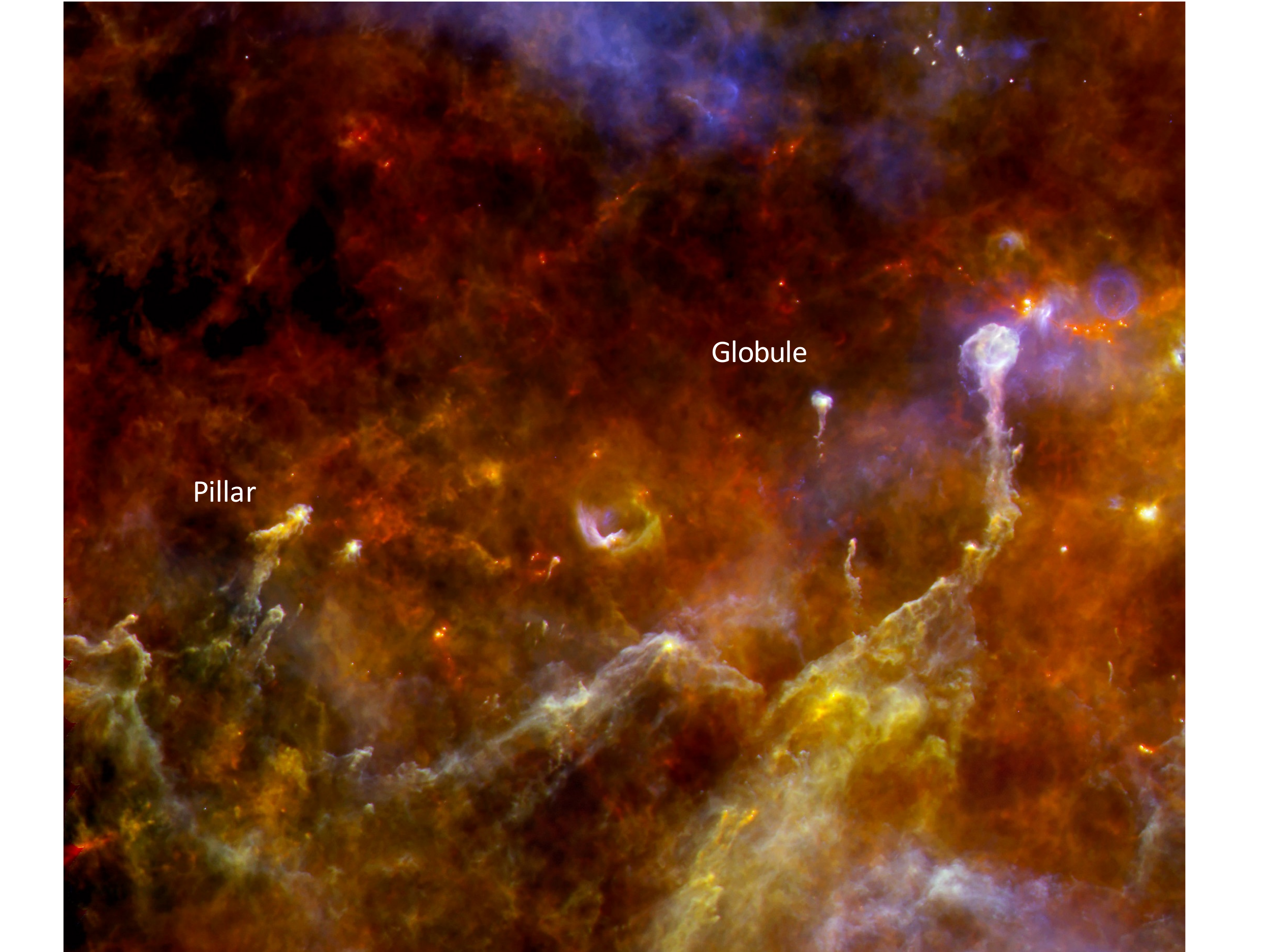}
\caption [] {Three-colour (blue: 70 $\mu$m, green: 160 $\mu$m, red: 250
  $\mu$m) image \citep{schneider2016} of the environment of Cyg OB2
  with the globule IRAS 20319+3958, labeled as 'globule'. The size of
  the image is $\sim$1.5$^\circ \times$1.4$^\circ$, which corresponds
  to $\sim$36 pc $\times$ 34 pc, assuming a distance of 1.4 kpc. The
  pillar indicated further east will be presented in another
  study. The most massive stars of the Cyg OB2 association are located
  in the northwest corner of the image (north is up, east to the
  left).}
\label{overview}
\end{center} 
\end{figure}

In the vicinity of massive stars, intriguing structures such as
column-shaped pillars and cometary-shaped globules are frequently
detected in optical as well as near- and far-infrared images
\citep[e.g.][]{schneps1980,hester1996,white1997,schneider2016}. They
are mostly the result of feedback processes -- ionisation and stellar
winds -- from massive stars and point toward the illuminating source.
Larger pillars, however, can also reflect the primordial cloud
structure \citep{white1999,lefloch2008,miao2009} or arise from eroded
convergent flows \citep{dale2015}.  Pillars still have a physical
connection to the gas reservoir of the molecular cloud, while globules
are isolated features.  Figure~\ref{overview} shows an example for
such objects in the Cygnus X region \citep{schneider2016}. \\
External UV-radiation creates photodissociation regions (PDRs) on the
surfaces of pillars and globules, often visible as a bright rim,
facing the ionising source. A rare example of a globule suggested to
be also illuminated internally by massive stars is IRAS 20319+3958 in
Cygnus X \citep{schneider2012,schneider2016,djupvik2017}.
This source (hereafter, 'the globule') is explored in this paper. Other
examples of star-formation activity within globules are found in
\citet{comeron1999}. The globule is also impacted by the radiation of
the Cyg OB2 cluster located at a projected distance of $\sim$30 pc
from the globule, assuming that this cluster is located at a distance
of 1.4 kpc. It is notoriously difficult to derive distances in Cygnus
X, see \citet{schneider2006,comeron2020} for more detailed
discussions. We use the value of 1.4 kpc, which is based on maser
observations \citep{rygl2012} of prominent star-forming sites in
Cygnus X.  An analysis of GAIA data \citep{lim2020} of Cyg OB2,
however, arrived at a larger distance of 1.6 kpc.

Many details of the formation and evolution of globules and pillars
have not yet been settled. Numerical modelling started out with simple models
of photo-ionisation \citep[e.g.][]{lefloch1994} that
explained basic properties such as shapes or lifetimes
\citep{johnstone1998}, while the concept of radiative-driven implosion
\citep{bertoldi1989} provides an explanation of how stars can form in
pillars and globules. More realistic models with a careful treatment
of heating and cooling processes \citep[e.g.][]{miao2006} that
consider the turbulent structure of the gas
\citep{gritschneder2009} and the curvature of the cloud surface
\citep{tremblin2012a,tremblin2012b} now have the capacity to explain additional properties,
such as the velocity field of the observed features. Furthermore, new
observations, including {\sl Herschel} imaging and spectroscopy in the
far-infrared (FIR) and SOFIA (Stratospheric Observatory for
Far-Infrared Astronomy) FIR spectroscopy, make it possible to establish a
classification scheme and a possible evolutionary sequence. In the
first part of this series of papers \citep{schneider2016}, we set up a
categorisation based on {\sl Herschel} 70 $\mu$m photometry and we
propose that pillars advance into globules, which, in turn, evolve into
evaporating gaseous globules (EGGs), dense gas condensations without
star-formation, or objects with protoplanetary disks (proplyds); or
they only resemble proplyds (proplyd-like), depending on density and
incident UV-field. In the second paper \citep{djupvik2017}, we carried
out optical and near-IR imaging and spectroscopy of the globule in
order to obtain a census of its stellar content and the nature of its
embedded sources.

In this work, we present spectroscopic observations of FIR cooling lines of the
globule in the southern part of Cygnus X \citep{reipurth2008}, where
the very massive and rich Cyg OB2 association illuminates the
molecular cloud.  The globule was mapped in the \CII\, line with SOFIA
\citep{schneider2012} as well as with {\sl Herschel}/HIFI, PACS, and SPIRE (this
paper). It was covered in {\sl Herschel} imaging observations of
Cygnus X within the Herschel imaging survey of OB Young
Stellar objects (HOBYS, \citealt{motte2010}) and was also shown and discussed in
\citet{schneider2016}. Figure~\ref{overview} displays a three-colour image
of the {\sl Herschel} photometry data with our source indicated.

The objective of this paper is to study the spatial emission
distribution of various PDR tracers and to perform a careful analysis
of line intensities and ratios using the KOSMA-$\tau$ PDR model
\citep{roellig2006} to disentangle external (Cyg~OB2) and internal
excitation sources.  We intend to show that it is possible to explain
most of the observed lines in a PDR model considering the geometry of
the source.  This approach is more sophisticated than studies
that assume plane-parallel, homogeneous layers of gas.  There are not
many sources holding such a large data set of cooling lines in the
mm- to FIR. All the line intensities are given in the accompanying tables and the maps can
be provided on demand (the HIFI \CII\ data is already provided via CDS, see
Sect.~\ref{obs:herschel-spectro}), offering the possibility for other
applications and studies. \\ Another goal is to study the dynamics of
the globule. The velocity-resolved extended \CII\ map suggests that
the globule rotates and that high-velocity outflowing gas escapes from
the globule head (some features were shown in \citet{schneider2012},
but not in such detail).
  
We present the various data sets ({\sl Herschel}, SOFIA, FCRAO, JCMT)
in Sect.~\ref{obs}, including a consistency check between FIR
  line intensities observed with {\sl Herschel} and SOFIA. We give
an overview of what is already known about the globule in
Sect.~\ref{studies}.  In Section~\ref{results}, we present the so far
unpublished \CII\ HIFI and \OI\ upGREAT maps of the
globule. Section~\ref{discuss} provides details about the PDR modelling
and discusses the PDR properties of the globule.
Section~\ref{summary} presents our summary.

\begin{table}[h]
  \begin{center}
    \caption{Summary of the HIFI, PACS, and SPIRE spectroscopic
  observations.}
\begin{small}
\begin{tabular}{ccccc}
\hline
\hline
Obsid & Date & $\alpha_{J2000}$ & $\delta_{J2000}$ & Map size\\
& & [$^{h}$:$^{m}$:$^{s}$] & [$^{\circ}$:$^{\prime}$:$^{\prime\prime}$] & [$^{\prime\prime}$]\\
\hline
\bf{HIFI}\tablefootmark{a} & & &  & \\
1342235079 & 12/2011 & 20:33:51.0  & 40:08:49.0  & 113.4$\times$110.0 \\
1342244954 & 04/2012 & 20:33:48.1  & 40:06:36.05 & 97.2$\times$159.5  \\
1342244957 & 04/2012 & 20:37:50.0  & 39:49:23.5  & 113.4$\times$258.5 \\
1342246343 & 05/2012 & 20:37:46.0  & 39:42:0.0   & 108.0$\times$132.0 \\
1342246344 & 05/2012 & 20:37:42.3  & 39:51:25.1  & 135.0$\times$132.0 \\
\hline
\bf{PACS}\tablefootmark{b} & & & & \\
1342211184 & 12/2010 & 20:33:50.0  & 40:08:36.0  & 47$\times$47\\
1342234987 & 12/2012 & 20:37:45.0  & 39:41:58.0  & 47$\times$47\\
1342235855 & 01/2012 & 20:37:42.0  & 39:51:27.0  & 47$\times$47\\
           &         &            &            & \\
\hline
\bf{SPIRE}\tablefootmark{c} & & & & \\
1342231073 & 04/2012 & 20:33:50.0 & 40:08:36.2 & 200$\times$200 \\   
1342231074 & 04/2012 & 20:33:49.0 & 40:06:40.2 & 200$\times$200 \\   
 &  &  &  &  \\   
\end{tabular}
\tablefoot{
\tablefoottext{a}{The HIFI map covers the whole globule.}
\tablefoottext{b}{The PACS maps focus on the globule head.}
\tablefoottext{c}{SPIRE pointed towards a position in the
  globule head (obsID 1342231073) and the tail (obsID 1342231074).}
}
\label{herschel}
\end{small}
\end{center}
\end{table}


\section{Observations} \label{obs}  

\subsection{Herschel spectroscopy} \label{obs:herschel-spectro} 

Far-infrared spectroscopic observations of the globule were performed
with the \emph{Herschel Space Observatory} \citep{pilbratt2010}, using
the instruments HIFI \citep{graauw2010}, PACS \citep{poglitsch2010},
and SPIRE \citep{griffin2010} within the framework of the Herschel
Open Time priority 1 project (ot1\_nschneid\_1) \emph{Pillars of
  creation: physical origin and connection to star formation} (PI
N. Schneider). Table~\ref{herschel} summarises the observational
parameters, such as observation date, central position, and map size.
All the data is available in the
\href{http://archives.esac.esa.int/hsa/whsa}{{\sl Herschel} science
  archive}.  For convenience, we provide the HIFI \CII\ data cube and
line integrated intensity at the
\href{http://cdsweb.u-strasbg.fr/cgi-bin/qcat?J/A+A/}{Centre de donn\'ees
  astronomiques de Strasbourg (CDS)}.

\subsubsection{HIFI} \label{obs:hifi}
The HIFI data consist of Nyquist-sampled, position switched on-the-fly
(OTF) maps of the \CII\ line at 158 $\mu$m with a beam size of
12.2$^{\prime\prime}$ in band 7b at 1910 GHz.  We employed the wide-band
spectrometer (WBS) with an local oscillator (LO) frequency of 1897.662
GHz. The WBS has a full intermediate frequency (IF)
bandwidth of 4 GHz at a spectral resolution of 1.1 MHz (corresponding
to a velocity resolution of 0.7 km s$^{-1}$), in both horizontal (H)
and vertical (V) polarisations. The frequency range covered in band 7b
is 1892.6 GHz to 1895.2 GHz in the lower sideband (LSB) and 1899.9 GHz
to 1902.5 GHz in the upper sideband (USB). The \emph{Herschel
  Interactive Processing Environment, (HIPE)} version 8.2 was used to
remove standing waves and to convert the observed data to CLASS
fits-format and the GILDAS
packages\footnote{http://www.iram.fr/IRAMFR/GILDAS/} were used for all
further procedures (baseline subtractions, line fittings etc.). In
order to obtain a better signal-to-noise ratio (S/N), we averaged the
H and V polarisations. The final processed Level 2 data is scaled in
$T^{'}_{A}$. To scale our data to $T_{mb}$ we multiplied $T^{'}_{A}$
by the factor of $\mathrm{\eta_{l}/\eta_{mb}}$ where
$\mathrm{\eta_{l}}$ is the forward efficiency (0.96) and
$\mathrm{\eta_{mb}}$ is the main beam efficiency (0.69) in band
7b. The overall calibration accuracy is $\sim$10\%
\citep{roelfsema2012}.

\subsubsection{PACS} \label{obs:pacs}
For the PACS range spectroscopy of the globule head, we used the
integral field spectrometer to investigate important cooling lines,
namely, the \OI\, 63 $\mu$m and 145 $\mu$m lines, the \NII\, 122 $\mu$m
line, and high-J CO lines. The data were observed in two wavelength
ranges: from 51 to 73 $\mu$m (blue side) and from 110 to 208 $\mu$m
(red side). The pipeline processes and all data reduction steps
(baseline subtraction, line fitting etc.) were done with HIPE version
7.0 via the built-in pipeline scripts. The line flux measurements were
done as described in \citet{schneider2012}, using the PACSman software
\citep{lebouteiller2012}. We note that because the PACS maps are
contaminated by emission in the off-position, the calibration was
derived only from on-source data and thus associated with a larger
error ($\sim$30\%). For more details, see also Sect.~\ref{compare}, where we compare
several FIR lines that were observed with {\sl Herschel} and SOFIA.
The PACS  maps of the globule head are displayed in Fig. A.1 in Appendix A. 

\subsubsection{SPIRE} \label{obs:spire}
The spectra were taken with the SPIRE Fourier Transform Spectrometer (FTS)
long and short wavelength receivers (SLW and SSW, respectively) at two
positions with sparse sampling. One position is located in the globule
head and one in the tail (see Table~\ref{herschel}). The SLW observes
a hexagonal pattern of 19 spectral pixels (spaxels) covering the
wavelength range 313-671 $\mu$m and the SWS observes 37 spaxels
covering 194-312 $\mu$m. The beam size at the receiver’s central
spaxels varies between 16$''$-20$''$ for SSW and 31$''$-43$''$ for SLW.

The SPIRE-FTS data were downloaded from the Herschel Science Archive,
processed using HIPE v14.1, with SPIRE calibration files
spire\_cal\_14\_3. The FTS extended source calibration
\citep{swinyard2014} was used, and in HIPE v14.1, this includes the
updates described in \citep{valtchanov2018} to align the absolute
brightness level to match the SPIRE photometer.

The lines were fitted using the Spectrometer Cube Fitting script in
HIPE v15, which carries out a simultaneous fit of the spectral lines
and continuum in each band for every spaxel in the spectral cube.
The SPIRE maps of the globule head and tail are displayed in Fig. A.2 and A.3
in Appendix A. 

\subsection{Herschel imaging} \label{obs:herschel-imaging} 
We used {\sl Herschel} imaging observation of PACS at 70 $\mu$m and 160
$\mu$m, and SPIRE at 250 $\mu$m, 350 $\mu$m, and 500 $\mu$m obtained
within the HOBYS guaranteed time Key Program \citep{motte2010}. The
angular resolution of the data varies between 6$''$ and 36$''$ (see
Table~\ref{observations}).  Column density and dust temperature maps,
both at an angular resolution of 36$'',$ were produced with a
pixel-by-pixel SED fit to the wavelengths 160 $\mu$m to 500 $\mu$m as
described in \citet{schneider2016}.

\begin{table}[] 
 \centering 
 \caption{Summary of the observational data sets of the globule.}
 \begin{tabular}{lccccc} 
\hline 
\hline 
 {\small Instrument}  &  {\small Species}                     & $\lambda$ &  $\nu$      & $\Delta$ {\rm v} & $\Theta$ \\ 
            &                             &  [$\mu$m] &  [GHz]      &   [km/s]     &  [$''$]   \\  
\hline 
\multicolumn{2}{l}{{\sl \bf {Herschel spectroscopy}}} &             &         &   \\
 {\small HIFI}        & {\small \CII}                & 157.7    & 1900.5    &  0.7    &  12.2      \\  
 {\small PACS}        & {\small \CII}                & 157.7    & 1900.5    &   -     &  $\sim$11  \\        
 {\small PACS}        & {\small \OI}                 & 145.5    & 2060.1    &   -     &  $\sim$10  \\        
 {\small PACS}        & {\small \OI}                 &  63.2    & 4744.8    &   -     &  $\sim$9.5 \\
 {\small PACS}        & {\small \NII}                & 121.9    & 2459.3    &   -     &  $\sim$9.4     \\        
 {\small PACS}        & {\small $^{12}$CO 16$\to$15}  & 162.8    & 1841.4    &  -      &  $\sim$11.5 \\          
 {\small PACS}        & {\small $^{12}$CO 14$\to$13}  & 186.0    & 1611.8    &  -      &  $\sim$12.5 \\          
 {\small PACS}        & {\small $^{12}$CO 13$\to$12}  & 200.3    & 1496.9    &  -      &  $\sim$13 \\         
 {\small SPIRE}       & {\small \CI}                 & 370.4     &  809.3    &  -     &  34.8     \\
 {\small SPIRE}       & {\small \CI}                 & 609.1     &  492.2    &  -     &  37.2     \\
 {\small SPIRE}       & {\small \NII}                & 205.2     & 1461.1    &  -     &  16.9     \\  
 {\small SPIRE}       & {\small $^{12}$CO 13$\to$12}  & 200.3     & 1496.9   &  -      &  16.8     \\  
 {\small SPIRE}       & {\small $^{12}$CO 12$\to$11}  & 216.9     & 1382.0   &  -      &  17.2      \\  
 {\small SPIRE}       & {\small $^{12}$CO 11$\to$10}  & 236.6     & 1267.0   &  -      &  17.6     \\ 
 {\small SPIRE}       & {\small $^{12}$CO 10$\to$9}   & 260.2     & 1152.0   &  -      &  17.7     \\   
 {\small SPIRE}       & {\small $^{12}$CO 9$\to$8}    & 289.1     & 1036.9   &  -      &  19.2     \\  
 {\small SPIRE}       & {\small $^{12}$CO 8$\to$7}    & 325.2     &  921.8   &  -      &  36.8     \\
 {\small SPIRE}       & {\small $^{12}$CO 7$\to$6}    & 371.7     &  806.7   &  -      &  34.8     \\
 {\small SPIRE}       & {\small $^{12}$CO 6$\to$5}    & 433.6     &  691.5   &  -      &  29.4     \\
 {\small SPIRE}       & {\small $^{12}$CO 5$\to$4}    & 520.3     &  576.3   &  -      &  32.6     \\   
 {\small SPIRE}       & {\small $^{12}$CO 4$\to$3}    & 650.3     &  461.0   &  -      &  40.4     \\       
 {\small SPIRE}       & {\small $^{13}$CO 9$\to$8}    & 302.4     &  988.8   &  -      &  36.1     \\  
 {\small SPIRE}       & {\small $^{13}$CO 8$\to$7}    & 340.2     &  881.3   &  -      &  36.1     \\  
 {\small SPIRE}       & {\small $^{13}$CO 7$\to$6}    & 388.7     &  771.2   &  -      &  34.0     \\  
 {\small SPIRE}       & {\small $^{13}$CO 6$\to$5}    & 453.5     &  661.1   &  -      &  30.0     \\  
 {\small SPIRE}       & {\small $^{13}$CO 5$\to$4}    & 544.2     &  550.9   &  -      &  32.9     \\  
\hline 
\multicolumn{2}{l}{{\sl \bf {Herschel photometry}}}    &             &        &   \\
 {\small PACS}        & {\small continuum}           &   70      &  4283       &   -    &   6.0      \\ 
 {\small PACS}        & {\small continuum}           &  160      &  1874       &   -    &  11.4      \\
 {\small SPIRE}       & {\small continuum}           &  250      &  1199       &   -    &  17.8      \\
 {\small SPIRE}       & {\small continuum}           &  350      &   857       &   -    &  25.0      \\
 {\small SPIRE}       & {\small continuum}           &  500      &   600       &   -    &  35.7      \\ 
\hline 
\multicolumn{2}{l}{{\sl \bf {SOFIA}}}                  &             &        &          \\   
 {\small GREAT}       & {\small \CII}              & 157.74      & 1900.5     &  0.23  &  15.1      \\  
 {\small GREAT}       & {\small $^{12}$CO 11$\to$10} & 236.61     & 1267.0     & 0.69   &  22.5      \\        
 {\small upGREAT}     & {\small \OI}               &  63.18      & 4744.8     &  0.25 &   6.1      \\  
 {\small upGREAT}     & {\small $^{12}$CO 16$\to$15} & 162.81     & 1841.4     & 0.64  &  15.3      \\        
\hline 
\multicolumn{2}{l}{{\sl \bf {FCRAO}}}                  &             &        &    \\   
 {\small SEQUOIA}     & {\small $^{13}$CO 1$\to$0}   & 2720.4     &  110.2    & 0.067   &  45       \\        
 {\small SEQUOIA}     & {\small CS 2$\to$1}          & 3059.1    &   98.0     & 0.075  &  48       \\        
\hline 
\multicolumn{2}{l}{{\sl \bf {JCMT}}}                   &             &         &            \\   
 {\small HARP}        & {\small $^{12}$CO  3$\to$2} & 869.0       &  345.8     &   0.42 &   15       \\
            &                              &           &             &         &          \\  
\end{tabular} 
\tablefoot{The first column gives the instrument, the second the line
  (or continuum), the third and forth the wavelength and frequency,
  the fifth the velocity resolution used in this paper, and the six
  column the angular resolution. We note that the PACS and SPIRE FIR
  lines are not spectrally resolved and that the SOFIA and FCRAO line
  observations were smoothed to a lower velocity resolution when used
  in displays of spectral line maps or channel maps. For
  PDR-modelling, we smoothed all data sets with an angular resolution
  higher than 20$''$ to a common value of 20$''$. }
\label{observations} 
\end{table}

\subsection{SOFIA} \label{obs:sofia} 

\subsubsection{GREAT: \CII\ 158 $\mu$m and CO 11$\to$10} \label{obs:sofia-great}
The \CII\, 1.9 THz line and the CO J=11$\to$10 molecular rotation line
at 1.267\,THz were observed with the PI-heterodyne receiver
GREAT\footnote{The German REceiver for Astronomy at Terahertz
  frequencies. GREAT is a development by the MPI f\"ur Radioastronomie
  and the KOSMA/Universit\"at zu K\"oln, in cooperation with the MPI
  f\"ur Sonnensystemforschung and the DLR Institut f\"ur
  Planetenforschung.} \citep{heyminck2012} on SOFIA during one flight on
November 10, 2011 from Palmdale, California. OTF maps of
the globule, with an angular resolution of $\sim$15$''$ for \CII\ and
23$''$ for CO, were produced. This data set was presented in
\citet{schneider2012}. In this paper, we compare the SOFIA \CII\ data
with what was obtained with HIFI and use the line intensity
information of the CO J=11$\to$10 line for PDR modelling.

\begin{figure*}[ht] 
\begin{center}  
\includegraphics[angle=0,width=14cm]{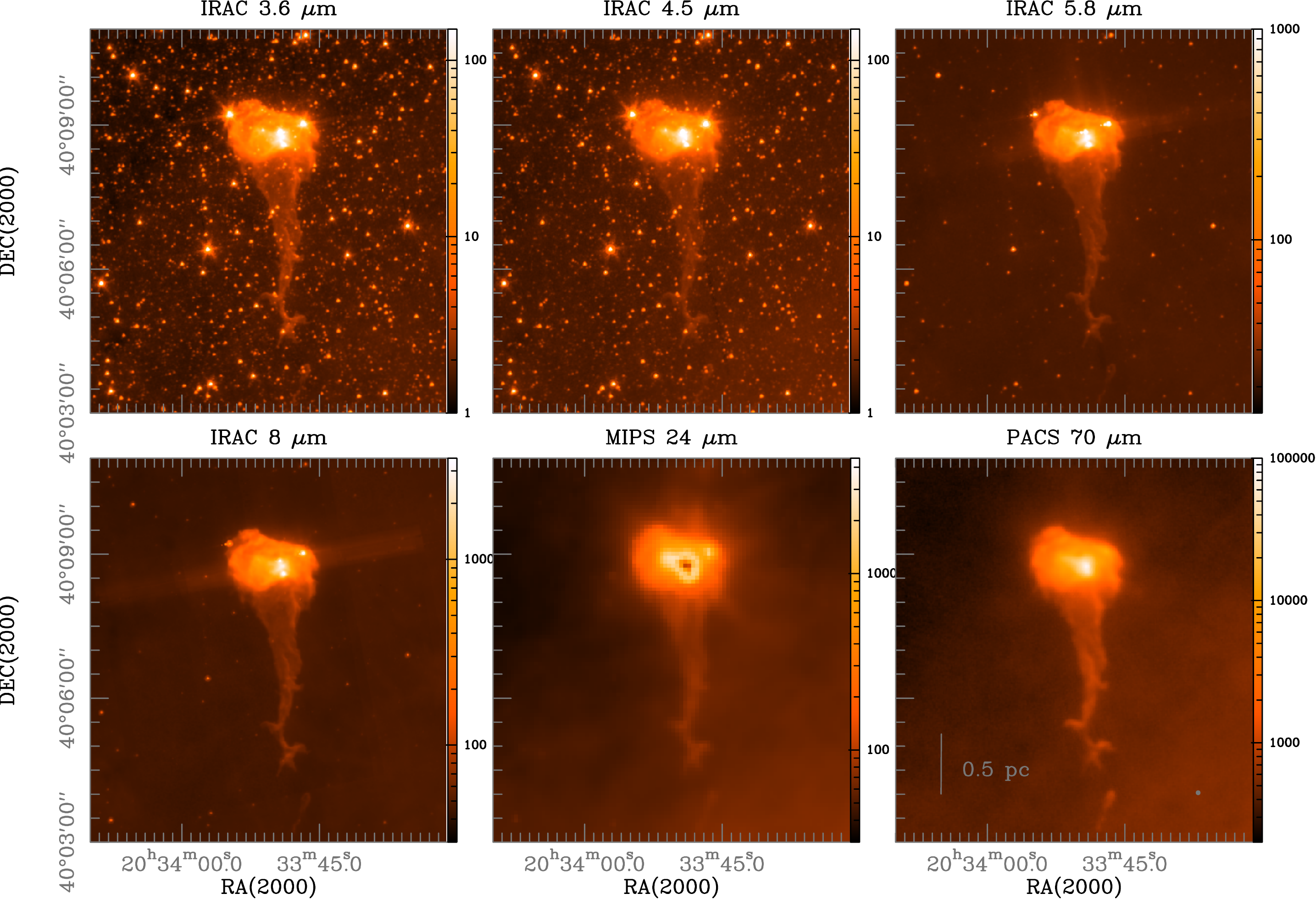}
\includegraphics[angle=0,width=14cm]{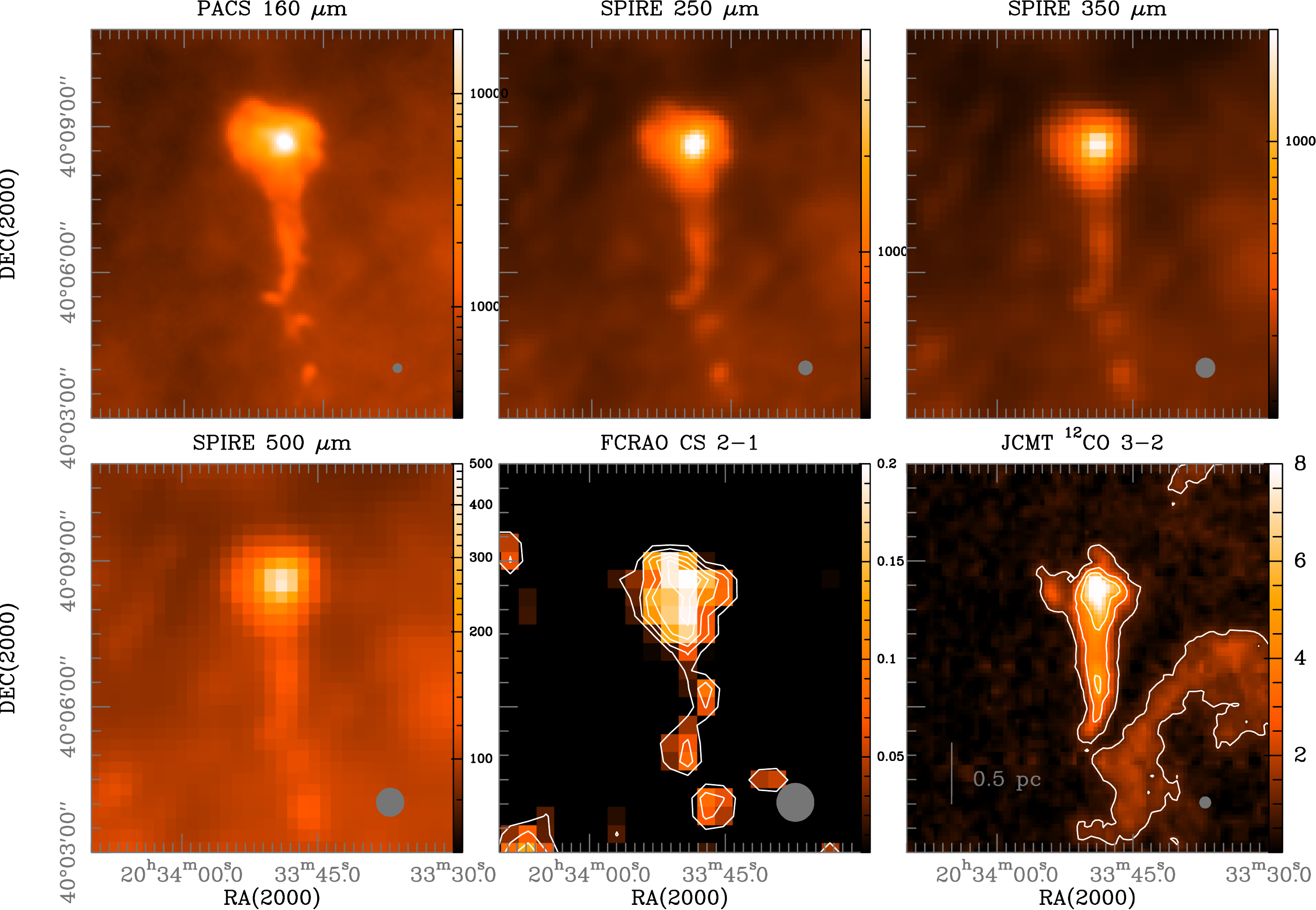}
\caption [] {Globule at IR- and FIR-wavelengths: {\sl
    Spitzer}/IRAC 3.6 to 8 $\mu$m, {\sl Spitzer}/MIPS at 24 $\mu$m,
  {\sl Herschel}/PACS 70, and 160 $\mu$m and SPIRE 250, 350, and 500
  $\mu$m (all units are MJy/sr). The two lower right panels show
  velocity integrated molecular line emission of CS 2$\to$1 and
  $^{12}$CO 3$\to$2 in [K km s$^{-1}$]. The beam is indicated in all
  panels with longer wavelength observations (starting with PACS 160
  $\mu$m) in the lower right corner. The {\sl Spitzer} data were
  already displayed in \citet{djupvik2017} and the {\sl Herschel} data
  in \citet{schneider2016}.}
\label{glob-1} 
\end{center}  
\end{figure*} 

\begin{figure*}[ht] 
\begin{center}  
\includegraphics[angle=0,width=15cm]{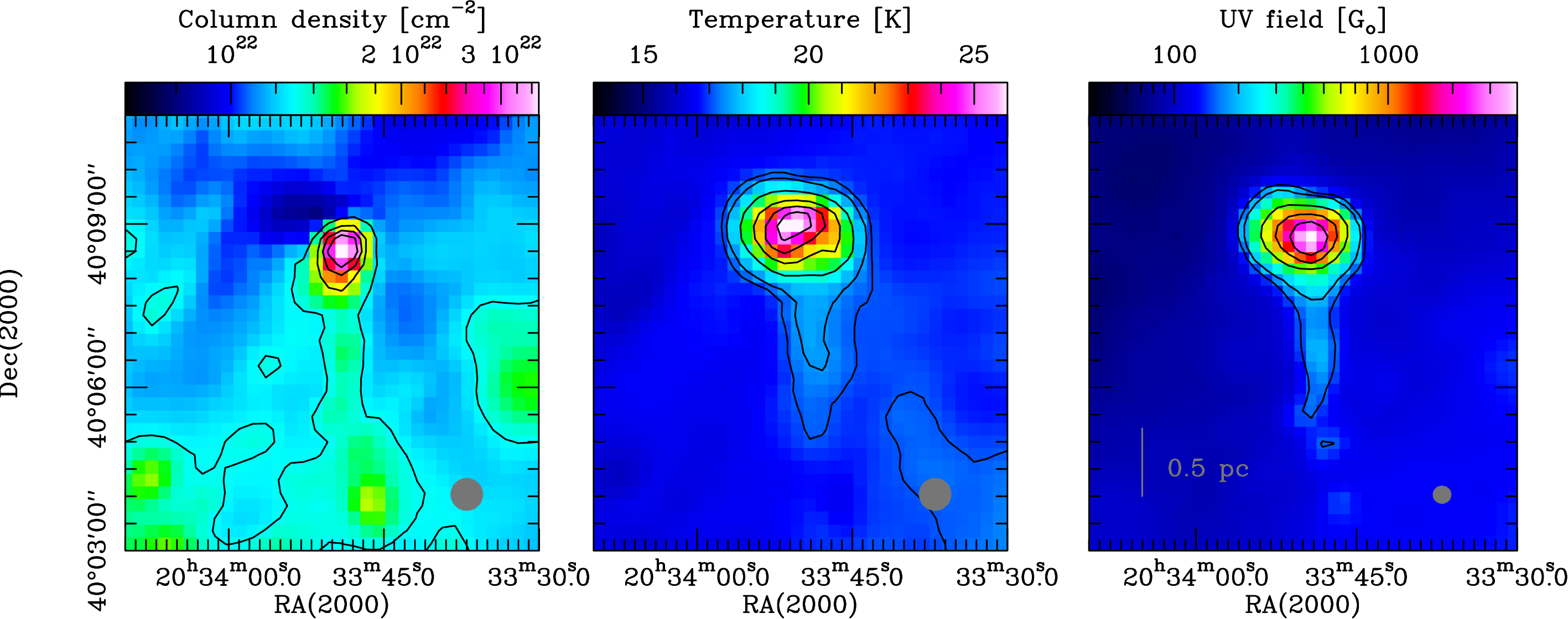}
 \caption [] {{\sl Herschel} view of the globule. From left to right:
   column density (contours 1.3, 1.8, 2.3, 2.8 10$^{22}$ cm$^{-2}$),
   temperature (contours 17.1, 17.5, 19, 21, 23, 25 K), and UV-flux
   map (contours 150, 200, 500, 1000, 2000 G$_\circ$) of the globule
   obtained from {\sl Herschel}. These images are cut-outs from
   figures shown in \citet{schneider2016}.  The column density and
   temperature maps have an angular resolution of 36$''$ and the UV
   map of 20$''$. These resolutions are indicated in the lower right
   corner of each panel.}
\label{glob-cd} 
\end{center}  
\end{figure*} 

\subsubsection{upGREAT: \OI\ 63 $\mu$m and CO 16$\to$15} \label{obs:sofia-upgreat}
The globule was observed on November 2, 2016, during one flight from
Palmdale, California with upGREAT \citep{risacher2016} on SOFIA. Only
the globule head was covered (map size $\sim$100$'' \times$80$''$),
the central position was RA(2000)=20$^h$33$^m$53.0$^s$,
Dec(2000)=40$^\circ$08$'$45$''$.  The seven-pixel HFA array was tuned on
the \OI\ 4.7 THz line, the single pixel L2 channel was tuned on the CO
16$\to$15 line. Both channels observed in parallel, optimised for the
\OI\ line. We note that the observed CO 16$\to$15 map thus has missing
data points because of the single pixel sampling. The observing mode was
chopping single phase A, with a chop amplitude of 100$''$ and a chop
frequency of 0.655 Hz with one cycle per dump. The OTF mapping was
performed with 1 slew per ref and 6 refs/load. The array orientation
was --19.1$^\circ$.  The bandpass averaged system temperature was 2657
K for the L2 channel and 3512 K for the H-array. All line intensities
are reported as main beam temperatures scaled with main-beam
effciencies of 0.69 and 0.68 for \OI\ and CO, respectively, and a
forward effciency of 0.97. The main beam sizes are 15.3$''$ for the L2
channel ($^{12}$CO 16$\to$15) and 6.1$''$ for the HFA channel (\OI,).

\subsection{FCRAO data} \label{obs:fcrao} 
We used molecular line data obtained with the 14m dish of the Five
College Radio Astronomy observatory (FCRAO), employing the single
sideband focal plane array receiver  Second Quabbin Optical
Imaging Array (SEQUOIA). The whole Cygnus X region ($\sim$35 square degrees)
was observed in the $^{13}$CO 1$\to$0 at 110.201 GHz, the C$^{18}$O
1$\to$0 at 109.782 GHz, and the CS line at 98.0 GHz. The beamwidth of
the FCRAO at 110 GHz is 46$''$. More details are found in
\citet{schneider2011}, where the CO data sets are presented.

\subsection{JCMT CO data} \label{obs:jcmt} 
The $^{12}$CO 3$\to$2 data used in this paper were obtained with the
16-pixel array HARP receiver in the B-band, and the ACSIS digital
autocorrelation spectrometer as the backend correlator system.  The
individual beams of HARP have a full width at half maximum (FWHM) of
15$''$ and the beams are spaced 30$''$ apart on a 4$\times$4 grid.
The map of the globule is part of the programs M07BU019 (PI R. Simon)
and M08AU018 (PI N. Schneider) that were carried out in 2007 and 2008
to map large parts of Cygnus X North and South. The data have a
velocity channel spacing of 0.42 km s$^{-1}$.

\begin{table}[h]
  \caption{Comparison of line integrated intensities.}
  \begin{tabular}{lccc}
\hline
\hline
                     & {\small \CII\ 158 $\mu$m}  & {\small \OI\ 63 $\mu$m} & {\small $^{12}$CO 11$\to$10}\\
                     &  [K km s$^{-1}$]   & [K km s$^{-1}$] & [K km s$^{-1}$]\\
\hline
SPIRE/{\sl Herschel} &   -                &   -            &  27.0 @17.6$''$ \\
PACS/{\sl Herschel}  &   68.0 @16$''$     &   4.0 @16$''$  &  - \\
HIFI/{\sl Herschel}  &  108.5 @12$''$     &   -            &  - \\
GREAT/{\sl SOFIA}    &  107.3 @15$''$     &   -            &  27.2 @22$''$\\
upGREAT/{\sl SOFIA}  &   -                &  10.6 @16$''$  &  - \\
 &  &  &  \\
\end{tabular}
\tablefoot{The line integrated intensities are given in main beam
  brightness temperatures with different angular resolutions at one
  common position at RA(2000)=20$^h$33$^m$49$^s$,
  Dec(2000)=40$^\circ$8$'$45$''$ in the globule head for each line in
  the globule observed with SPIRE, PACS, HIFI ({\sl Herschel}), and
  GREAT or upGREAT on SOFIA. }
\label{line-comp}
\end{table}

\subsection{Consistency check between FIR lines observed with PACS, SPIRE, and HIFI ({\sl Herschel}) and (up)GREAT (SOFIA) } \label{compare}  
The globule is a rare example of a source that was observed in various
FIR lines with different instruments on {\sl Herschel} and SOFIA over
the last seven years and thus offers the possibility to compare the
observed line intensities. The estimated total calibration
uncertainties are $\sim$20\% for GREAT and upGREAT
\citep{heyminck2012,risacher2016}, and 10\% for HIFI
\citep{roelfsema2012}.
The SPIRE calibration uncertainty for extended sources was estimated
to be 7\% \citep{swinyard2014}, although when there is structure in
the beam, the uncertainty is larger and dominated by source-beam
coupling \citep{wu2013}. Our data include the latest corrections to
match the FTS extended calibration to the SPIRE Photometer
\citep{valtchanov2018}.
The PACS data suffer from contamination in the off-position so that
the calibration was done only using the on-source data. We thus
estimate that the error on the flux is high ($>$30\%) and the observed
values are upper limits. 

Table~\ref{line-comp} shows a comparison
between the \CII\ 158 $\mu$m line, the \OI\ 63 $\mu$m line, and the CO
11$\to$10 $\mu$m lines, determined at one position in the globule head
at RA(2000)=20$^h$33$^m$49$^s$, Dec(2000)=40$^\circ$8$'$45$''$.  The
flux values obtained for \CII\ for HIFI and GREAT as well as the CO
11$\to$10 line for SPIRE and GREAT agree very well. In contrast, the
PACS values for the \CII\ and the \OI\ line are significantly lower
than those obtained with HIFI/GREAT and upGREAT, respectively, which
cannot be explained by the PACS contamination problem in the
off-position because the PACS values are upper limits. One explanation
can be attributed to positional uncertainties because the \OI\ line emission show a
large spatial pixel-to-pixel variation and pointing differences can
lead to different values. We take the SOFIA \OI\ 63 $\mu$m data for
PDR modelling, but we need to use the PACS \OI\ 145 $\mu$m data since
this line was only observed with {\sl Herschel}. However, this caveat
needs to be kept in mind.

\begin{table}[h]
\caption{Physical properties of the globule in total (column 2) and
  the head and tail (column 3 and 4), respectively.}
  \begin{tabular}{lccc}
\hline
\hline
                                       & {\small Globule}  & {\small Globule head} & {\small Globule tail}\\
\hline
$\langle {\rm N}\rangle$ [10$^{21}$ cm$^{-2}$]       &   14.2            &   14.3               &  14.0 \\
Mass  [M$_\odot$]                       &   238             &   166                &  73 \\
Density [10$^3$ cm$^{-3}$]              &   6.5             &   5.2                &  7.7 \\
$\langle {\rm T}\rangle_{dust}$ [K]                  &  19.7             &   19.7               &  17.0 \\
Length [pc]                            &   1.78            &   0.85               &  0.93 \\
Width  [pc]                            &   0.3-0.9         &   0.9                &  0.3 \\
Area   [pc$^2$]                        &   0.9             &   0.62               &  0.28 \\
average UV-field  [G$_\circ$]          &                  &    550               &  170 \\
peak    UV-field  [G$_\circ$]          &                  &   4300               &  220 \\
incident UV-field [G$_\circ$]          & 150-600           & 150-600              & 150-600 \\
  &  &  &  \\
\end{tabular}
\tablefoot{The values were derived from the {\sl Herschel} column
  density and temperature maps \citep{schneider2016}. A distance of
  1.4 kpc was assumed, so that 1$'$ corresponds to 0.4 pc. The area is
  the equivalent area of a polygon used to derive the mass, density
  etc. $\langle {\rm N} \rangle$ is the average H$_2$ column
  density. The average and peak flux values were derived from the {\sl
    Herschel} fluxes, explained in \citet{schneider2016}, the incident
  UV-flux from a census of the stars from Cyg OB2.}
\label{globule-para}
\end{table}

\section{Multiwavelengths observations of the globule} \label{studies} 

In the following, we shortly present previous works on the globule and 
summarise the most important physical properties in Table~\ref{globule-para}. \\

\noindent {\bf IR- and FIR-data} \\ Figure~\ref{glob-1} displays the
globule in the IR- to FIR-wavelength range (3.6 $\mu$m to 500 $\mu$m,
observed with {\sl Spitzer} and {\sl Herschel}), and in the molecular
lines of CS 2$\to$1 and $^{12}$CO 3$\to$2. The IR data show the
stellar content of the globule and its environment, while the FIR
observations at 70 $\mu$m and 160 $\mu$m have a too low angular
resolution (6$''$ and 11$''$, respectively) to resolve the
(proto)-stars.  The head and tail are well visible at all wavelengths,
indicating that there is warm and cold dust present in both. The IR
traces the hot PAH dust features while the FIR data longer than 160
$\mu$m trace the warm to cold dust.  The tail/head ratio flux peaks
around 160 $\mu$m, but the globule head clearly dominates the emission
at all wavelengths. We note that the tail contains no pre- or
protostellar sources. \\

\noindent {\bf Molecular line data} \\ The two lower right panels of
Fig.~\ref{glob-1} show that the line-integrated $^{12}$CO 3$\to$2 and
CS 2-1 emission arises from the whole globule though the peak of
emission is found in the head. The CS emission is more beam diluted
because of the lower resolution of 46$''$ with respect to CO 3$\to$2
with 15$''$. Intererestingly, there is a lack of CO 3$\to$2 emission in
the northeastern globule head, giving the impression that gas was
blown out of the centre, leaving this hole. The density of the globule
is at least $\sim$10$^4$ cm$^{-3}$ if we assume that the molecular
line emission is thermalised \citep{shirley2015}. \\
 
\noindent {\bf Herschel column density and temperature maps and the UV
  field} \\ High (column) densities are confirmed by the {\sl
  Herschel} dust column density map (Fig.~\ref{glob-cd}, left panel),
which indicates peak values of a few 10$^{22}$ cm$^{-2}$ for the
globule head, \citep[see also][]{schneider2016}, where the column density
map, temperature map, and UV-field map were already shown for a larger
area and the globule was labelled 'g1' in region 1-3. The values for
average H$_2$ column density $\langle {\rm N} \rangle$, mass M,
average density n, length and width, and dust temperature T are given
in Table~\ref{globule-para}.  We note that we averaged across the whole
head as it was originally defined \citep{schneider2016} by 70 $\mu$m
emission, which corresponds to the area seen in the temperature
map. The average values for column density and density are thus lower
than as if we would have only taken the high column density area seen
in Fig.~\ref{glob-cd}. The temperature, also determined from an SED
fit to the {\sl Herschel} data, is 19.7 K, but shows a strong variation
from 17.2 K south of the globule centre to 26.3 K at the centre of the
globule head (Fig.~\ref{glob-cd}, middle panel).

From the {\sl Herschel} 70 $\mu$m and 160 $\mu$m flux,
\citet{schneider2016} calculated an average UV flux across the globule
head of $\sim$550 G$_\circ$ (in units of the Habing
field \footnote{Note that the Habing field G$_\circ$ relates to the
  Draine field $\chi$ by $\chi=1.71~{\rm G}_\circ$ where G$_\circ$ is
  the mean interstellar radiation field from
  \citet{habing1968,draine1978}. We use both measures in this paper.})
and a peak value (in a 20$''$ beam) of $\sim$4300 G$_\circ$.
From a census of the exciting stars of the Cyg OB2 cluster, the
incident UV-field on the globule is only 150-600 G$_\circ$ (not
accounting for extinction by the molecular clouds of the Cygnus X
region and ignoring projection effects). We thus anticipate that
internal sources must also contribute to the measured UV field from
the {\sl Herschel} fluxes. \\

\noindent{\bf Stellar content inside the globule } \\ Earlier studies
reported nine cluster members within a projected radius of $\sim$0.5
pc \citep{kronberger2006,kumar2006} and two visible stars had been
estimated to have mid-B spectral types \citep{cohen1989}. The scenario
of embedded stars was further explored in \citet{djupvik2017}, where
we found that the globule contains an embedded aggregate of about
30-40 young stellar objects within one arcmin (or 0.4 pc with the
adopted distance of 1.4 kpc). Based on the high ratio of Class I to
Class II objects, the small cluster was estimated to have an age $<$1
Myr.  The most massive members were designated stars A, B, and C. Star
A was discovered to be a binary with one component being a Herbig Be
star with an estimated mass of 13 M$_\odot$. Star B was found to have
spectral type B0.5 to B1.5 and an estimated mass of 23 M$_\odot$. The
bright mid-IR Class I source, Star C, was resolved in a binary, of
which at least one is a massive YSO of spectral type late O or early B
with 8.1 M$_\odot$.  Optical spectroscopy of the nebula next to these
stars revealed clear signs of a low-excitation \HII\ region, as one
would expect from early B-type stars rather than the harder radiation
from O stars in the nearby Cygnus OB2 association.  Furthermore, the
morphology seen in high-angular-resolution images of H$_2$ line
emission tracing the PDR, and Br-$\gamma$ line emission tracing the
ionised gas, was interpreted as additional evidence that the globule
is illuminated from the inside.

\section{Results and analysis} \label{results} 

Here, we present a study of many cooling lines in the mm- to
FIR-wavelength range, which all arise from photodissociation regions
(PDRs).  The hot (T$>$100 K) PDR component is best traced in the
cooling lines of atomic oxygen (\OI\, at 63 and 145 $\mu$m) and high-J
CO rotational lines. The warm (T$\sim$100 K) layer of the PDR is seen
in the 158 $\mu$m line of ionised carbon, followed by the
fine-structure lines of neutral carbon (\CI\ 1$\to$0 and 2$\to$1 at
609 $\mu$m and 370 $\mu$m, respectively). The cool (T$<50$ K)
molecular cloud is traced in low-J CO lines. Apart from the gas
temperature and UV field, it is also the density in the PDR that
determines which of the lines is the dominant cooling line. It is a
major challenge to correctly reproduce the observed line intensities
in PDR models. Some models focus on establishing a careful chemical
network, while others emphasise geometrical effects, such as
considering the inhomogeneous structure of the PDR
\citep[e.g.,][]{tielens1985,black1987,lebourlot1993,kaufman1999,sternberg1995,wolfire2003,meijerink2005,roellig2006,bisbas2015}.
\citet{roellig2007} and \citet{bisbas2015} give an overview of the
various PDR codes with a comprehensive reference list.

\begin{figure}[ht] 
\begin{center}  
\includegraphics[angle=0,width=8.5cm]{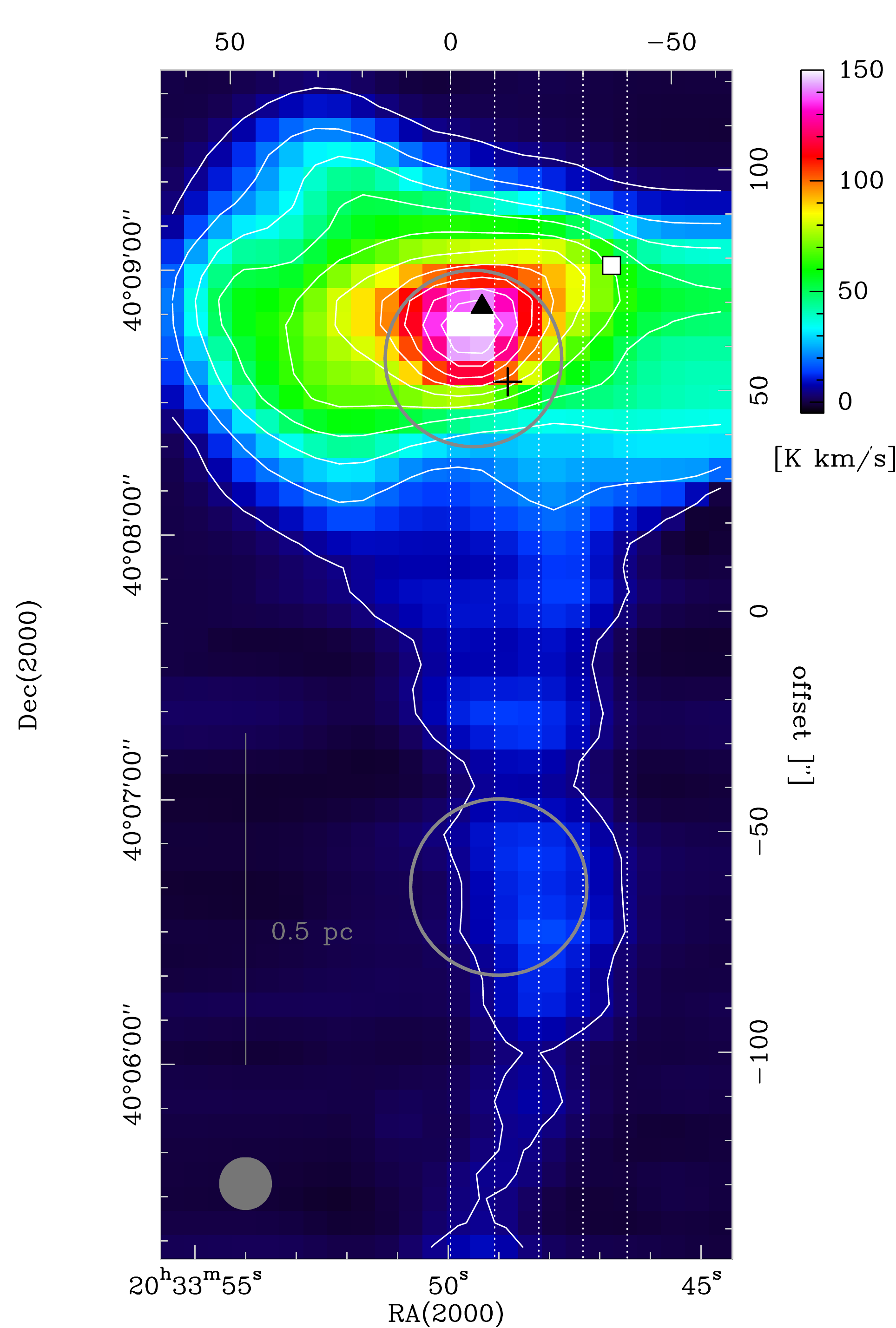}
\caption [] {Line-integrated \CII\ 158 $\mu$m emission of the globule
  obtained with HIFI on {\sl Herschel} in the velocity range 0 to 15
  km s$^{-1}$. The black triangle indicates the double system (Star A)
  of which at least one is a Herbig Be star, the white rectangle
  points to Star B with a B0.5 - B1.5 spectral type, and the black
  cross marks Star C, a resolved binary of which one is a late O or
  early B star. The centre position (0,0) is at
  RA(2000)=20$^h$33$^m$49.95, Dec(2000)=40$^\circ$07$'$42.75$''$. The
  \CII\ beam is indicated in the lower left corner.  The dashed lines
  indicate the vertical (north-south) cuts where we performed
  position-velocity maps and the dark grey circles mark the positions
  and the extend of the SPIRE beam for the longest wavelengths (FWHM
  $\sim$40$''$) we used for PDR modelling. The position in the north
  is the globule 'head' and the one in the south the globule 'tail'.}
\label{cii_cut} 
\end{center}  
\end{figure} 

\begin{figure}[ht] 
\begin{center}  
  \includegraphics[angle=0,width=11cm]{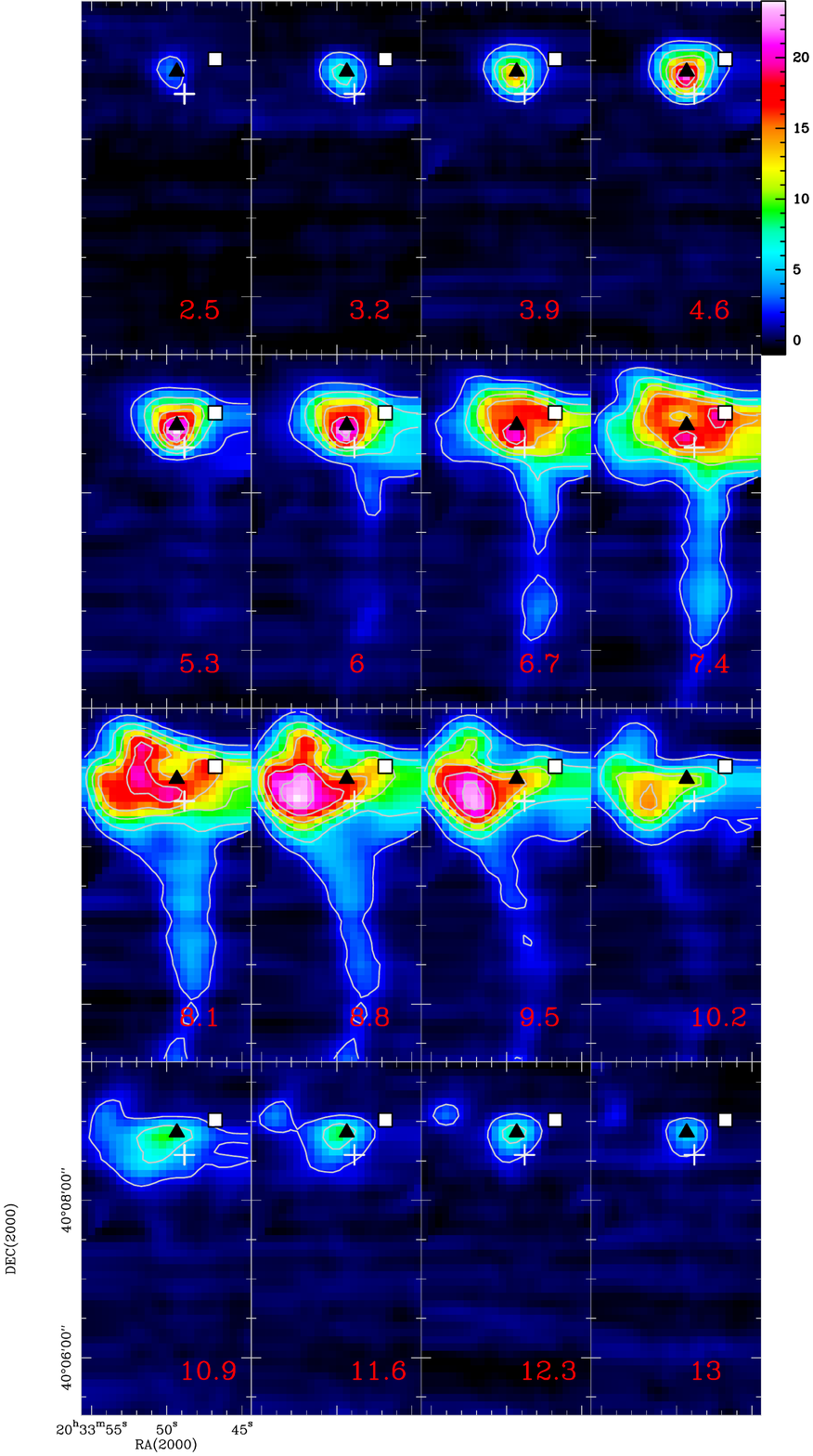}
\caption [] {Channel maps of \CII\ emisson obtained with HIFI on {\sl
    Herschel} from 2.5 km s$^{-1}$ to 13 km s$^{-1}$. The symbols
  indicate the three stellar systems, similar to Fig.~\ref{cii_cut},
  except that here Star C is indicated with a white cross for better
  visibility.}
\label{cii_channel} 
\end{center}  
\end{figure}

\begin{figure}[ht] 
\begin{center}  
  \includegraphics[angle=0,width=7cm]{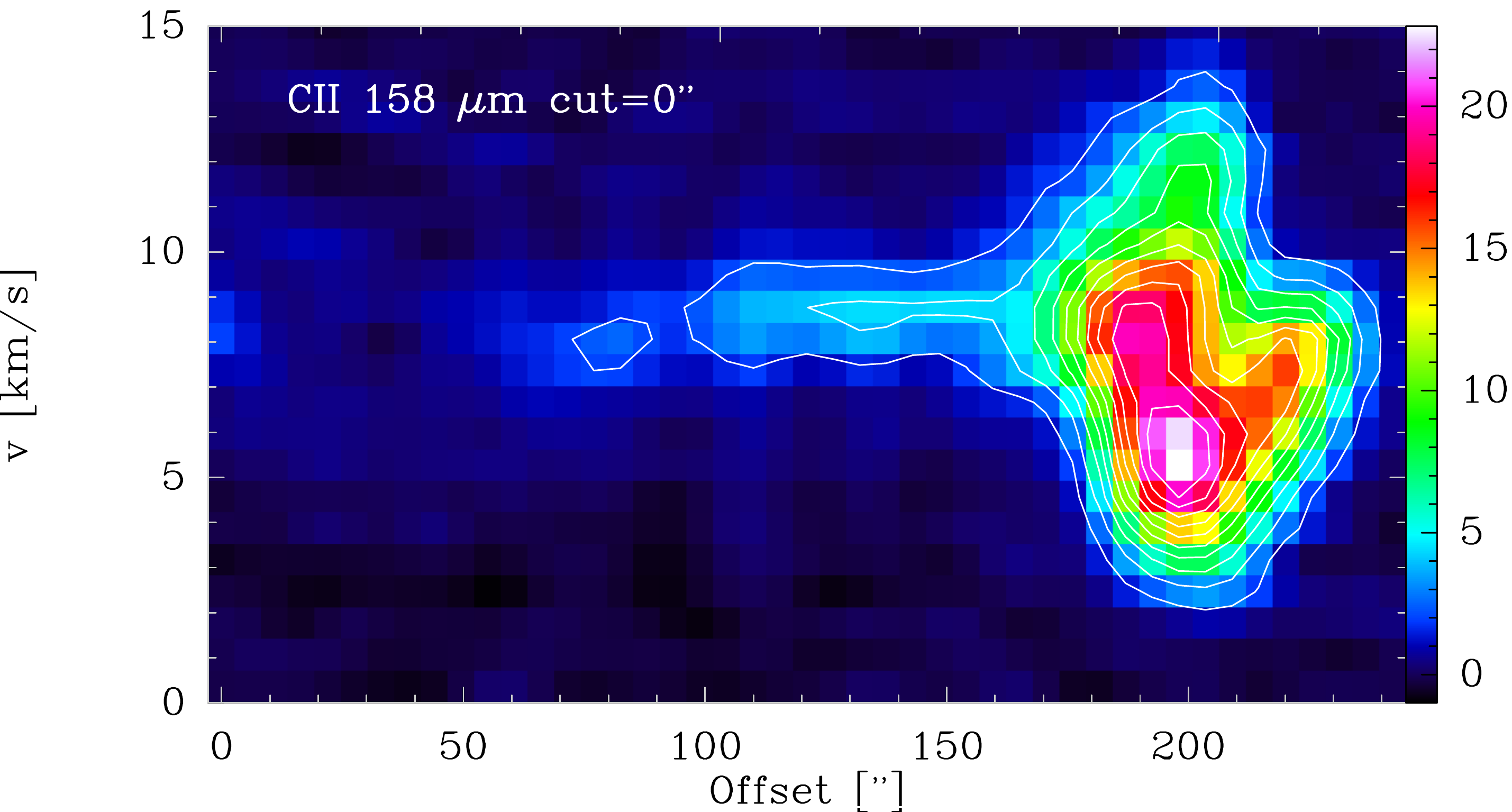}
  \includegraphics[angle=0,width=7cm]{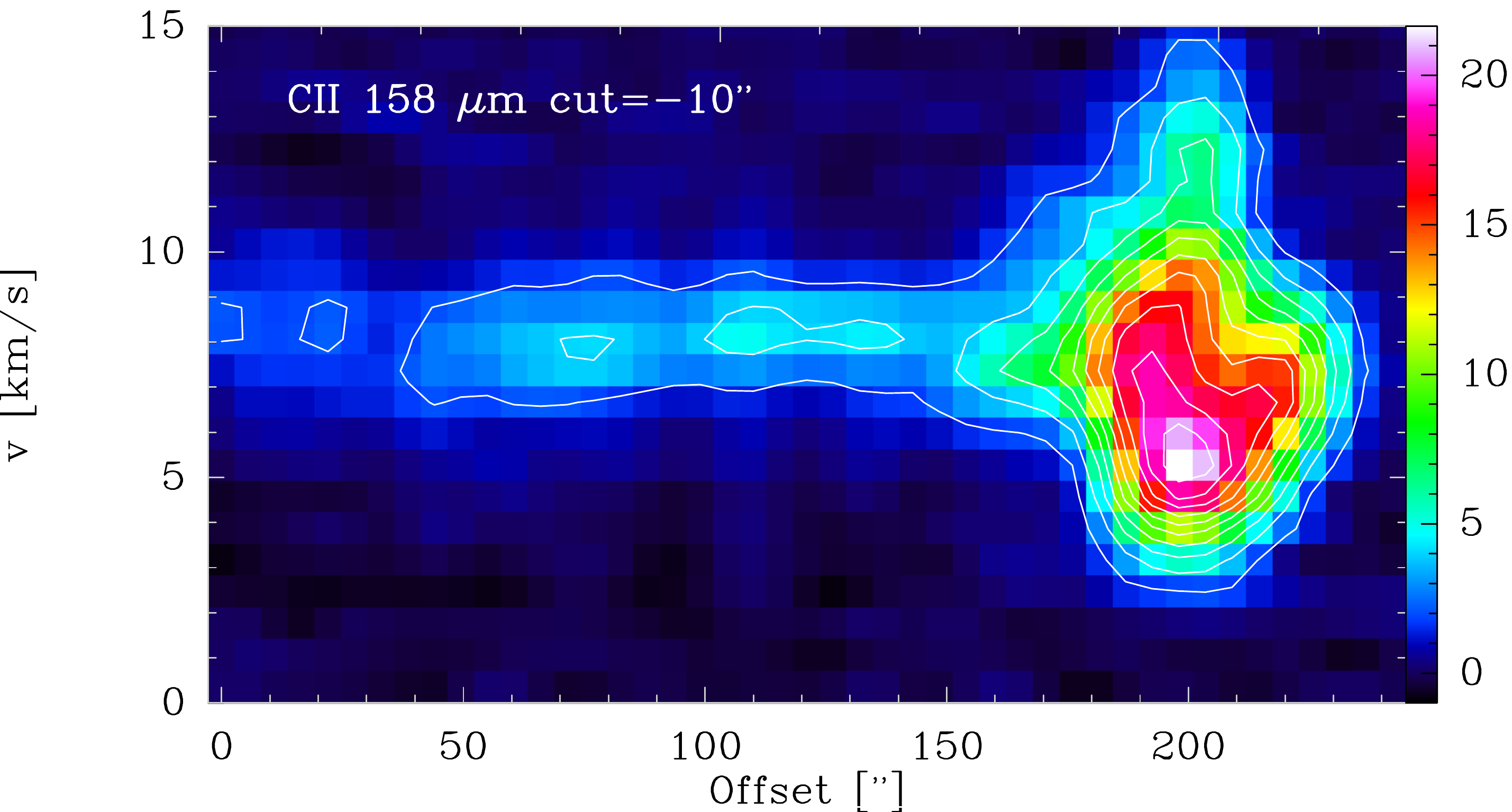}
  \vspace{-1.0cm} 
  \includegraphics[angle=0,width=7cm]{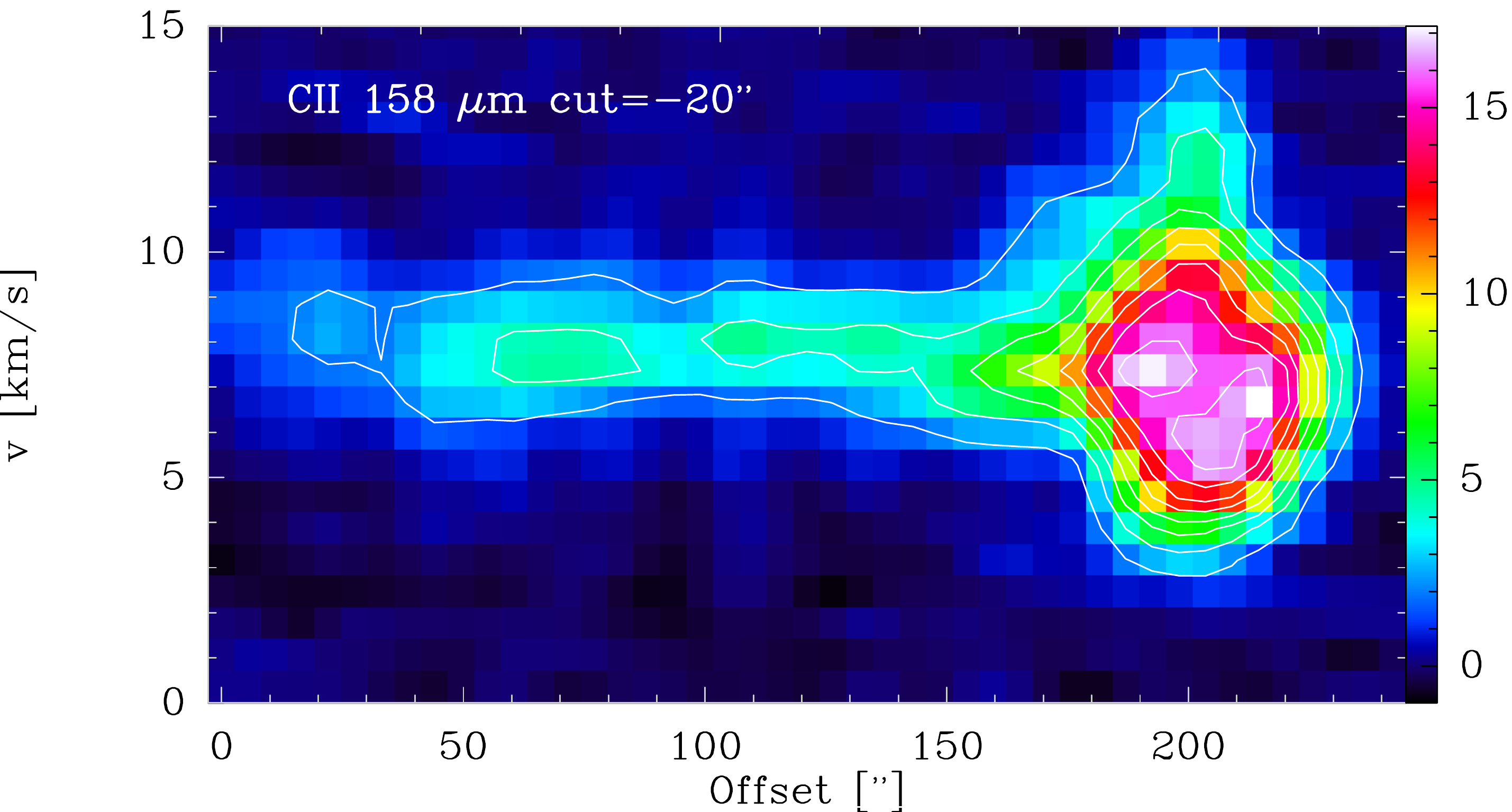}
  \includegraphics[angle=0,width=7cm]{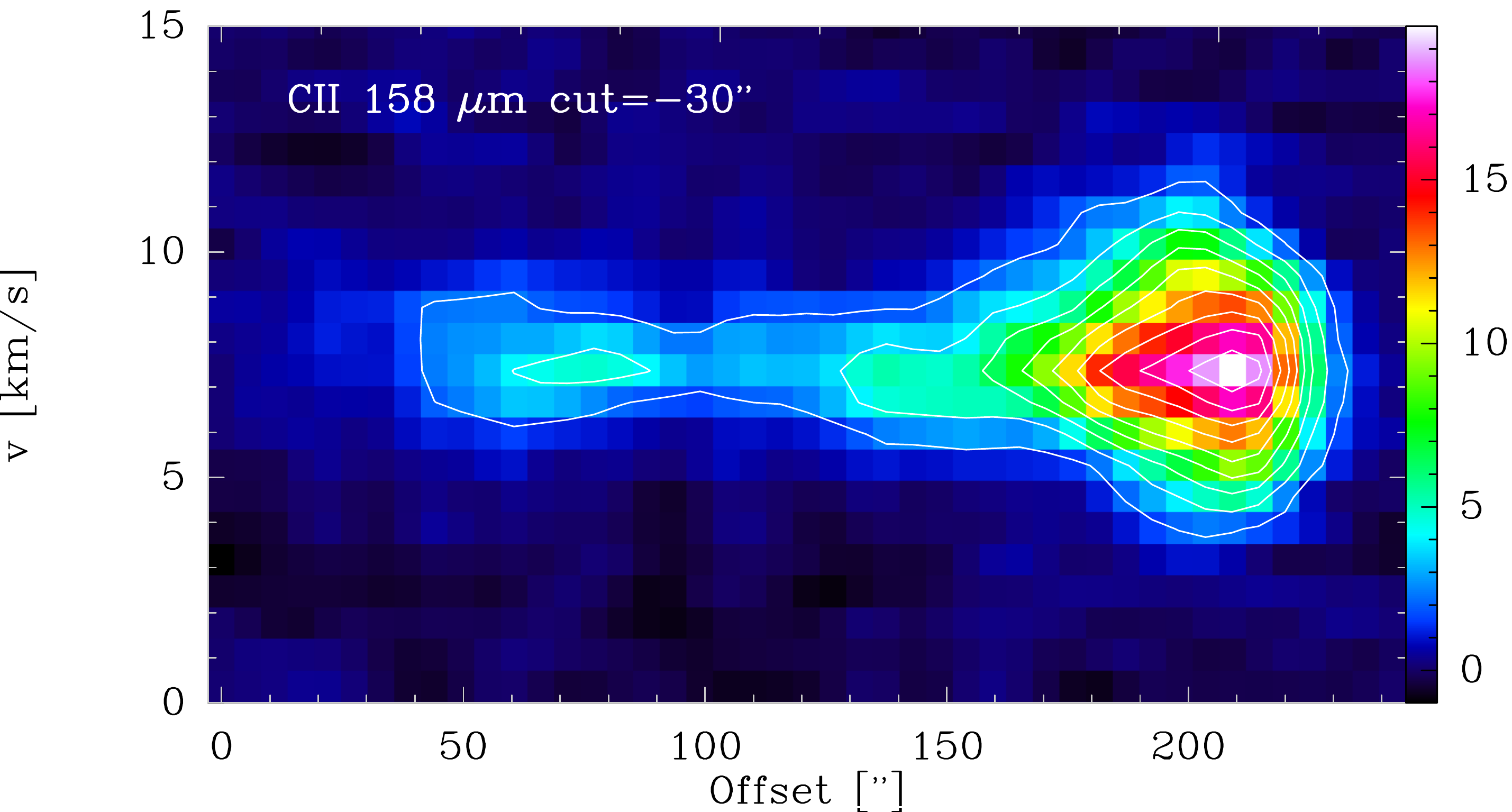} 
  \includegraphics[angle=0,width=7cm]{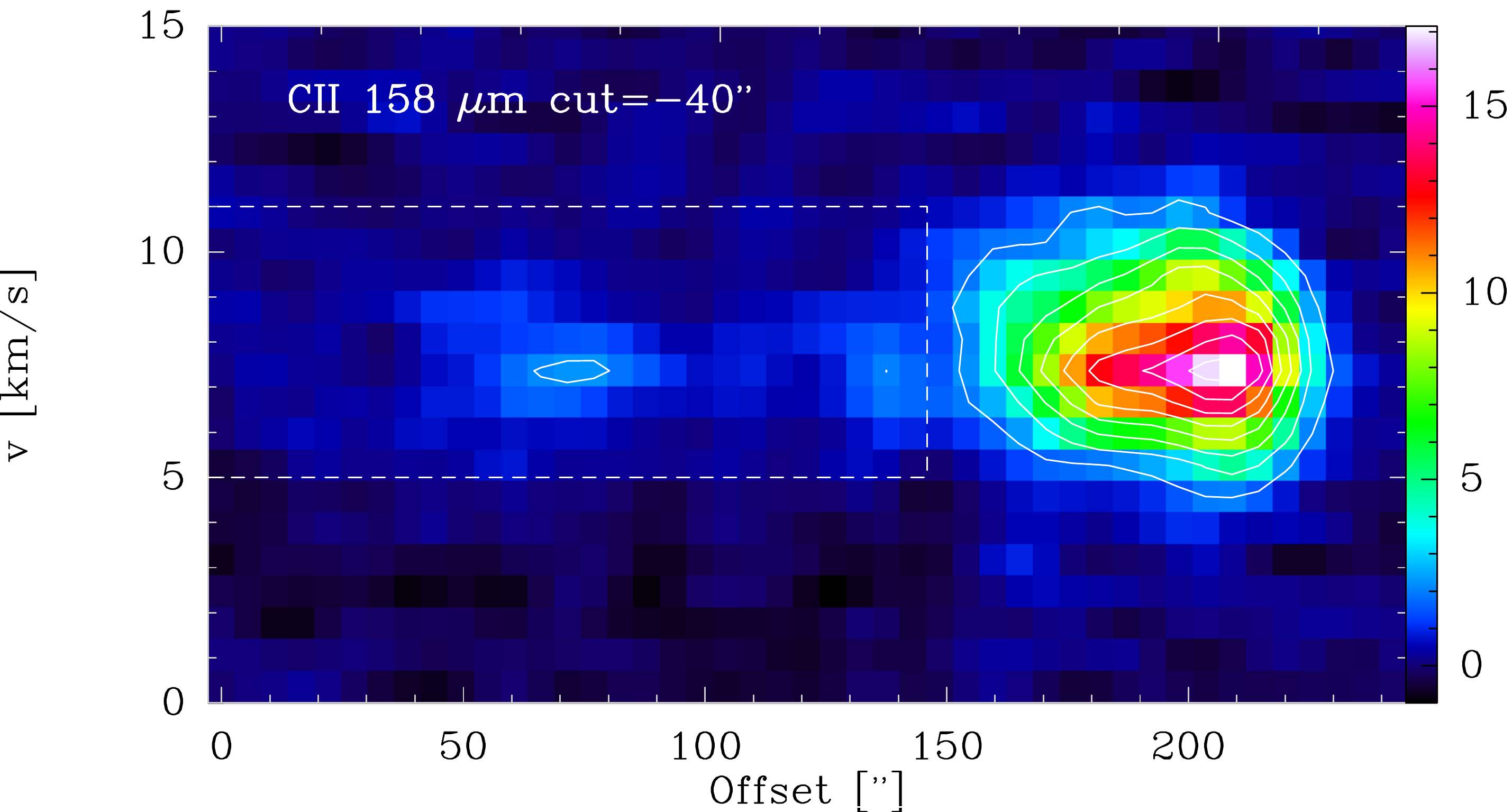}
\caption [] {Position-velocity maps of the globule in \CII\ for 5
  vertical cuts cuts in declination at different RA offsets, indicated
  in Fig.~\ref{cii_cut}. The zero-offset is at RA(2000) =
  20$^h$33$^m$50$^s$ and goes then in steps of 10$''$.  The Dec (2000)
  range is 40$^\circ$05$'$15$''$ to 40$^\circ$09$'$45$''$.  The bottom
  panel for cut=-40$''$ outlines with a white dashed rectangle the
  globule's tail for which we show an overlay between two PV cuts in
  Fig.~\ref{cut-overlay}.}
\label{cii-pv} 
\end{center}  
\end{figure} 

\begin{figure}[ht] 
\begin{center}  
\includegraphics[angle=0,width=8cm]{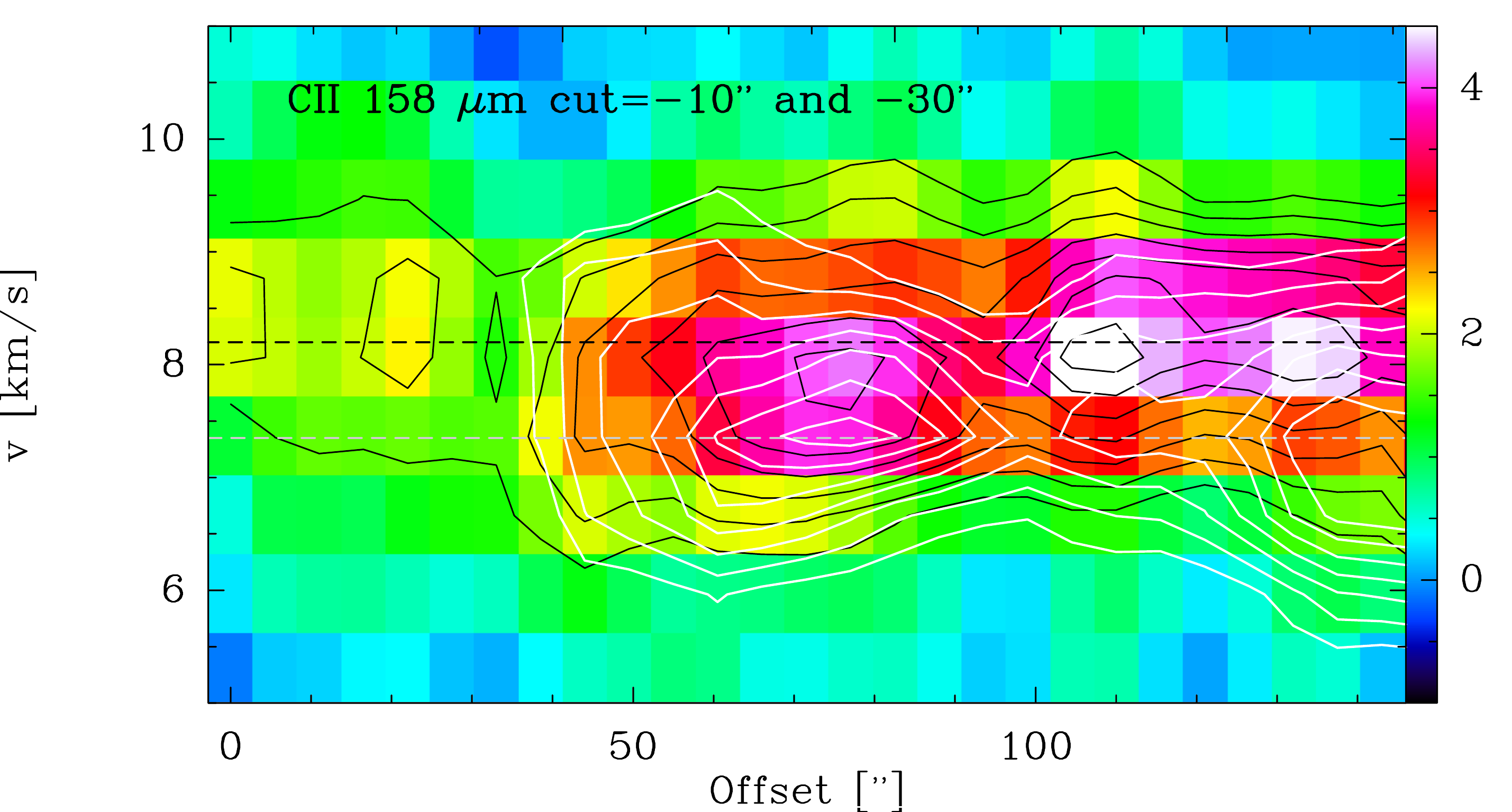}
\caption [] {Overlay between PV maps of the globule in \CII\ for the
  -10$''$ (black contours) and -30$''$ (white contours) vertical
  cuts, indicated in Fig.~\ref{cii_cut}. The dashed lines show the
  centre velocity of the globule's tail for the cuts, i.e. 8.20 km
  s$^{-1}$ for the -10$''$ cut and 7.35 km s$^{-1}$ for the -30$''$
  cut, respectively.}
\label{cut-overlay} 
\end{center}  
\end{figure}

\begin{figure}[ht] 
\begin{center}  
\includegraphics[angle=0,width=8cm]{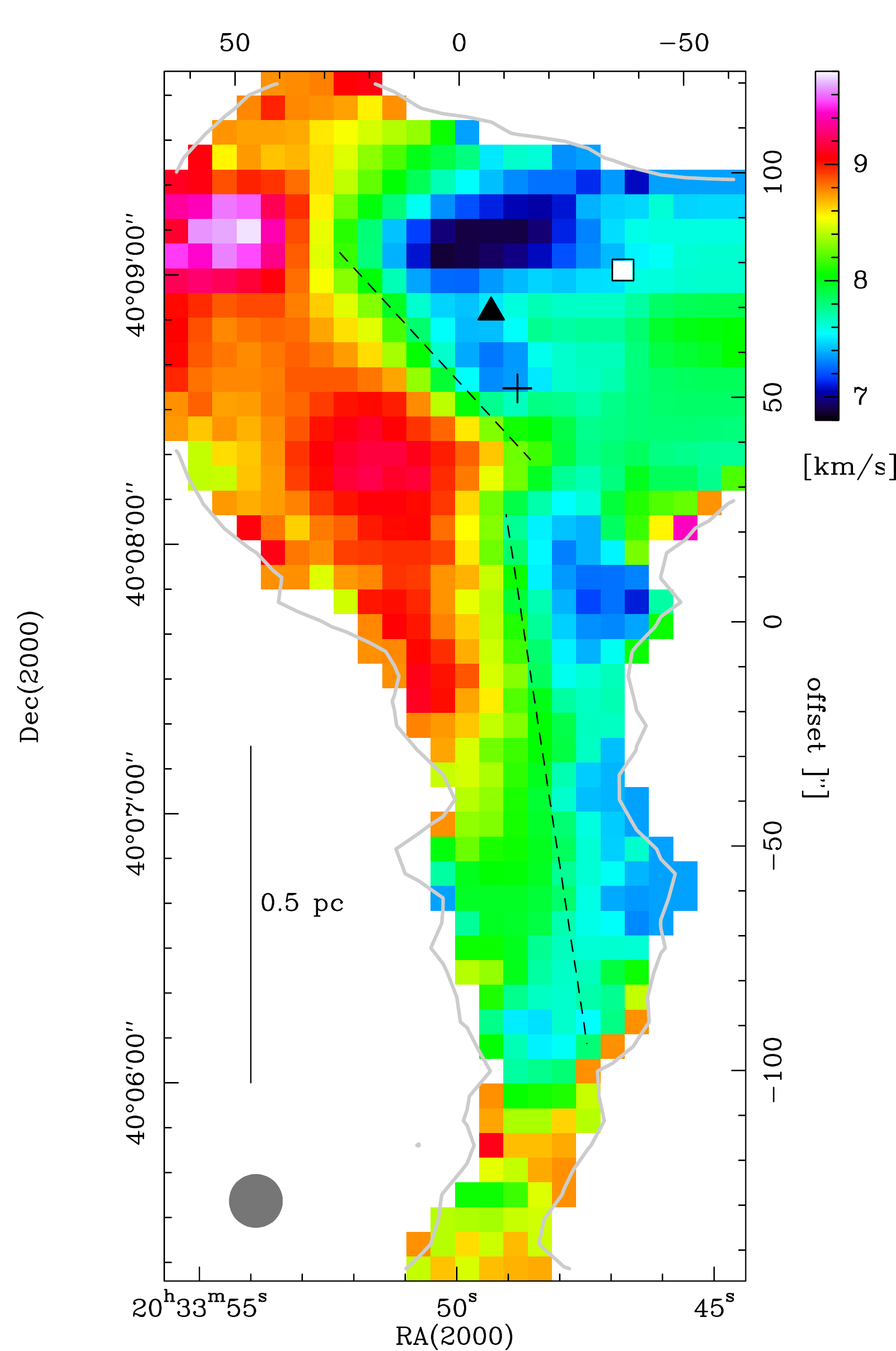}
\caption [] {Velocity map (first moment) of the \CII\ emission,
  showing possible patterns of counter-clockwise rotation with the
  approximate axes indicated with dashed lines.  The centre
    position (0,0) is at RA(2000)=20$^h$33$^m$49.95,
    Dec(2000)=40$^\circ$07$'$42.75$''$. The \CII\ beam is indicated in
    the lower left corner.  We note that the axes for the globule head,
  where most of the mass resides, and the tail have not the same
  inclinations, and the velocity difference is larger for the head
  ($\sim$9.5 to $\sim$6.9 km s$^{-1}$) than for the tail ($\sim$8.3 to
  $\sim$7.4 km s$^{-1}$). The grey contour outlines the 2 K km
  s$^{-1}$ contour level of \CII\ emission, the symbols are the same
  as in Fig.~\ref{cii_cut}.}
\label{rotate} 
\end{center}  
\end{figure} 

\subsection{{\sl Herschel}/HIFI \CII\ 158 $\mu$m emission distribution} \label{cii-hifi} 

Figures~\ref{cii_cut} and \ref{cii_channel} show the spatial and
velocity distribution of the \CII\ emission in the globule observed
with HIFI on {\sl Herschel}. The overall emission in the globule head
is similar to what was found in \citet{schneider2012}, based on
GREAT/SOFIA observations. We note that the globule tail was not observed
with SOFIA in 2012. The channel maps (Fig.~\ref{cii_channel}) reveal
the correlation between \CII\ emission and the stellar content. Star A
(the Herbig Be star, black triangle) displays a clear correlation
between high velocity \CII\ emission in the blue (2.5-5.3 km s$^{-1}$)
and red (11.6-13 km s$^{-1}$) velocity range. We observe outflowing
gas from the inner PDR region around Star A that had already created a
small cavity because of its stellar wind and radiation
\citep{schneider2012}. This point is discussed in more detail in
Sect.~\ref{cii-outflow}.  Interestingly, the two other star systems do
not display a strong correlation with \CII\ emission. Star B (the
single early B-star, white rectangle) lies outside of significant
\CII\ emission for all velocity channels, while Star C (binary with a
late O star, grey cross) can also serve as an exciting source for
ionising carbon in the PDR region. The channel maps show that the bulk
emission of the globule between $\sim$6.7 km s$^{-1}$ and 10.2 km
s$^{-1}$ has first a prominent peak enclosing the three stellar
systems, then forms an arc-like structure (v=8.1 km s$^{-1}$) and then
develops a single peak south-east of the stars. \\ The globule tail is
fainter in \CII\ emission. The integrated intensity in the tail
(Fig.~\ref{cii_cut}) is typically $\sim$10 K km s$^{-1}$, which is a
factor of 10-15 smaller than what is found in the globule head.
Nevertheless, this level of \CII\ emission indicates that there is
some external heating (there are no exciting sources in the tail) from
the overall FUV field around the globule, mostly caused by the Cyg OB2
cluster.

\subsection{\CII\ column density and mass} \label{cii-mass} 

We calculate the mass associated to the gas traced by \CII\ emission
using a simplified version of the formula for the \CII\ column density
in the limit of optically thin emission given by Eq.~2 in
\citet{langer2010} or Eq.~26 in \citet{goldsmith2012} that is
\begin{equation}
  N(C^+) = 2.9 \times 10^{15} \, (1 + 0.5 \, e^{91.25/T_{kin}} (1 + \frac{2.4 \times 10^{-6}}{R_{ul} \, n})) \, I(C^+) \, [cm^{-2}],
\end{equation}   
with $R_{ul}$ as the collisional de-excitation rate coefficient at a
kinetic temperature $T_{kin}$ and density $n$ of the collisional partner, which may be either electrons or atomic or
molecular hydrogen. In addition, $I(C^+)$ is the line integrated observed \CII\ intensity.
As a first-order approximation, we assume high kinetic temperatures and high densities, so that this
equation is simplified to:
\begin{equation}
  N(C^+) = 4.38 \times 10^{15} \, I(C^+) [cm^{-2}].
\end{equation}
We distinguish between the gas entrained in the outflow in the globule
head, namely, \CII\ emission in the blue (v=0 to 5 km s$^{-1}$) and red
(v=12 to 15 km s$^{-1}$) velocity range, and the bulk emission (v=5 to
12 km s$^{-1}$).  For the bulk emission, we derive an average
\CII\ column density of 2.0$\times 10^{17}$ cm$^{-2}$ and 0.4$\times
10^{17}$ cm$^{-2}$ for the globule's head and tail,
respectively. Assuming an abundance of C/H=1.2$\times$10$^{-4}$
\citep{wakelam2008}, we estimate a mass of $\sim$13.63 M$_\odot$ and
$\sim$1.55 M$_\odot$ for the globule head and tail, respectively. The
gas mass in the outflow is 1.57 M$_\odot$ and 0.68 M$_\odot$ for the
blue and red velocity range, respectively.  We note that the masses
derived from \CII\ are lower limits and lower than what was derived
from the dust.  However, the values are not directly comparable. While
dust emission is optically thin and traces all gas along the
line-of-sight, the \CII\ emission can be optically thick and arise
mostly from the PDR surface.

\subsection{Large-scale dynamics of the globule} \label{cii-posv} 

The dynamics of the globule head was already discussed in
\citet{schneider2012}. We confirm with the HIFI \CII\ data the
detection of a velocity gradient and differences in line position and
width within the globule head (cf. Fig.~\ref{cii_channel}).
%
Figure~\ref{cii-pv} displays position-velocity (PV) plots for five
vertical cuts through the whole globule that are indicated in
Fig.~\ref{cii_cut}. The cuts at offsets 0 and -10$''$ cross Star A and
we observe - similar to the channel maps - high velocity blue- and
red-shifted emission and an opening in the globule head with an
arc-like structure. The cuts further away from Star A show more
confined emission spatially and kinematically and reflect the
primordial velocity structure of the globule with gas around 7 km
s$^{-1}$. The globule tail does not show a velocity gradient along its
main axis (north-south orientation) in the individual cuts, but there
is an east-west velocity gradient. This becomes obvious in an overlay
between the emission in the cuts at -10$''$ and -30$''$
(Fig.~\ref{cut-overlay}) and is also seen in the channel maps of
Fig.~\ref{cii_channel}.  The 'blue' part of the \CII\ tail at 7.4 km
s$^{-1}$ is located further west compared to the 'red' part of the
\CII\ tail visible at higher velocities around 8.1-8.8 km s$^{-1}$.

Figure~\ref{rotate} displays the velocity pattern of the whole globule
in more detail, this time including the tail, which was not observed
in \CII\ before. The globule head shows a possible rotation feature
with an inclined north-south axis, similar what was seen in
\citet{schneider2012}.  The globule tail also shows a possible
rotation, but the axis is less inclined and the velocity difference is
smaller. The rotation is clock-wise around an axis located between the
blue and red part of the tail if we view the globule from above.

Possible rotation features in pillars were observed before
\citep{gahm2006,sofue2020}. \citet{gahm2006} proposed for 'elephant
trunks' in several sources observed in CO lines that rotation can be
provoked by the formation of compressed magnetic filaments that were
present in the parent molecular cloud and are now impacted by the
expanding \HII\ region. They developed a double helix model in which a
pillar rotates as a solid body with the same angular speed along the
major axis.  This picture would be consistent with the observations
for our globule.  In this case, it is possible that the globule
started as a pillar (proposed in \citet{schneider2016}), which was
linked to the bulk emission of the molecular cloud and had the same
velocity. After the pillar detaches from the cloud, it becomes a
globule that floats freely into space but still carries the initial
momentum of the pillar. Simulations \citep{tremblin2012b} predict that
pillars and globules in the same close environment have a velocity
difference of typically a few km s$^{-1}$. This applies also to our
globule, as shown in \citet{schneider2012}. Other models of UV
radiation impacting a turbulent cloud \citep{gritschneder2009} display a
slightly different picture. The UV-radiation clears out and pushes
away the low-density material of the cloud but leaves dense
structures, such as pillars (see their Fig.~1).
The extent to which magnetic fields and the impact of the external
UV-radiation can also influence the velocity field seen in \CII\ is
not clear and this area requires further investigation.

\begin{figure}[ht] 
\begin{center}  
\includegraphics[angle=0,width=4.4cm]{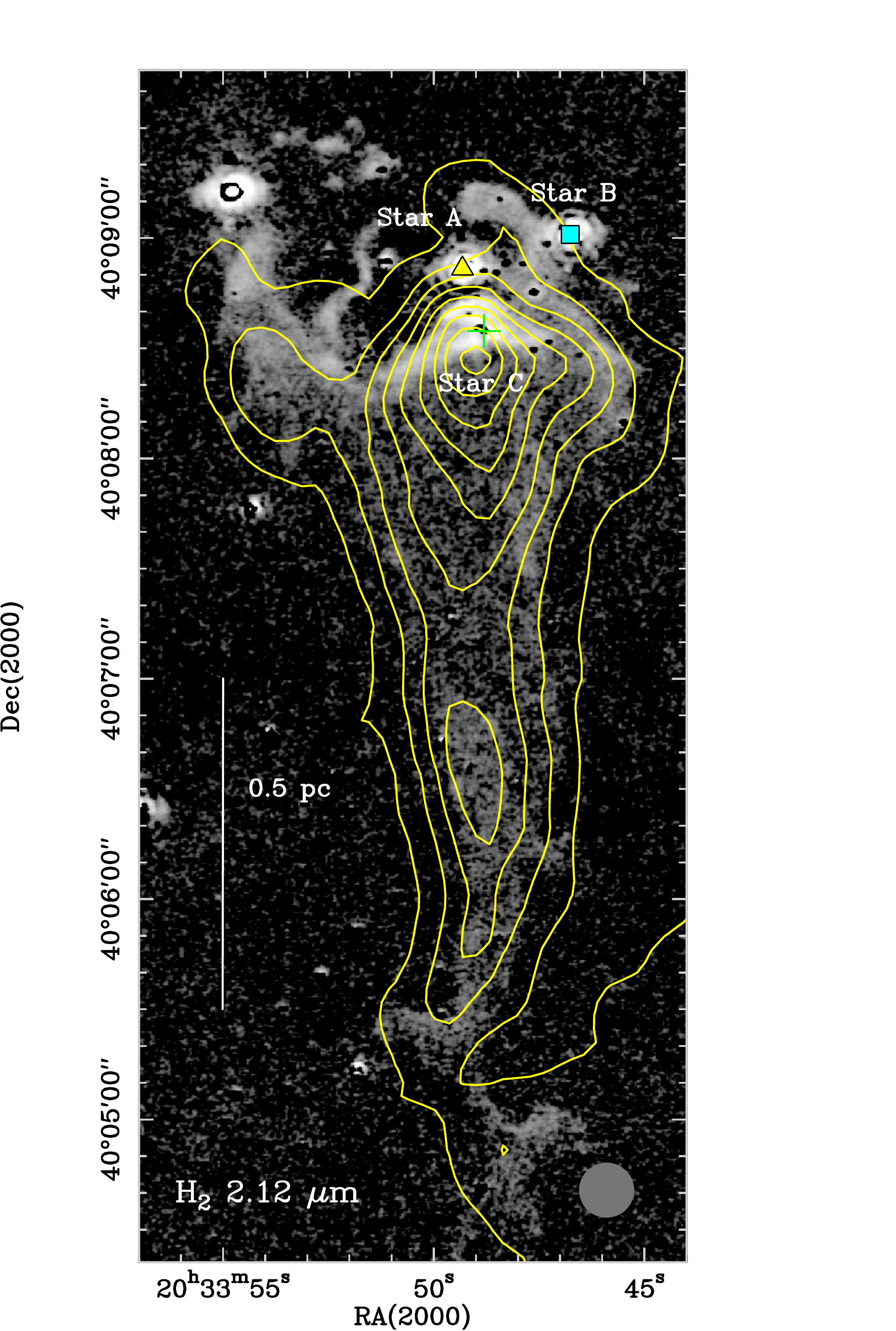}
\includegraphics[angle=0,width=4.4cm]{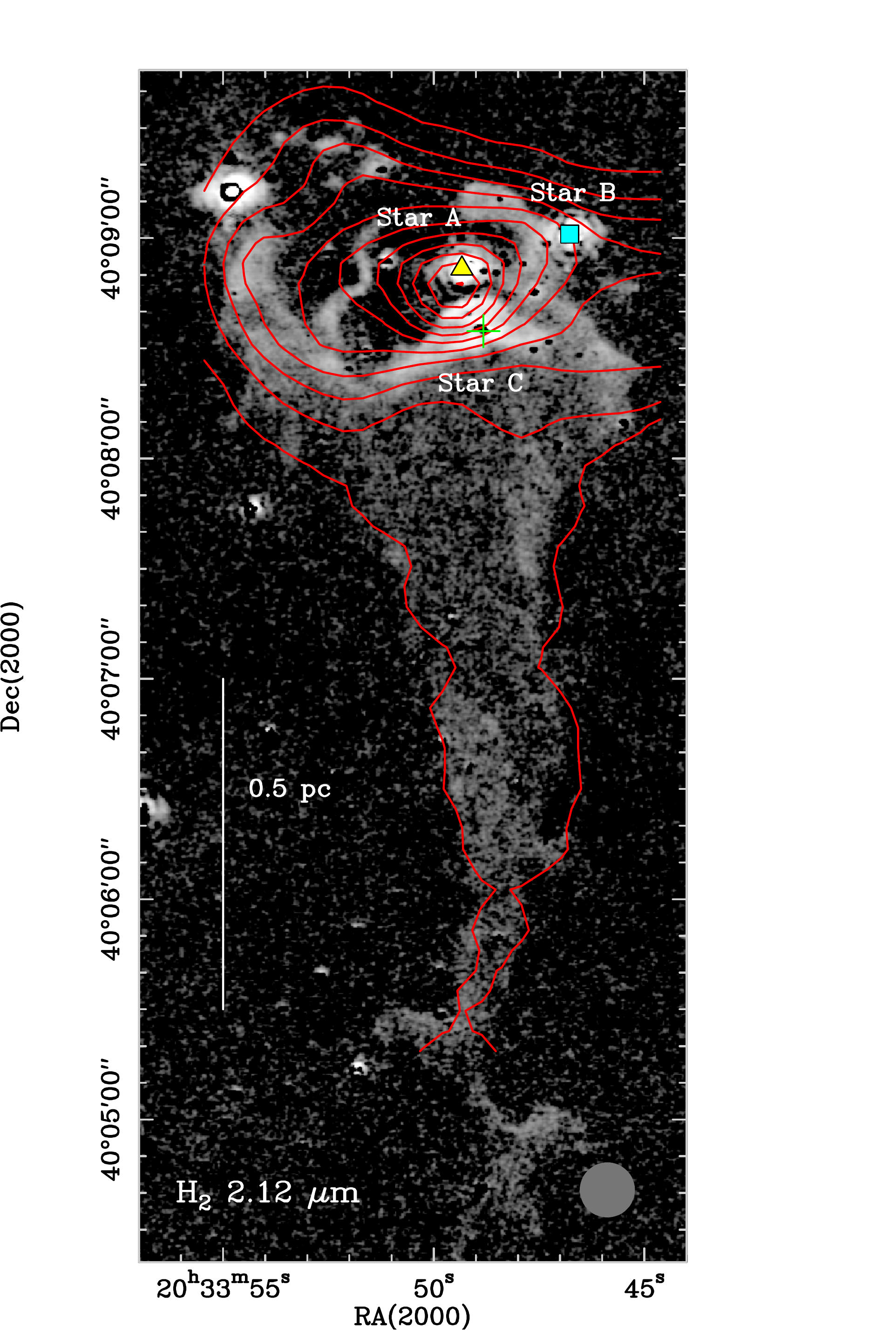}
\caption [] {Continuum subtracted narrow-band image that contains the
  H$_2$ 1-0 S(1) ro-vibrational line at 2.122 $\mu$m
  \citep{djupvik2017} at $\sim$2$''$ resolution with contours of
  velocity-integrated $^{12}$CO 3$\to$2 emission (yellow, left panel)
  and velocity-integrated \CII\ emission (red, right panel), both at
  15$''$ resolution, overlaid. The CO contours go from 1 to 13 by 1.5
  K km s$^{-1}$ and the \CII\ contours go from 5 to 155 by 15 K km
  s$^{-1}$. The embedded stars A, B, and C are indicated.}
\label{h2-overlay-a} 
\end{center}  
\end{figure}

\begin{figure}[ht] 
\begin{center}  
\includegraphics[angle=0,width=8.5cm]{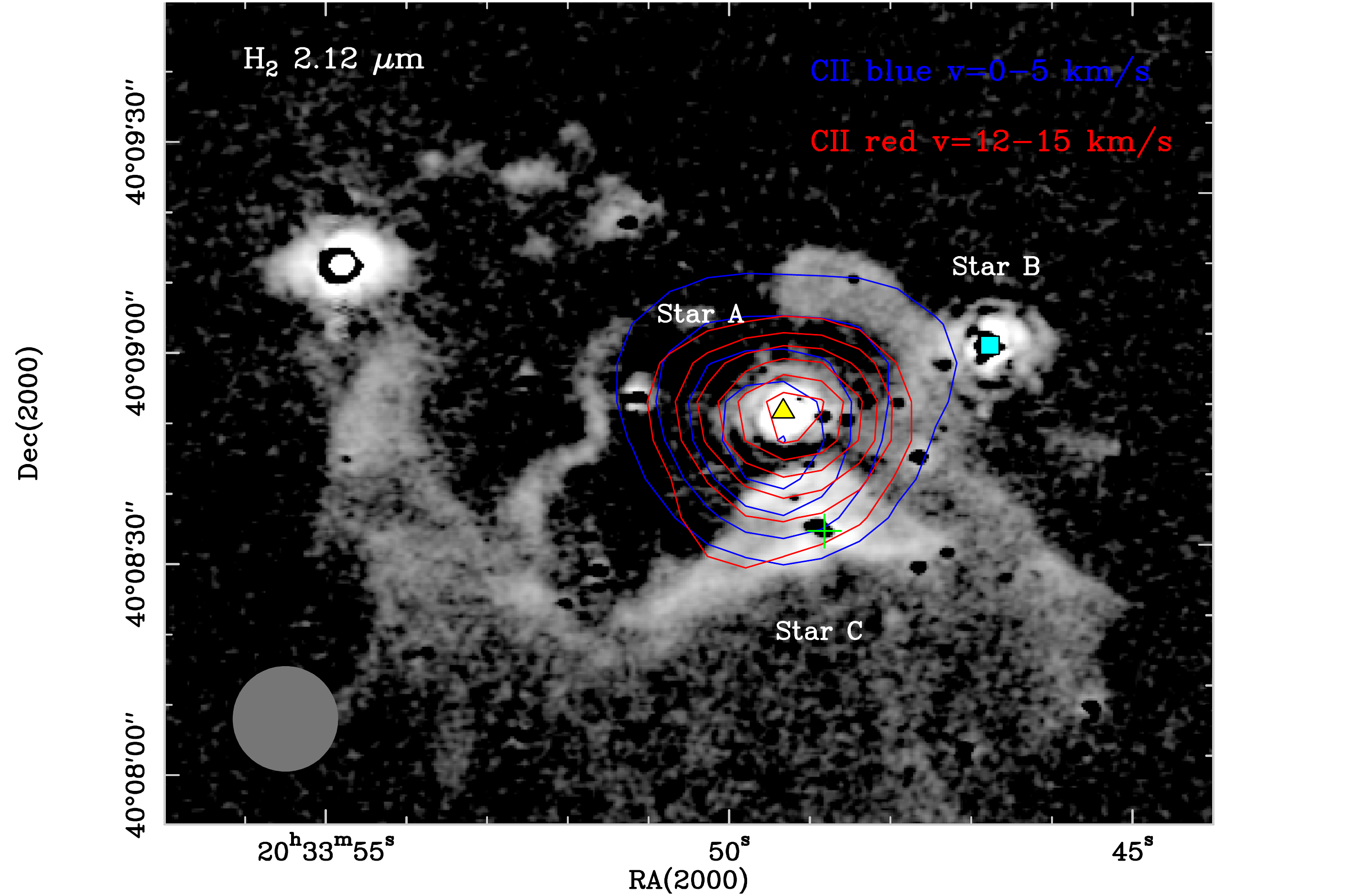}
\caption [] {Map of H$_2$ 2.12 $\mu$m emission \citep{djupvik2017}
  with contours of \CII\ outflow emission. The blue velocity range of
  the \CII\ line ranges from 0 to 5 km s$^{-1}$, contours go from 10
  to 50 by 10 K km s$^{-1}$. The red velocity range is 12 to 15 km
  s$^{-1}$ and contours go from 3 to 13 by 2 K km s$^{-1}$. The
  embedded stars and the \CII\ beam size are indicated.}
\label{h2-overlay-b} 
\end{center}  
\end{figure}

\begin{table}[htbp] 
 \centering 
 \caption{Line fluxes and ratios for one position in the globule head.} 
     \begin{tabular}{llccc} 
\hline 
\hline 
{\small Instrument}  & {\small Species}           & {\small I($\Theta$=20$''$)} &  {\small I($\Theta$=20$''$)}               & {\small Ratio}  \\ 
                     &                            & {\tiny [K km s$^{-1}$]}     & {\tiny [erg s$^{-1}$ cm$^{-2}$ sr$^{-1}$]} &                  \\  
\hline 
{\small HIFI}        & {\small \CII\, 158 $\mu$m}   &  {\small 153.5}         &   {\small 1.08E-03}                         &   -          \\  
{\small PACS}        & {\small \OI\,  145 $\mu$m}   &  {\small  25.5}         &   {\small 2.29E-04}                         &   -          \\        
{\small upGREAT}     & {\small \OI\,   63 $\mu$m}   &  {\small  6.3}          &   {\small 6.85E-04}                         &   -         \\  
{\small SPIRE}       & {\small  \CI\, 2-1}          &  {\small 7.4}          &   {\small 4.04E-06}                          &  -      \\
{\small SPIRE}       & {\small  \CI\, 1-0}          &  {\small 10.2}         &   {\small 1.25E-06}                          &  -      \\
{\small upGREAT}     & {\small $^{12}$CO 16$\to$15} & {\small 1.1}            &   {\small 7.03E-06}                          &   -          \\
{\small SPIRE}       & {\small $^{12}$CO 13$\to$12} & {\small 9.3}            &   {\small 3.19E-05}                          &   -          \\
{\small SPIRE}       & {\small $^{12}$CO 12$\to$11} & {\small 14.9}           &   {\small 4.02E-05}                          &   -          \\
{\small SPIRE}       & {\small $^{12}$CO 11$\to$10} & {\small 20.5}           &   {\small 4.27E-05}                          &   -          \\
{\small SPIRE}       & {\small $^{12}$CO 10$\to$9}  & {\small 26.4}           &   {\small 4.13E-05}                          &   -          \\
{\small SPIRE}       & {\small $^{12}$CO  9$\to$8}  & {\small 31.7}           &   {\small 3.62E-05}                          &   -         \\
{\small SPIRE}       & {\small $^{12}$CO  8$\to$7}  & {\small 45.1}           &   {\small 3.62E-05}                          &   -         \\
{\small SPIRE}       & {\small $^{12}$CO  7$\to$6}  & {\small 51.3}           &   {\small 2.76E-05}                          &   -         \\
{\small SPIRE}       & {\small $^{12}$CO  6$\to$5}  & {\small 68.2}           &   {\small 2.31E-05}                          &   -         \\
{\small SPIRE}       & {\small $^{12}$CO  5$\to$4}  & {\small 91.8}           &   {\small 1.80E-05}                          &   -         \\
{\small SPIRE}       & {\small $^{12}$CO  4$\to$3}  & {\small 104.6}          &   {\small 1.05E-05}                          &   -         \\
{\small SPIRE}       & {\small $^{13}$CO  9$\to$8}  & {\small  2.6}           &   {\small 2.61E-06}                          &   -         \\
{\small SPIRE}       & {\small $^{13}$CO  8$\to$7}  & {\small  5.5}           &   {\small 3.86E-06}                          &   -         \\
{\small SPIRE}       & {\small $^{13}$CO  7$\to$6}  & {\small  9.0}           &   {\small 4.24E-06}                          &   -         \\
{\small SPIRE}       & {\small $^{13}$CO  6$\to$5}  & {\small 11.1}           &   {\small 3.30E-06}                          &   -         \\
{\small SPIRE}       & {\small $^{13}$CO  5$\to$4}  & {\small 19.9}           &   {\small 3.41E-06}                          &   -         \\
{\small SPIRE}       & {\small  \CI\, 2-1/1-0}       &     -                  &        -                                    &  {\small 3.23}      \\
{\small SPIRE}       & {\small  $^{12}$CO  8-7/7-6}   &     -                 &         -                                    &  {\small 1.19}      \\
{\small SPIRE}       & {\small $^{12}$CO  6-5/5-4}    &     -                 &         -                                    &  {\small 1.71}      \\
{\small SPIRE}       & {\small $^{12}$CO  5-4/4-3}    &     -                 &         -                                    &  {\small 1.94}      \\
{\small SPIRE}       & {\small  $^{13}$CO  8-7/7-6}  &     -                  &        -                                     &  {\small 0.91}      \\
{\small SPIRE}       & {\small  $^{13}$CO  6-5/5-4}  &     -                  &        -                                     &  {\small 0.97}      \\
            &                     &                   &                                            &                  \\  
\end{tabular} 
 \tablefoot{The fluxes are given in erg s$^{-1}$ cm$^{-2}$ sr$^{-1}$,
   the ratios are determined from these fluxes for the SPIRE central
   pixel in the globule head (RA(2000)=20$^h$33$^m$50$^s$,
   Dec(2000)=40$^\circ$08$'$36$''$). All data points have the same
   angular resolution of 20$''$. The absolute error for the SPIRE data
   for the \CI\ lines and CO J$\le$8 is 2.5-3.1E07 erg s$^{-1}$
   cm$^{-2}$ sr$^{-1}$ and for J$>$8 $\sim$9E07 erg s$^{-1}$
   cm$^{-2}$.  The low- and mid-J CO data have a resolution of
   $\sim$30-40$''$ so that we only use line ratios.}
\label{pdr-fluxes} 
\end{table} 

\begin{figure*}[ht] 
\begin{center}  
\includegraphics[angle=0,width=14cm]{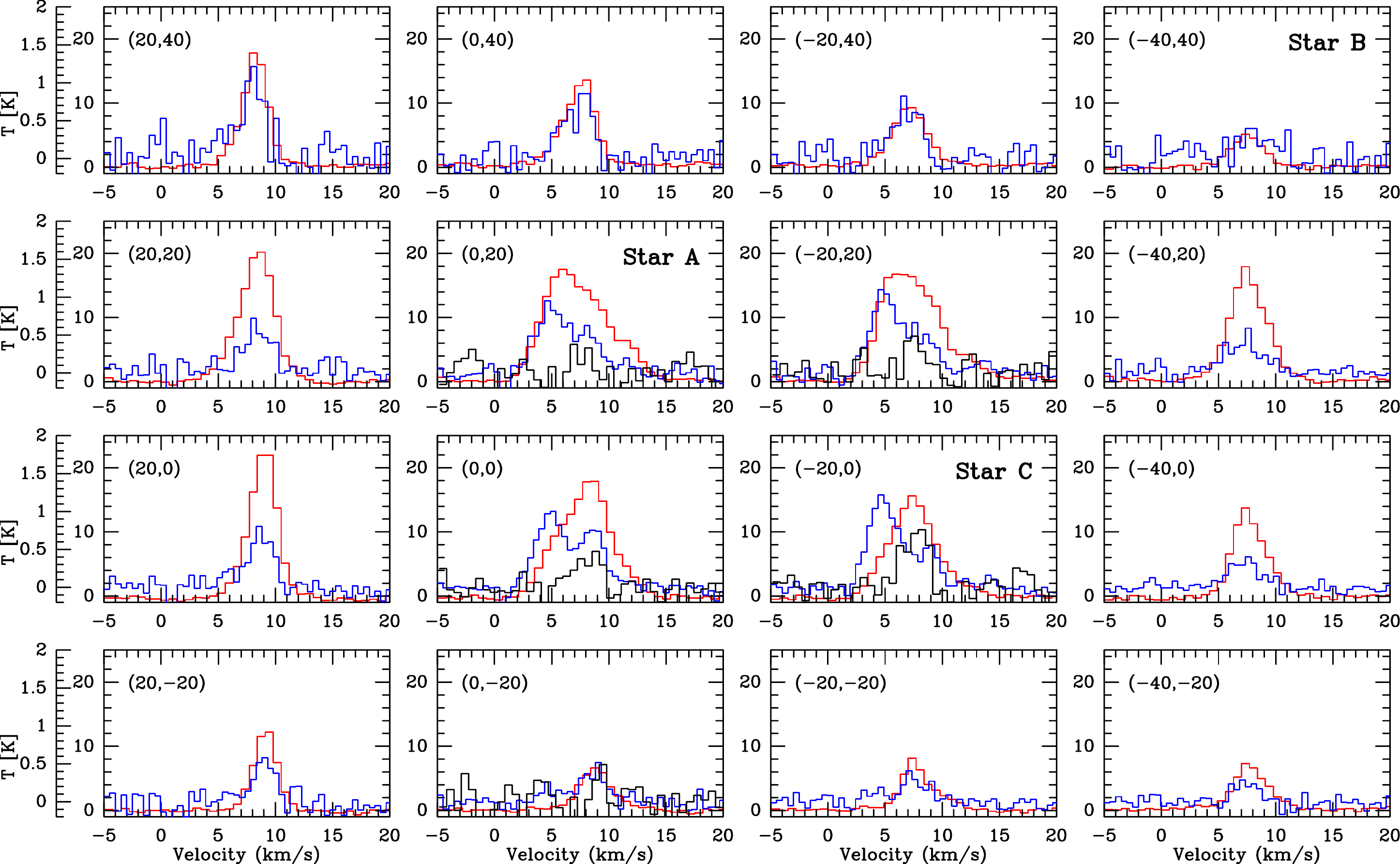}
\caption [] {Spectral maps of the \OI\ 63 $\mu$m line (blue), the
  \CII\ 158 $\mu$m line (red) and the CO 16$\to$15 line (black) of the
  globule head in the velocity range $-$5 to 20 km s$^{-1}$. The
  leftmost main beam brightness temperature scale ranges from $-$0.2
  to 2 K and is valid for \OI\ and CO 16$\to$15. The temperature
  scales at each panel are valid for \CII. All data are smoothed to an
  angular resolution of 20$''$ and sampled on a grid of 20$''$ in
  order to increase the S/N. However, the CO 16$\to$15
  line was not observed at all positions (Sect.~\ref{obs}) so that we
  only plot the few spectra with observed emission above the 5$\sigma$
  level. The velocity resolution is 0.5, 0.7, and 0.6 km s$^{-1}$ for
  \OI, \CII, and CO 16$\to$15, respectively.  The map centre position
  is RA(2000)=20$^h$33$^m$50$^s$, Dec(2000)=40$^\circ$08$'$36$''$. The
  approximate locations of the star systems are indicated.}
\label{oi} 
\end{center}  
  \end{figure*}

\begin{table}[htbp] 
 \centering 
  \caption{Line fluxes and ratios for one position in the globule tail.} 
 \begin{tabular}{llccc} 
\hline 
\hline 
{\small Instrument}  & {\small Species}           & {\small I($\Theta$=40$''$)} &  {\small I($\Theta$=40$''$)}               & {\small Ratio}  \\ 
                     &                            & {\tiny [K km s$^{-1}$]}     & {\tiny [erg s$^{-1}$ cm$^{-2}$ sr$^{-1}$]} &                  \\  
\hline 
{\small HIFI}        & {\small \CII\, 158 $\mu$m}   & {\small 10.0}          &   {\small 7.03E-05}                         &   -          \\  
{\small SPIRE}       & {\small $^{12}$CO 11$\to$10} & {\small 0.78}           &   {\small 1.60E-06}                          &   -          \\
{\small SPIRE}       & {\small $^{12}$CO 10$\to$9}  & {\small 1.46}           &   {\small 2.29E-06}                          &   -          \\
{\small SPIRE}       & {\small $^{12}$CO  9$\to$8}  & {\small 4.38}           &   {\small 5.00E-06}                          &   -         \\
{\small SPIRE}       & {\small $^{12}$CO  8$\to$7}  & {\small  6.99}          &   {\small 5.61E-06}                          &   -          \\
{\small SPIRE}       & {\small $^{12}$CO  7$\to$6}  & {\small 12.50}          &   {\small 6.72E-06}                          &   -          \\
{\small SPIRE}       & {\small $^{12}$CO  6$\to$5}  & {\small 18.81}          &   {\small 6.37E-06}                          &   -         \\
{\small SPIRE}       & {\small $^{12}$CO  5$\to$4}  & {\small 29.23}          &   {\small 5.73E-06}                          &   -         \\
{\small SPIRE}       & {\small $^{12}$CO  4$\to$3}  & {\small 31.68}          &   {\small 3.18E-06}                          &   -         \\

{\small SPIRE}      & {\small $^{13}$CO  8$\to$7}  & {\small 0.96}            &   {\small 0.67E-06}                          &   -         \\
{\small SPIRE}      & {\small $^{13}$CO  7$\to$6}  & {\small 1.25}            &   {\small 0.59E-06}                          &   -         \\
{\small SPIRE}      & {\small $^{13}$CO  6$\to$5}  & {\small 3.51}            &   {\small 1.04E-06}                          &   -         \\
{\small SPIRE}      & {\small $^{13}$CO  5$\to$4}  & {\small 7.53}            &   {\small 1.29E-06}                          &   -         \\

{\small SPIRE}       & {\small  \CI\ 2-1}           & {\small 4.66}           &   {\small 2.53E-06}                           &               \\
{\small SPIRE}       & {\small  \CI\ 1-0}           & {\small 7.38}           &   {\small 9.01E-07}                           &               \\
{\small SPIRE}       & {\small  \CI\, 2-1/1-0}       &     -                  &        -                                    &  {\small 2.81}      \\
{\small SPIRE}       & {\small  $^{12}$CO  8-7/7-6}   &     -                 &         -                                    &  {\small 0.83}      \\
{\small SPIRE}       & {\small $^{12}$CO  6-5/5-4}    &     -                 &         -                                    &  {\small 1.11}      \\
{\small SPIRE}       & {\small $^{12}$CO  5-4/4-3}    &     -                 &         -                                    &  {\small 1.80}      \\
{\small SPIRE}       & {\small  $^{13}$CO  8-7/7-6}   &     -                  &        -                                     &  {\small 1.13}      \\
{\small SPIRE}       & {\small  $^{13}$CO  6-5/5-4}   &     -                  &        -                                     &  {\small 0.81}      \\
            &                     &                   &                                            &                  \\  
\end{tabular} 
\tablefoot{The fluxes are given in erg s$^{-1}$ cm$^{-2}$ sr$^{-1}$,
  the ratios are determined from these fluxes for the SPIRE central
  pixel in the globule tail (RA(2000)=20$^h$33$^m$49$^s$,
  Dec(2000)=40$^\circ$06$'$40.2$''$). All data points have the same
  angular resolution of 40$''$. The absolute error for the SPIRE data
  varies between 2.5 and 3.1E07 erg s$^{-1}$ cm$^{-2}$ sr$^{-1}$. Note
  that there are less observations for the globule tail position
  compared to the head and the lines are weaker.}
\label{pdr-fluxes-tail} 
\end{table}

\subsection{Comparison to H$_2$ emission and small-scale dynamics of \CII\ emission} \label{cii-outflow} 

Figure~\ref{h2-overlay-a} shows how the emission distributions of cool
molecular gas (traced by CO) and warm PDR gas (traced by \CII) compare
to the narrow-band imaging of the H$_2$ 1-0 S(1) at 2.122 $\mu$m.
This line is excited either by shocks, driven by stellar winds, but
can also be associated with dense PDRs.  A large opening towards the
north-east becomes obvious in the H$_2$ and CO 3$\to$2 map and
suggests that the internal \HII\ region breaks out of the globule. The
deficit of molecular gas is also clearly seen in the {\sl Herschel}
column density map in Fig.~\ref{glob-cd}. In contrast, the peak in
column density seen in Fig.~\ref{glob-cd} corresponds also to peak
emission in CO and H$_2$, and Star C is centred on this dense clump.
Interestingly, the \CII\ emission shows no decline in the
north-eastern corner of the globule (right panel in
Fig.~\ref{h2-overlay-a}) and the emission peak is clearly centred on
Star A. In fact, we observe high-velocity blue- and red-shifted emission
in \CII, shown in Fig.~\ref{h2-overlay-b}, suggesting an outflow
oriented along the line-of-sight of the observer. The outflow is very
collimated and could thus originate from a young stellar object (YSO)
(the \CII\ beam is 15$''$ so that the emission is beam diluted).  It
is unlikely that this very localised outflow interacts with the
external UV field.  The red wing (velocities $>$12 km s$^{-1}$) is
fully visible in the \CII\ spectra taken around the position of Star A
and displayed in Fig.~\ref{oi}. There is no high-velocity emission
detected in CO 16$\to$15 and only a very weak velocity component in
\OI\ 63 $\mu$m around 12 km s$^{-1}$.

In \citet{djupvik2017}, we found that Star A has two components of
which one is an early B-type star, a Herbig Be star. These objects are
intermediate-mass pre-main-sequence stars and are divided into three
categories \citep{fuente2002}: The youngest ($\sim$0.1 Myr) Type I
stars are embedded in a dense molecular clump and have associated
bipolar outflows that are detected in CO. Type II stars are also
associated to molecular material, but not immersed in a dense clump,
their ages are between a few 0.1 to a few Myr. Type III stars (typical
age $>$1 Myr) have fully dispersed the surrounding material and
created a cavity inside the molecular cloud. In addition,
\citet{diaz1998} showed that only stars with spectral type earlier
than B5 can create significant PDRs.  Our observations are, thus, fit best
with an Herbig Be Type III star since we observe that the star is
located in a cavity and not associated with a dense clump, that there
is no CO outflow, but high-velocity \CII\ emission, tracing the PDR
surfaces of the inner cavity walls, namely, the interface between the
\HII\ region and the molecular gas. This sort of \CII\ dynamics was
also observed - and interpreted in a similar way - for the bipolar
nebula S106 \citep{schneider2018}. We note that we exclude shock
excitation as a significant origin for the outflow because firstly,
\CII\ is not a good shock tracer and its origin is mostly PDRs, and
secondly, the \OI\ 63 $\mu$m line does not show prominent
high-velocity wings, which would be the case if there were shocks.

It is out of the scope of this paper to go into more detail what is
the driving source for the \CII\ outflow, but we note that it must be
associated with the Herbig Be star.  In the literature on Herbig Ae
and Be stars, accretion and outflow signatures were detected
\citep{cauley2014,moura2020,rodriguez2014}, and stellar winds are
commonly promoted as the most likely outflow mechanism, although
magneto-centrifugally driven outflows from the star–disk interaction
region can also occur. Higher angular resolution cm-observations and
spectroscopy of lines from the stellar atmosphere of the star may help
to investigate in more detail the accretion and outflow properties of
the source.

In any case, we confirm the conclusion from \citet{djupvik2017} that
the emission distribution of H$_2$ indicates that the sources of
ionisation are the B stars of the embedded aggregate, rather than the
external UV field caused by the O-stars of Cyg OB2.  We note that
another line of evidence in \citet{djupvik2017} showed that the visible
spectrum of the \HII\ region is that of a soft ionising radiation,
typical of an early-B star.  In Sec.~\ref{pdr}, we will give further
evidence for this proposal by modelling the observed FIR lines.

\subsection{\OI\ 63 $\mu$m and CO 16$\to$15 line mapping with upGREAT/SOFIA and PACS/SPIRE spectroscopic maps} \label{oi-sofia} 
Figure~\ref{oi} shows a spectra map of the globule head observed in
the \OI\ 63 $\mu$m, the \CII\ 158 $\mu$m and CO 16$\to$ 15 lines with
upGREAT on SOFIA. The \OI\ spectra also show that there are 
several velocity components and not a single Gaussian line profile but
comparing to the \CII\ and CO 16$\to$15 emission reveals that the line
profile is moreover due to self-absorption.  The CO 16$\to$15 line
peaks at a velocity of $\sim$9 km s$^{-1}$ where there is a dip in
\OI\ emission and where the \CII\ line also reveals a decrease.  This
indicates that the \CII\ line can also be slightly self-absorbed at
the peak positions though this is difficult to tell because of the
broad red wings due to the \CII\ outflow.  The velocity component at
$\sim$9 km s$^{-1}$ seen in \OI, CO 16$\to$15 and \CII\ is mostly
associated with the dynamics caused by the impact of the Herbig Be star
on the surrounding molecular cloud (see Sec.~\ref{cii-outflow}).

The SOFIA data confirm the PACS \OI\ 63 $\mu$m map (Fig.~\ref{pacs} in
the Appendix), indicating that the velocity integrated \OI\ emission
is very localised. Interestingly, the peak \OI\ emission is not found
at the position of Star A (as is the case for \CII\ emission) or at
the position of Star C. Moreover, it peaks in between the two stars
and correlates partly with the extended clump seen in H$_2$ emission
(Fig.~\ref{h2-overlay-b}).
Furthermore, we note that the \NII\ lines at 122 $\mu$m and 205 $\mu$m
(Figs.~\ref{pacs} and A.2 in the appendix), best tracing  the
\HII\ region, have their emission peak close to Star A, further north
than the \OI\ and CO 16$\to$15 peaks.

\begin{figure*}[hbt]
\resizebox{\hsize}{!}{\includegraphics{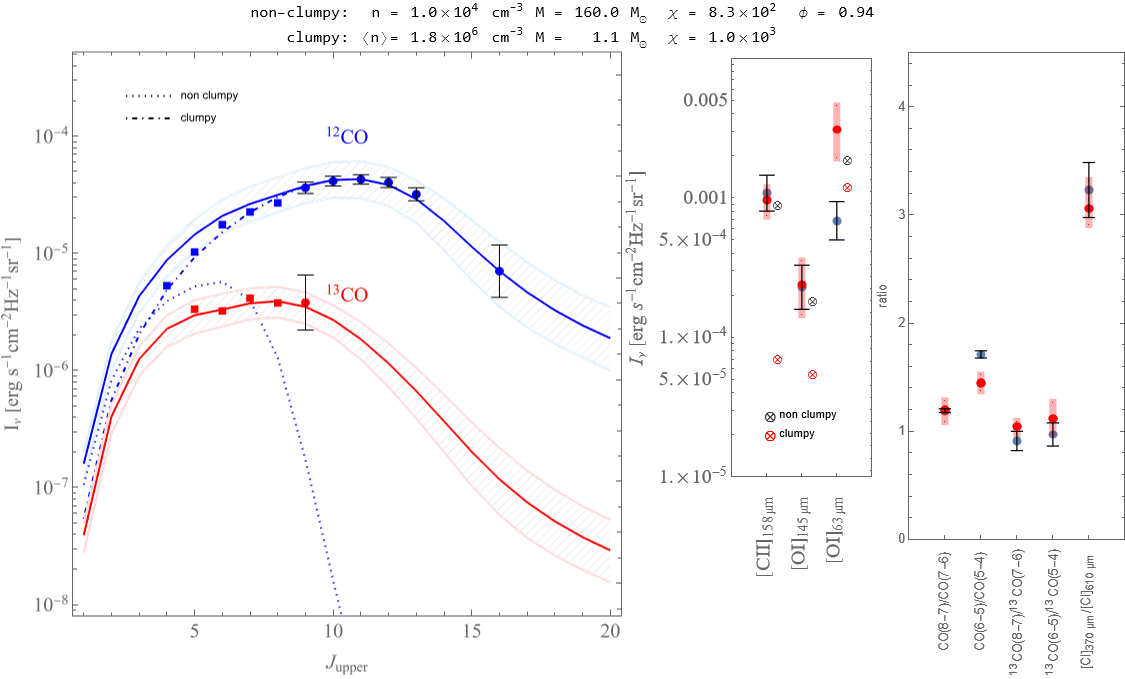}}
\caption{PDR model results of the two-component globule model. The left
  panel shows the $^{12}$CO (blue lines) and $^{13}$CO (red lines)
  model SLED (spectral line energy distribution) together with the
  observed data (blue for $^{12}$CO and red for $^{13}$CO,
  respectively). We note that data points without error bars are excluded
  from the fit but are shown as a consistency check. The non-clumpy
  and clumpy contribution to the total SLED are displayed with dotted
  and dash-dotted lines, respectively. The hatched areas around the
  SLED's indicate the model sensitivity to 20\% variations of the
  model parameters.  The centre panel shows the fine-structure line
  data. We note that the \OI\ 63$\mu$m line was excluded from the
  fit. The blue dots (with error bars) are the observations and the
  red dots are the model derived values with the red bands indicating
  the model response to 20\% parameter variations, For each line
  intensity, we show for information their contribution from the
  non-clumpy and clumpy components with black and red crossed circles,
  respectively (with a slight offset to the right for easier
  reading). The right right panel displays the behavior of the various
  line ratios with the same coding. }
\label{fig:globule_PDR_fit}
\end{figure*}

\begin{figure*}[hbt]
\centering
\includegraphics[angle=0,width=12cm]{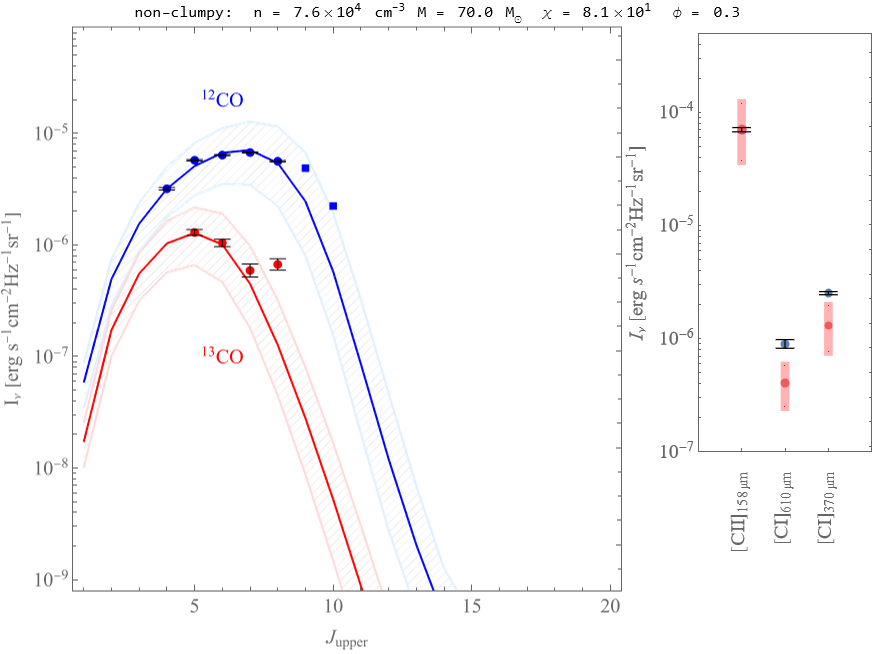}
\caption{PDR model results of the non-clumpy globule tail model. The
  left panel shows the $^{12}$CO (blue points and line) and $^{13}$CO
  (red points and line) model SLED (spectral line energy
  distribution).  This is the best-fitting non-clumpy model. The
  colour-shaded areas give the range of models by varying all parameters by
  $\pm 20\%$.  The right panel shows the observed (blue) and modelled
  (red) fine-structure line fluxes for this model, with the model
  variations indicated by the vertical red bands. Data points without
  error bars, namely, the two highest-J $^{12}$CO lines J=10$\to$9 and
  9$\to$8), are excluded from the fit, but shown to provide further information.}
\label{fig:globule_PDR_fit_tail}
\end{figure*}

\section{Discussion} \label{discuss} 

Summarising the observations presented in Sec.~\ref{results}, it
becomes obvious that we detect different gas components in the
globule. The \CII\ emission revealed widespread, extended emission in
the globule head and tail at bulk velocities ($\sim$8 km
s$^{-1}$). The carbon in this component is probably mostly excited by
the external Cyg OB2 cluster that impacts the globule from a
north-western direction. Cosmic-ray (CR) excitation can also
contribute, but the UV-field at the location of the globule is still a
few 100 G$_\circ$ and thus dominates over CRs. The tail is exclusively
externally heated, but the globule head contains intermediate stars
that created a cavity with an internal PDR surface that also emits in
\CII\ and in other typical cooling lines (high-J CO, \OI).  Red- and
blue-shifted high-velocity \CII\ outflow emission is caused by the
Herbig Be star A in the globule head. In the following, we will
disentangle the different gas components and determine their physical
properties in the globule head and tail, using PDR modelling.

\subsection{PDR modelling} \label{pdr}

We compare the observed line intensities and ratios with predictions
from the KOSMA-$\tau$ PDR model \citep{roellig2006,roellig2013}. This
model is able to compute line and continuum emission arising from
spherical clouds as well as for clumpy PDR ensembles
\citep{cubick2008,labsch2017}. The full model parameters are
summarised in Table~B.1 in Appendix B, we here vary the most important
variables that are density $n$ [cm$^{-3}$], mass $M$ [M$_\odot$], and
FUV field strength $\chi$ in units of the Draine field.  KOSMA-$\tau$
can model single spherical clumps (non-clumpy PDR model) and ensembles
of clumps (clumpy PDR model), according to a clump-mass distribution
law \citep[for details see][]{cubick2008,labsch2017}. Because the
model has a finite mass and the volume of different chemical species
(and thus the corresponding angular filling factors) are
self-consistently considered, KOSMA-$\tau$ is able to compute absolute
intensities that are directly comparable to observations.
Summarising, we apply the following modeling strategy:

We modelled a single position for the globule head
and tail, respectively.  The head position is at
RA(2000)=20$^h$33$^m$50$^s$, Dec(2000)=40$^\circ$8$'$36$''$.  This is
the centre position of the SPIRE map of the globule head and indicated
in Figs.~\ref{cii_cut} and ~\ref{pacs}. It is not a peak emission
position for many lines, but we have the largest data set for this
point. For the tail, the data set is even smaller and contains mostly
SPIRE lines. We take the central position of the map at
RA(2000)=20$^h$33$^m$49$^s$, Dec(2000)=40$^\circ$06$'$40.2$''$.

In order to account for the different beam sizes,
the velocity integrated line intensities of \CII\ 158 $\mu$m, \OI\ 63
$\mu$m and 145 $\mu$m, CO 16$\to$15, CO 13$\to$12, CO 12$\to$11, CO
11$\to$10, CO 10$\to$9, CO 9$\to$8, and $^{13}$CO 9$\to$8, were all
smoothed to a common angular resolution of 20$''$ for the globule head
and 40$''$ for the tail.  A distance of 1.4 kpc is adopted.

The CO transitions lower than J=8$\to$7 and the
two \CI\ transitions have a larger beam size of typically
30$''$-45$''$ (principally the SPIRE observations). For the globule
head, we thus compared the line ratios (CO 8$\to$7/7$\to$6, CO
6$\to$5/5$\to$4, $^{13}$CO 8$\to$7/7$\to$6, $^{13}$CO 6$\to$5/5$\to$4,
and \CI\ 2$\to$1/1$\to$0) to cancel out beam size effects to the first
order.  For the globule tail, we used absolute intensities because all
line intensities are smoothed to 40$''$, but give the ratios in
Table~\ref{pdr-fluxes-tail} for information. The best-fitting model
parameters are summarised in Table~\ref{tab:pdr_fit}.

\begin{figure*}[hbt]
\begin{center}  
 \includegraphics[angle=0,width=\textwidth]{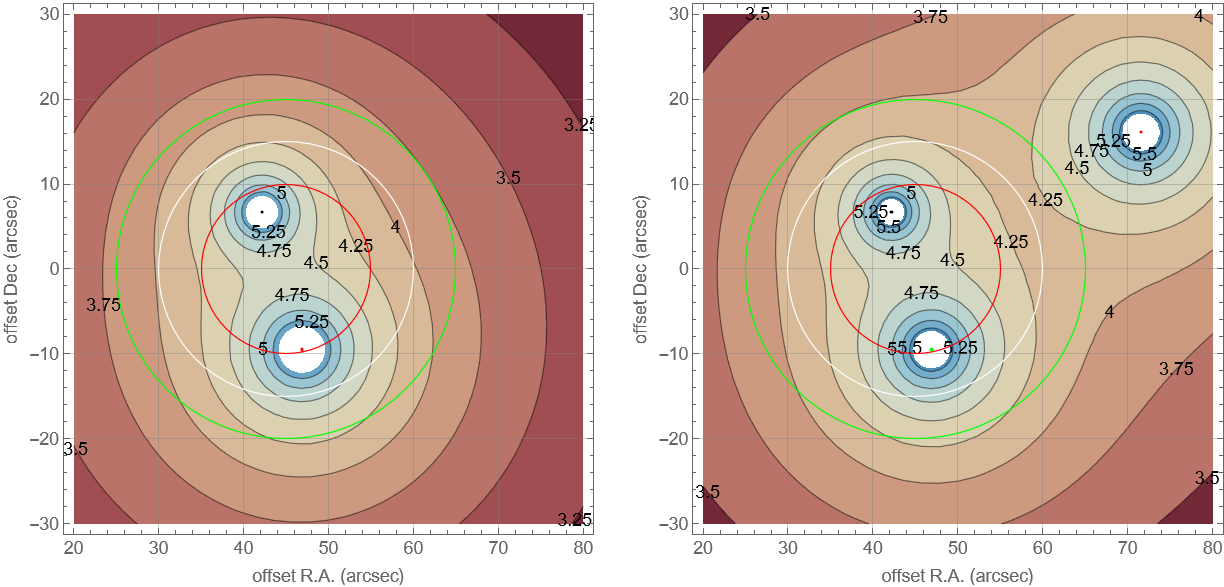}
\caption{FUV estimate for the globule. The three main stars A,B, and C
  are assumed to reside in the same plane on the sky. Spatial
  variations are shown in units of arcsec. The values of the contours
  are $\log \chi$ where the Draine field $\chi$ has been computed from
  pure geometrical dilution. The two panels show $\chi$ excluding
  (left) and including (right) Star B. For information on the size
  scale, we plot three circles with radii of 10$''$ (red), 15$''$
  (white), and 20$''$ (green) around a central position between Star A
  and C. }
\label{fig:FUV_estimate_2}
\end{center}  
\end{figure*}

\begin{figure}[hbt]
\begin{center}  
  \includegraphics[angle=0,width=8cm]{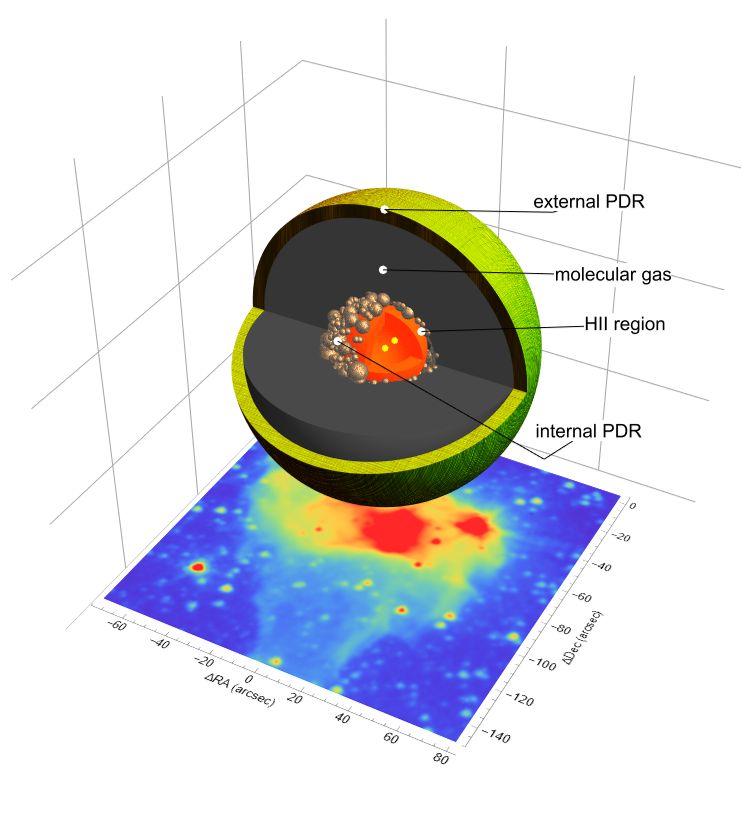}
\caption{Model geometry of the observed globule hosting two embedded
  YSO and an internal embedded cavity/\HII\ region. The relative sizes
  of the individual shells are not shown to scale. The external
  non-clumpy PDR is shown as yellow outer shell, the internal, clumpy
  PDR is shown as golden clump ensembles between the embedded
  \HII\ region and the molecular cloud shown as gray spherical
  shell. The image projected on the bottom plane is {\sl Spitzer} 8
  $\mu$m emission (for orientation, the flux values are not of
  interest here). The offsets are given in arcsec referring to the
  position RA(2000)=20$^h$33$^m$49.95$^s$,
  Dec(2000)=40$^\circ$07$'$42.75$''$. UV-radiation impacts externally
  via the Cyg OB2 cluster and internally via the YSOs. The cavity
  radius is approximately 15$''$-20$''$, determined from modelling and
  consistent with the region of brightest IR emission.}
        \label{fig:model}
\end{center}  
\end{figure}

\begin{table}[]
    \centering
    \caption{Summary of the best-fit models for the globule head and tail positions.}
    \begin{tabular}{lll}
         \hline \hline 
         parameter& value&description\\
         \hline
         \multicolumn{3}{c}{\bf globule head}\\
         \hline
         \multicolumn{2}{l}{non-clumpy component}&\\
         
         $n_{n-c}$&$1.0\times 10^4$~cm$^{-3}$& total gas density\\
         $M_{n-c}$&$160$~M$_\odot$& total clump mass\\
         $\chi_{n-c}$&$830$~$\chi_\mathrm{Draine}$& FUV field strength\\
         $\phi$&0.943&beam filling factor\\
         \multicolumn{2}{l}{clumpy component}&\\
         
         $\langle n_{c}\rangle$&$1.8\times 10^6$~cm$^{-3}$& mean ensemble gas density\\
         $M_{c}$&$1.1$~M$_\odot$& ensemble mass\\
         $\chi_{c}$&$1.0\times 10^3$~$\chi_\mathrm{Draine}$& FUV field strength\\
         \hline
         \multicolumn{3}{c}{\bf globule tail}\\
         \hline
         $n$&$7.6\times 10^4$~cm$^{-3}$& total gas density\\
         $M$&$70$~M$_\odot$& total clump mass\\
         $\chi$&$81$~$\chi_\mathrm{Draine}$& FUV field strength\\
         $\phi$&0.3&beam filling factor\\
         \hline
    \end{tabular}
    \label{tab:pdr_fit}
\end{table}

\subsection{Clumpy versus non-clumpy models for the globule head}
We showed in the previous sections that the globule head is subject to
an internal FUV field produced by the embedded star system and an
external FUV field produced by the Cyg OB2 association. The observed
line intensities are thus the superposition of the two PDRs and
consequently we assume a two-component PDR model. Two non-clumpy
PDR components are ruled out due to the strong emission for the high-J
CO ($J>10$) transitions which requires larger amounts of hot (surface)
CO than can be explained in non-clumpy models. Similarly, two clumpy
PDR components are unable to explain the observed high levels of
\CII\ and \OI\ emission.  We thus set up a two-component model that
consists of a non-clumpy external PDR component illuminated by the Cyg
OB2 cluster and an internal clumpy PDR illuminated by the embedded
stars (see Sect.~\ref{uv-field}).  This scenario is the one we already
proposed in the sections before for explaining the spatial and
kinematic emission distributions of the FIR lines.

The external PDR (non-clumpy) corresponds to a single spherical clump
with the surface density $n_\mathrm{n-c}$, mass $M_\mathrm{n-c}$,
UV-field $\chi_\mathrm{n-c}$ and considering a beam filling factor,
$\phi$. This model component corresponds to the yellow and gray
spherical shells shown in Fig.~\ref{fig:model}.

The internal PDR (clumpy) corresponds to a clumpy PDR ensemble, with
the ensemble averaged density clump density $\langle
n_\mathrm{c}\rangle$, mass $M_\mathrm{c}$, UV field $\chi_\mathrm{c}$
and a beam size of 20$''$. This component is depicted as ensemble of
golden clumps in Fig.~\ref{fig:model}.

In addition to these components, there is also the \HII\ region cavity
around star A (indicated in red in Fig.~\ref{fig:model}). This one is
rather small ($\sim$15$''$-20$''$) as can be inferred from the extend
of \NII\ emission (Fig.~A.1) and from our UV-field estimate
(Sect.~\ref{uv-field}). We did not model the \HII\ region (using a
different code since KOSMA-$\tau$ is not designed for that) to explain
the \NII\ lines because this is out of the scope of this paper.

We numerically minimise the reduced chi-square function for the seven free parameters:
\begin{equation}
\chi^2=\frac{1}{N_I+N_R-7} \left( \sum_{i=1}^{N_I} p_i \frac{I_\mathrm{obs,i}-I_\mathrm{mod,i}}{\epsilon_{I,i}} +
\sum_{i=1}^{N_R} p_i \frac{R_\mathrm{obs,i}-R_\mathrm{mod,i}}{\epsilon_{R,i}}\right),
\end{equation}
summing over all $N_I$ line transitions $I$ and $N_R$ line ratios $R$
to be included.  The error depends on the observed tracer, we used the
absolute error for the SPIRE observations (see Tables~\ref{pdr-fluxes}
and \ref{pdr-fluxes-tail}), a 30\% error for the PACS data, 20\% for
(up)-GREAT/SOFIA and 10\% for HIFI observations (see
Sec.~\ref{compare}).  We introduce a penalty factor, $p_i$, to allow for
weaker or stronger weighting of individual transitions or ratios in the
numeric fit.  $\chi^2$ is minimised in logarithmic space using the
Nelder-Mead method assuming a shrink and contract ratio of 0.85, and a
reflect a ratio of 3 using the software {\sl Mathematica} \citep{math}.

The best fitting model parameters, assuming $p_i=1$ for all $i$ with
the exception $p_ {(16-15)}=1000$, are $n_\mathrm{n-c}=1.0\times 10^4
\,{\rm cm}^{-3}, M_\mathrm{n-c}=160 \,{\rm M}_\odot,
\chi_\mathrm{n-c}=830, \phi=0.94$, and $\langle n_\mathrm{c}\rangle =
1.8 \times 10^6 \,{\rm cm}^{-3}, M_\mathrm{c}=1.1 \, {\rm M}_\odot,
\chi_\mathrm{c}=10^3$ with a $\chi^2=5.8$. To assess the sensitivity
of the result to the model parameters we varied all parameters by 20\%
($\phi$ was varied by $\pm 0.1$) and present the resulting intensity
variations as coloured bands around the best fit result. Across all
intensities and ratios a 20\% parameter variation changes the model
intensities and intensity ratios by $46-110\%$. The total model
mass of 160 M$_\odot$ is a good fit to the mass determined from the
dust (166 M$_\odot$), while the FUV field strengths are somewhat
different from complementary estimates. These differences are
discussed in Sect.~\ref{uv-field}.

Overall, the model intensity fit is very good with the exception of
the \OI\ 63 $\mu$m line, where the model intensities are about a
factor of 10 too high, while the \OI\ 145 $\mu$m line is well
reproduced. The over-prediction of \OI\ 63 $\mu$m intensity is a
notorious problem that we attribute to an absorbing foreground layer
(between the clumpy PDR and the observer), resulting in a significant
optical depth along the line of sight. This is in agreement with
recent SOFIA observations of star-forming regions were the \OI\ 63
$\mu$m line was found to be heavily affected by foreground absorption
while the upper \OI\ 145 $\mu$m line is mostly unaffected
\citep{schneider2018,guevara2020}. Figure~\ref{oi} also shows that the
\OI\ 63 $\mu$m line arising from the PDR of the globule head suffers
from significant self-absorption. A factor of $\sim$10 in missing
intensity is fully reasonable, although we cannot estimate the exact
value, which is the reason for excluding the line from the numerical
fit.

In order to asses the relative contribution of the clumpy and the
non-clumpy component we separately plot the predicted emission in the
left panel of Fig.~\ref{fig:globule_PDR_fit}. The three fine-structure
lines show a different fraction of their emission coming from either
component, resulting in a relatively sensitive probe to the degree of
clumpiness in the region. We note that the \CII\ emission receives a large
contribution from the non-clumpy PDR. This points towards a scenario
that the \CII\ emission at velocities of the bulk emission of the
globule is mostly caused by the external excitation from Cyg OB2. This
is what we also concluded from the extended emission distribution seen
in the \CII\ maps. We note that ionised carbon is probably well mixed
within the non-clumpy PDR component -- and not only a thin external
surface layer because we detect the rotation of the globule in
\CII. It is unlikely that it is only the external surface layer that
is rotating. In addition, we presume that there is little
\CII\ emission coming from the ionised phase because the PDR model
alone already well explains the observed intensities.  However, we
only modelled one point and cannot thus conclude over the full globule
head.

\subsection{Non-clumpy model for the globule tail}

The position we model in the globule tail (Fig.~\ref{cii_cut}) is a
more quiescent location than the one in the globule head since there
are no internal sources and excitation happens only externally via the
OB-cluster and by cosmic rays.  We performed the fit with data
smoothed to a larger beam size (40$''$) and tested again various models
and found that a non-clumpy model with a single model component gives
the best fitting results. These are displayed in
Fig.~\ref{fig:globule_PDR_fit_tail} and show the SLEDs for the globule
tail position with the observed CO line fluxes in $^{12}$CO and
$^{13}$CO. The $^{12}$CO 10$\to$9 and 9$\to$8 and the $^{13}$CO
8$\to$7 transitions have to be treated with care because the lines are
weak and just above the noise level.  The mass of the model was fixed
to 70 M$_\odot$, following the value determined from the dust column
density (Sec.~\ref{studies}). All model parameters were varied and the
colour-shaded areas in the left panel and coloured band in the right
panel of Fig.~\ref{fig:globule_PDR_fit_tail} show how much the model
intensities change. The illumination of the tail is assumed to be
one-side only.  The right panel displays the observational (blue) and
model (red) fluxes for the \CII\ and \CI\ lines. Overall, a model with
a UV field of around $\chi$ = 80 (corresponding to 137 G$_\circ$) and
a density of $\sim$8 10$^4$ cm$^{-3}$ with a filling factor of 0.3
fits our observations {\bf($\chi^2=1.9$)}. The low model UV field is
interesting, it is the lower limit from what was determined from the
census of Cyg OB2 stars or the {\sl Herschel} flux. Nevertheless, the
non-clumpy fit for the globule tail is less convincing compared to the
globule head model. The \CII\ model intensities match the observed
value and the $^{12}$CO lines are well reproduced up to 8$\to$7. The
two lowest $^{13}$CO lines fit the observations well, the upper lines
are underestimated as well as the higher-J $^{12}$CO lines. Both
\CI\ fine-structure lines are significantly underestimated. The fit
returned a filling factor of 0.3, namely, only 1/3 of the tail at the
assumed position is supposedly illuminated by FUV. Comparing
the\CII\ contours in Fig.~\ref{cii_cut} with the beam size at the tail
position would suggest a significantly larger filling of about
2/3. This emphasises the limits of the non-clumpy, single-component
model that we applied. The next step could be a multi-component model,
but this would require more data, for instance, the \OI\ fine-structure lines. In addition,
the high-J $^{13}$CO lines show a non-monotonous trend after the SLED
peak, which is difficult to explain in a simple model.

\subsection{UV field estimate for the globule head} \label{uv-field}

We assume that the FUV field affecting the gas in the globule has two
components. Firstly, an external radiation field, created by the
massive stars of the Cyg OB2 association, and secondly, an internal
radiation field created by the YSO embedded in the globule.  In
\citet{schneider2012} and \citet{schneider2016}, we already presented
an estimation of the FUV field based on the number of O-stars in Cyg
OB2 and on the {\sl Herschel} fluxes at 70 and 160 $\mu$m.  We arrived
to a value of G$_\circ \simeq 313$ ($\chi \simeq 183$), considering 50
O-stars at the position of the globule at a distance of $\sim$30 pc
from Cyg OB2.  These are upper limits since no extinction but only
1/r$^2$ distance dilution was considered. We also did not take into
account possible shadowing effects from the globule's head.  The
number of O-stars in Cyg OB2, however, is uncertain and estimates
range between $\sim$50 \citep{comeron2002,wright2015} and $\sim$120
\citep{knoedl2000}.  The FUV field derived from the {\sl Herschel}
fluxes (right panel in Fig.~\ref{glob-cd}) is $\sim$150-200 G$_\circ$
for the globule tail, where we can assume that the illumination is
only caused by the external radiation field.  In summary, a value of
150-300 G$_\circ$ (88-176 $\chi$) is probably a reasonable assumption
for the total external radiation field impacting the globule.
However, the field strength necessary to explain the non-clumpy PDR
emission is about two to three times stronger than that. Possible reasons for
this discrepancy could come from a significantly higher number of OB stars in
the cluster, as suggested by \citet{knoedl2000}. This would still be in
conflict with the FUV estimates for the tail and it is unclear whether
those can be explained by geometrical effects, for instance
shielding or shadowing by the globule head. Alternative explanations for
the higher FUV illuminating the external PDR could be an additional,
possibly closer source such as the YSO B (see discussion below) or a
much stronger fragmentation of the molecular gas in the head that
allows the internally generated FUV to escape and also affect the
external PDR.

Inside the globule head, the internal sources produce an internal
Str\"omgren sphere embedded in the globule and illuminate the inner
surface of the remaining spherical shell. Here, we estimate the
strength for this internal radiation field for comparison and as a
constraint for the PDR model.  The FUV field of the YSOs is dominated
by the internal sources, named Star A, B, and C in
\citet{djupvik2017} and we compute the FUV intensity by assuming
stellar black-body emission with T$_\mathrm{eff}$ = 22600, 26200,
26200~K and stellar luminosities $\log L$ = 3,72, 4.04, 4.04,
respectively, integrating over the FUV range from 910 to 3000
$\AA$. The flux is diluted with 1/$r^2$ and superposed in
Fig.~\ref{fig:FUV_estimate_2}. Any additional attenuation, for instance, by
dust is neglected, hence the result is an upper limit to the FUV field
strength. Star B is slightly offset with regard to the \HII\ region
and the peaks of \CII\ emission, so it is not clear whether this YSO
is still embedded in the globule or whether it only appears related
due to projection effects. However, the contribution of Star B to the
radiation field close to our model position is relatively weak due to
the larger distance.  We thus assume that the internal FUV field
is created by Stars A and C only. The PDR model fit gives a radiation
field $\chi_\mathrm{c} \sim 1000$ for the clumpy component.
A comparison with Fig.~\ref{fig:FUV_estimate_2} shows that the FUV field
estimated from the census of the stars and assuming no extinction is
higher, typically a factor of 2-3. On the other hand, the FUV field from
the {\sl Herschel} fluxes is $\chi \sim 2500$ in a 20$''$ beam at the
peak position and $\chi \sim 1500$ at the position where we perform
the PDR modelling\footnote{This estimation assumes that the dust in
  the PDR region and in the \HII\ region cavity fully absorbs the
  emitted UV photons and re-emits in the FIR.}. These values are in
agreement with the PDR model estimates of the total FUV field
(external and internal) which both contribute to the total continuum
flux. The FUV derived from the {\sl Herschel} fluxes and our model
results differs from the census of the embedded stars. This cannot be
explained via the dust attenuation of the FUV because any significant
amount of dust in the \HII\ region cavity that absorbs UV photons would
still contribute to the IR continuum emission. The \HII\ region cavity
has a radius of $\sim$15$''$-20$''$, which is consistent with the
extent of the area of brightest IR and H$_2$ and Br$_\gamma$ 
emission (Figs.~\ref{glob-1} and~7 in \citet{djupvik2017}). A
significantly larger cavity and, therefore, a lower FUV at the clumpy,
internal PDR is unlikely. Most likely, our estimate of the FUV
brightness of the embedded YSOs is too high due to lower
$T_\mathrm{eff}$ and $\log L$.

\subsection{Discussion of the model results}

From the previous sections, we can see that the observed emission
requires at least two PDR components: a non-clumpy, high mass
component with a FUV illumination of $\chi_\mathrm{n-c}\approx 850$,
and a less massive clumpy PDR component that is about two orders of
magnitude denser and requires a stronger FUV illumination of
$\chi_\mathrm{c} \approx 1000$.  Given the geometrical constrains of
the source, we propose the scenario outlined in Fig.~\ref{fig:model}.
The embedded YSOs are creating a cavity/\HII\ region embedded in the
globule head. The inner surface of the remaining shell is compressed
by the expanding \HII\ region and possibly fragments into clumps, and
is heated by the strong radiation of the YSOs.

The external surface of the globule is irradiated by the ambient FUV
field and emits as a spherical (non-clumpy) PDR at a density of
$\sim$10$^4$ cm$^{-3}$. The non-clumpy PDR component has a clump
radius of $\sim$60$''$, which is consistent with the observed extended
FIR line emission of \CII\ that traces mostly the outer PDR
layer. Other lines with critical densities around 10$^4$ cm$^{-3}$ and
excitation temperatures around 50-100 K are the \CI\ 2$\to$1 and
1$\to$0 lines and the mid-J CO lines. Their spatial emission
distribution is also more extended than the emission lines of tracers
that require higher densities and temperatures (such as the \OI\ lines
and the high-J CO lines, see Figs.~A.1 and A.2).  The latter have
their origin in the PDR created at the internal surface of the cavity,
which is clumpy and dense ($\sim$2$\times$10$^6$ cm$^{-3}$) and
covers a relatively small volume due to its small mass.  To test how
realistic this scenario may be, we computed the thickness of the internal
clumpy PDR layer because the clumpy PDR model fit returns the total
PDR mass and volume ($V=7.2\times10^{50}$ cm$^{3}$) and this can be
converted to a thickness as function of
$R_\mathrm{cavity}$. Accounting for irregularities and turbulent
structures in the cavity surface, we can also apply a volume filling
factor that describes how efficiently the clumpy PDR fill up the
available volume.

\begin{figure}[hbt]
\resizebox{\hsize}{!}{\includegraphics{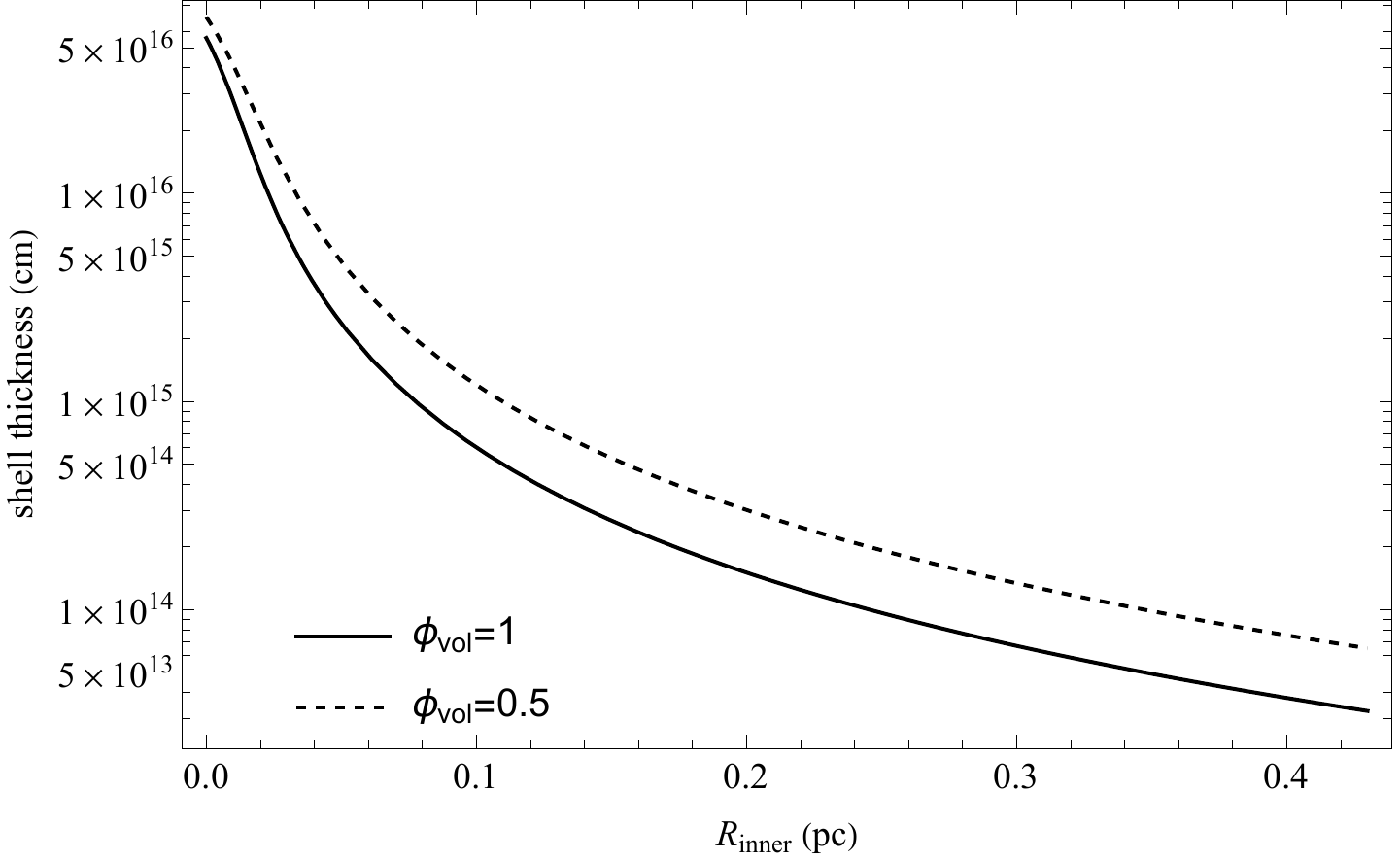}}
\caption{Thickness of the internal PDR layer as function of cavity
  radius. Different lines correspond to variations in the volume
  filling factor of the ensemble. }
        \label{fig:shell-thickness}
\end{figure} 

Figure~\ref{fig:shell-thickness} shows how the PDR thickness varies as
a function of $R_\mathrm{cavity}$. We compare two volume filling
scenarios. We find that the internal PDR layer is relatively thin with
a thickness of $\approx 3-5\times 10^{14}$~cm
(0.9-1.6$\times$10$^{-4}$ pc) only. We used the radius of the
\HII\ region cavity of $\sim$15$''$, corresponding to 0.14 pc, to
derive the thickness. Geometrically, this is consistent with our
picture of the internal PDR surface.  However, the spherical shell
picture would decrease the volume and mass of the remaining globule
and is inconsistent with the assumption of a full spherical
(non-clumpy) PDR. Naturally, this mostly affects the molecular cloud
tracers. In our model, however, the non-clumpy PDR contributes mostly
to the \CII\ emission and other surface tracers. We can, therefore,
ignore this inconsistency at this point. We want to stress that the
PDR model results were performed for all parameters independently. The
proposed geometry also explains the over-predicted \OI\ 63 $\mu$m
intensity. In the model fit we simply add the clumpy and non-clumpy
contribution. Geometrically, however, the non-clumpy PDR shell around
the clumpy PDR is optically thick against the \OI\ 63 $\mu$m line
because the higher \OI\ levels are not excited.  We conclude that the
resulting model components very nicely fit into the proposed geometry
scenario and agree with complementary constraints such as the strength
of the internal FUV field, the globule mass, and the observed
\OI\ self-absorption.

\section{Conclusions and Summary} \label{summary} 

We presented new spectroscopic FIR data for the globule IRAS
20319+3958 in Cygnus X South, located at 1.4 kpc distance, obtained
with HIFI, PACS, and SPIRE on {\sl Herschel}, and with upGREAT on
SOFIA. The observations include all important FIR cooling lines in the
interstellar medium, namely, the \CII\ 158 $\mu$m line, the \OI\ 63
$\mu$m and 145 $\mu$m lines, the \CI\ 2$\to$1, 1$\to$0 lines, the mid-
to high-J CO ladder (16$\to$15 down to 3$\to$2, and the \NII\ lines at
205 $\mu$m and 122 $\mu$m. These tracers cover a large range of
excitation temperatures and densities. \\

The \CII\ line is the only FIR line that covers the full globule and
is spectrally resolved. The kinematic \CII\ distribution revealed
several features.  Firstly, the \CII\ velocity map is consistent with
rotation, that we attribute as a relic from the initial momentum the
globule carried away while it was detaching from the molecular
cloud. A comparison with the simulations would help in exploring this
possibility.  Secondly, we detected a rather collimated high-velocity
blue- and red-shifted \CII\ outflow, associated with an embedded
Herbig Be star. This star, together with two other systems of B-stars,
is located inside the globule head and created an internal
\HII\ region.  The outflow is not visible in the \OI\ 63 $\mu$m line
or in CO and we cannot discern the driving source, namely, the stellar
wind of the Herbig Be star or the disk wind in case of an
accretion disk.

We performed careful PDR modelling using the large observational data
set of FIR lines (see above) for one position in the globule head and
one in the tail. The objective was to determine the physical
properties of the PDR components that are responsible for the emission
of the various cooling lines and to establish a geometrical model for
the globule head. \\
The best-fitting model is one with an extended
($\sim$60$''$ or $\sim$0.4 pc), external non-clumpy PDR layer where
most of the \CII\ emission originates. The UV radiation of the
$\sim$30 pc distant Cyg OB2 cluster estimated from the stellar census
of a few hundred G$_0$ seems to be insufficient to account for the
model FUV intensities of G$_0\approx1500$. A much larger stellar
content of the OB cluster and/or additional possibly closer FUV
sources may explain this discrepancy.  The total mass from the PDR
model is $\sim$160 M$_\odot$, which corresponds well to the mass
determined from dust (166 M$_\odot$), and an average density of 10$^4$
cm$^{-3}$.  Between the shell and the \HII\ region cavity is a thin
PDR layer ($<$0.1 pc) that is clumpy, dense ($\sim$2$\times$10$^6$
cm$^{-3}$), but not very massive ($\sim$1 M$_\odot$) and illuminated
by the embedded YSOs that create a radiation field of
G$_0>10^3$. \\
The tail position has no complex structure, the best
fitting model is the one of a non-clumpy PDR with a mass of $\sim$70
M$_\odot$, illuminated by an external UV field of $\sim$140 G$_\circ$
which corresponds to the lower UV field limit derived from the census
of the stars and the {\sl Herschel} flux estimate and may hint at
additional shadowing of the tail by the globule head.

With this study, we establish evidence in support of our proposal from
\citet{schneider2012} and \citet{djupvik2017} that the globule is an
example of a region where intermediate-mass stars form in isolation
within a single dense clump. We also show that PDR modelling of many
cooling lines and a consideration of a complex geometry allows us to successfully
explain the observed intensities.

\begin{acknowledgements} 
This work was supported by the Agence National de Recherche
(ANR/France) and the Deutsche Forschungsgemeinschaft (DFG/Germany)
through the project "GENESIS" (ANR-16-CE92-0035-01/DFG1591/2-1).
N.S. acknowledges support from the BMBF, Projekt Number 50OR1714 (MOBS
- MOdellierung von Beobachtungsdaten SOFIA). This work is based on
observations made with the NASA/DLR Stratospheric Observatory for
Infrared Astronomy (SOFIA). SOFIA is jointly operated by the
Universities Space Research Association, Inc. (USRA), under NASA
contract NAS2-97001, and the Deutsches SOFIA Institut (DSI) under DLR
contract 50 OK 0901 to the University of Stuttgart. This work was
supported by the German \emph{Deut\-sche
  For\-schungs\-ge\-mein\-schaft, DFG\/} project number SFB 956.\\ GJW
gratefully acknowledges the receipt of an Emeritus Fellowship from The
Leverhulme Trust.\\ SPIRE has been developed by a consortium of
institutes led by Cardiff University (UK) and including
Univ. Lethbridge (Canada); NAOC (China); CEA, LAM (France); IFSI,
Univ. Padua (Italy); IAC (Spain); Stockholm Observatory (Sweden);
Imperial College London, RAL, UCL-MSSL, UKATC, Univ. Sussex (UK); and
Caltech, JPL, NHSC, Univ. Colorado (USA). This development has been
supported by national funding agencies: CSA (Canada); NAOC (China);
CEA, CNES, CNRS (France); ASI (Italy); MCINN (Spain); SNSB (Sweden);
STFC (UK); and NASA (USA).  PACS has been developed by a consortium of
institutes led by MPE (Germany) and including UVIE (Austria); KU
Leuven, CSL, IMEC (Belgium); CEA, LAM (France); MPIA (Germany);
INAF-IFSI/OAA/OAP/OAT, LENS, SISSA (Italy); IAC (Spain). This
development has been supported by the funding agencies BMVIT
(Austria), ESA-PRODEX (Belgium), CEA/CNES (France), DLR (Germany),
ASI/INAF (Italy), and CICYT/MCYT (Spain).
\end{acknowledgements}

 
\begin{appendix}  

\section{PACS and SPIRE spectroscopy}
  
\begin{figure*}[ht] 
\begin{center}  
\includegraphics[angle=0,width=7cm]{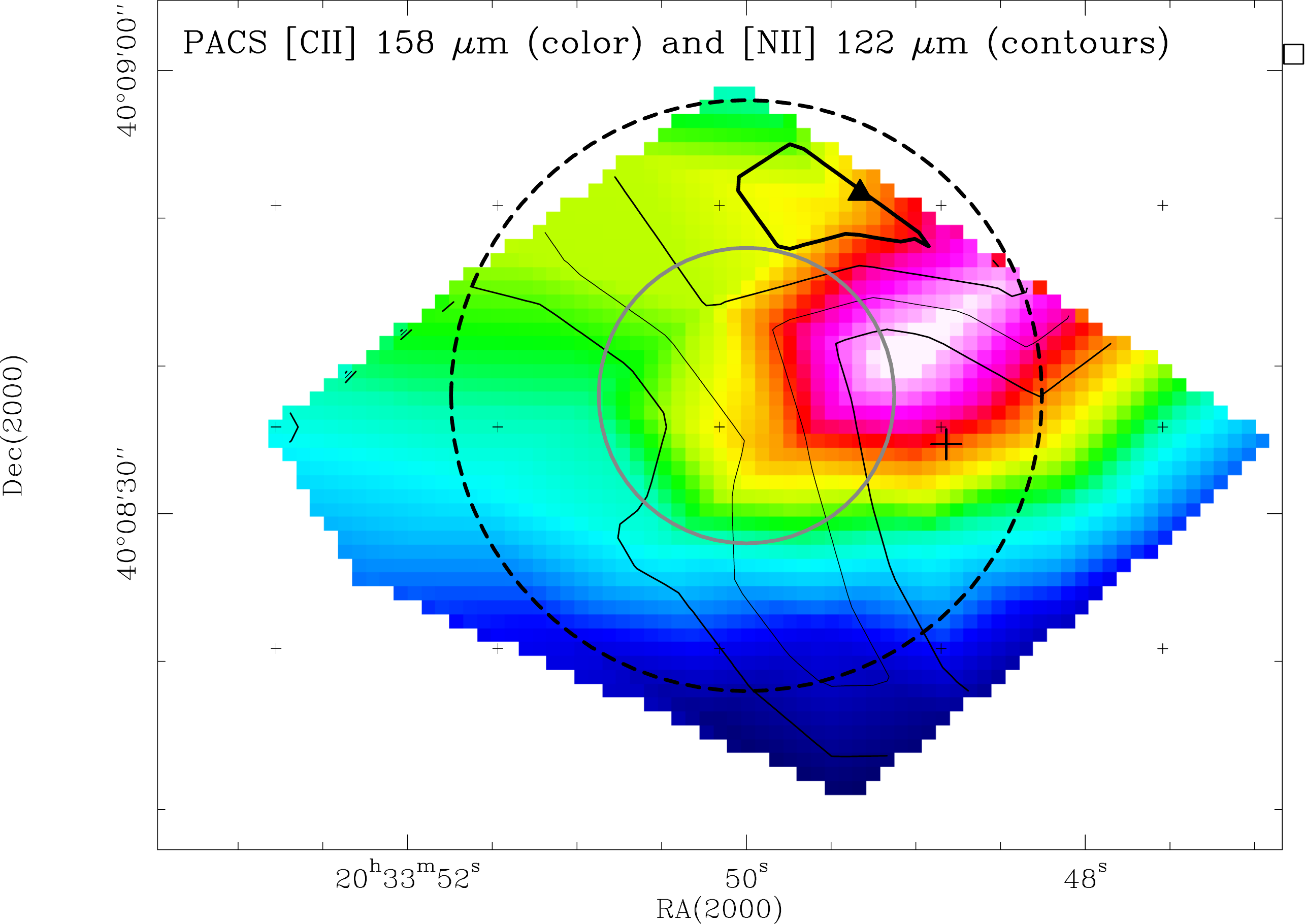}
\includegraphics[angle=0,width=7cm]{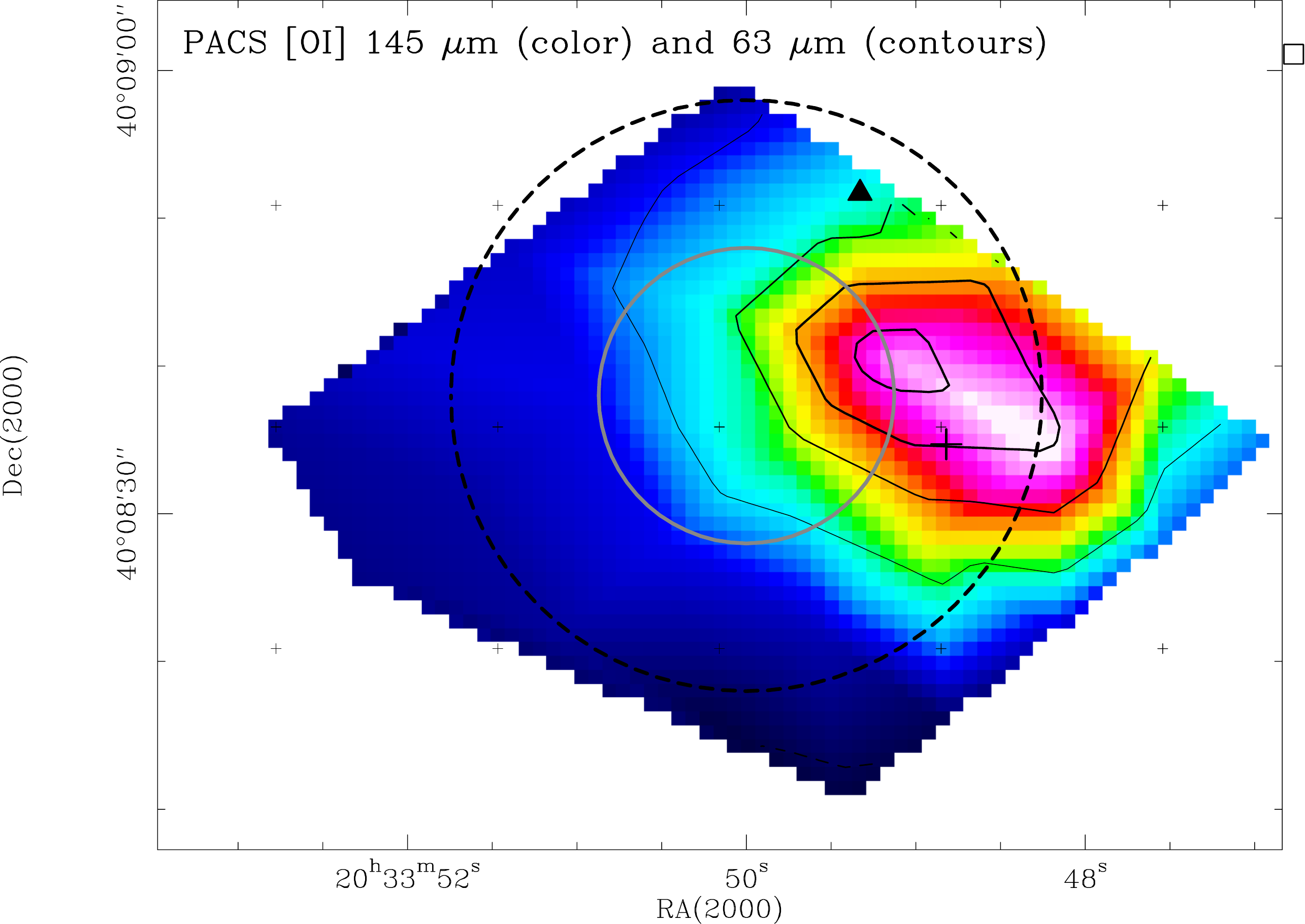}
\includegraphics[angle=0,width=7cm]{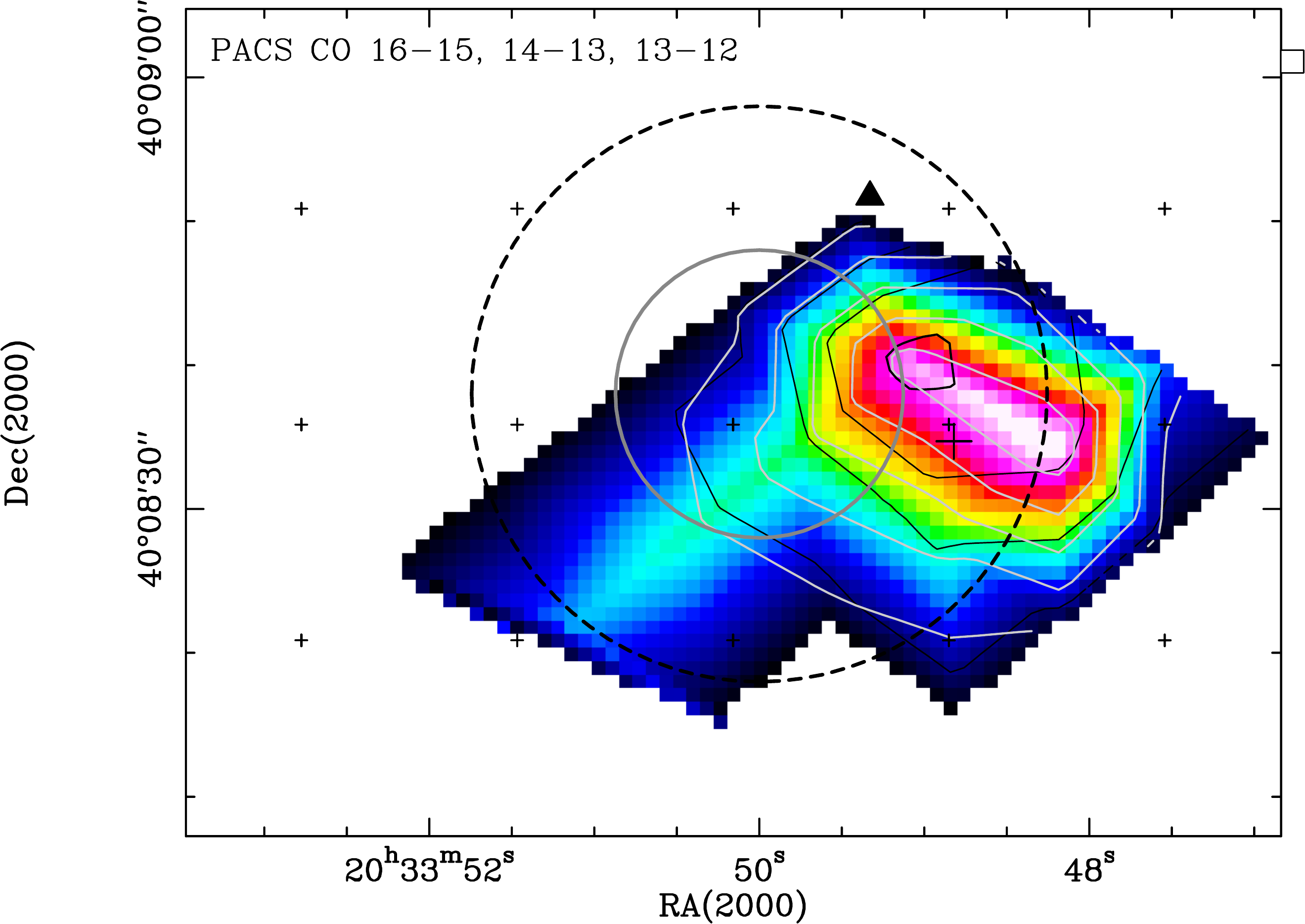}
\caption [] {Various overlays of PDR lines observed with PACS. The
  colour range for the PACS \CII\ data (top left) is 20 to 211 K km
  s$^{-1}$, contours (1, 2, 3, 4 K km s$^{-1}$) of \NII\ emission are
  overlaid.  The colour range for the PACS \OI\ 145 $\mu$m data (top
  right) is 0 to 85 K km s$^{-1}$, contours (10, 50, 90, 130, 170 K km
  s$^{-1}$) of PACS \OI\ 63 $\mu$m emission are overlaid.  The colour
  range for the PACS CO 13$\to$12 data (bottom) is 0 to 16 K km
  s$^{-1}$, black contours (1, 4, 7, 11 K km s$^{-1}$) of PACS CO
  16$\to$15 emission, and grey contours (5 to 25 by 5 K km s$^{-1}$)
  of PACS CO 14$\to$13 emission are overlaid. The 'finger' of emission
  in CO 13$\to$12 emission is probably an artefact since it is not
  visible in the CO 16$\to$15 and 14$\to$13 lines. The black triangle
  indicates the double system (Star A) of which at least one is a
  Herbig Be star, the white rectangle points to Star B with a B0.5
  B1.5 spectral type, and the large black cross marks Star C, a
  resolved binary of which one is late O or early B star. The solid
  grey circle has a size of 20$''$ and the dashed one of 40$''$. This
  is the position for the flux determination for PDR modelling.}
    \label{pacs}
\end{center}  
  \end{figure*}

\begin{figure*}[ht] 
\begin{center}  
\includegraphics[angle=0,width=8cm]{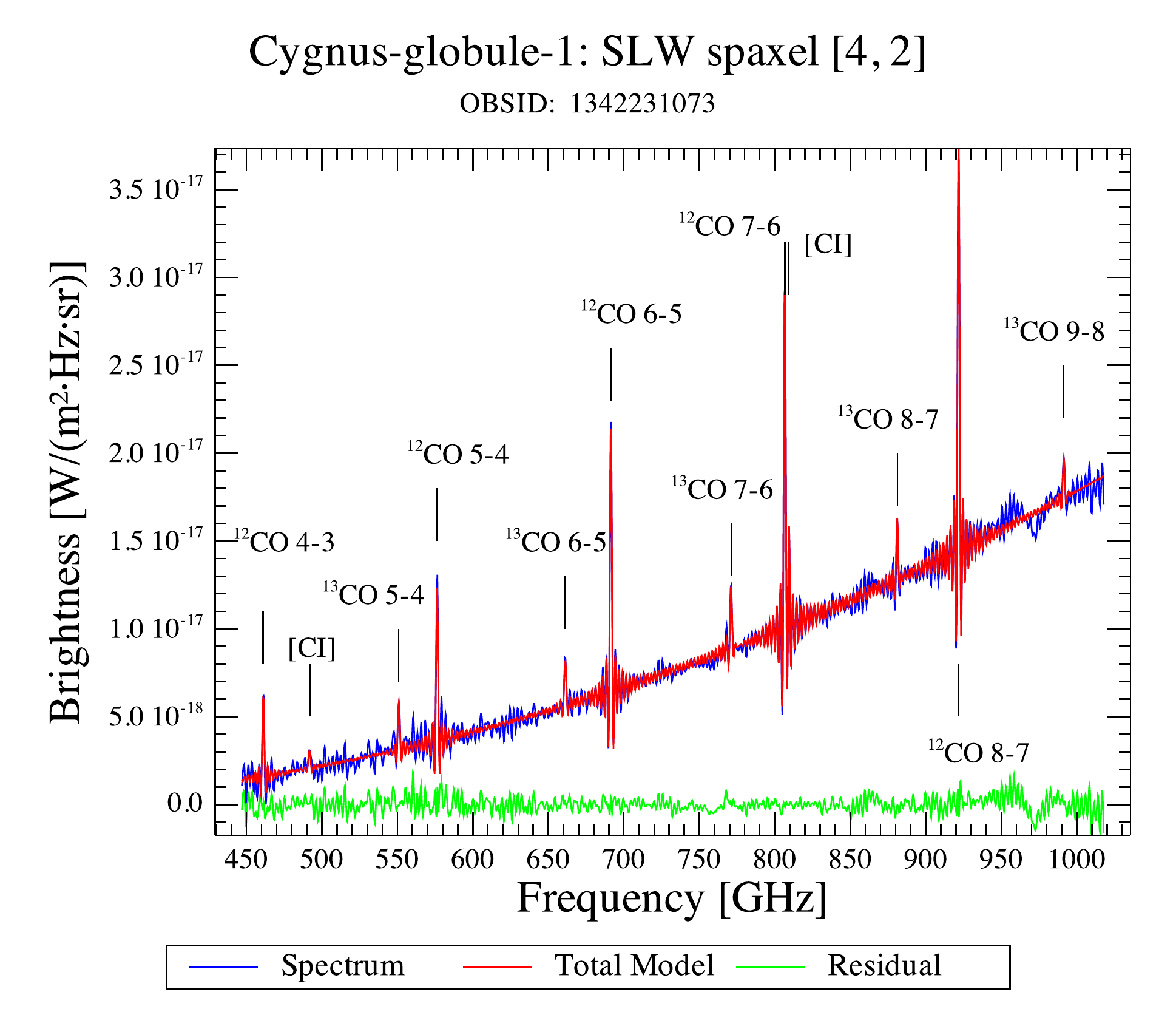}
\includegraphics[angle=0,width=8cm]{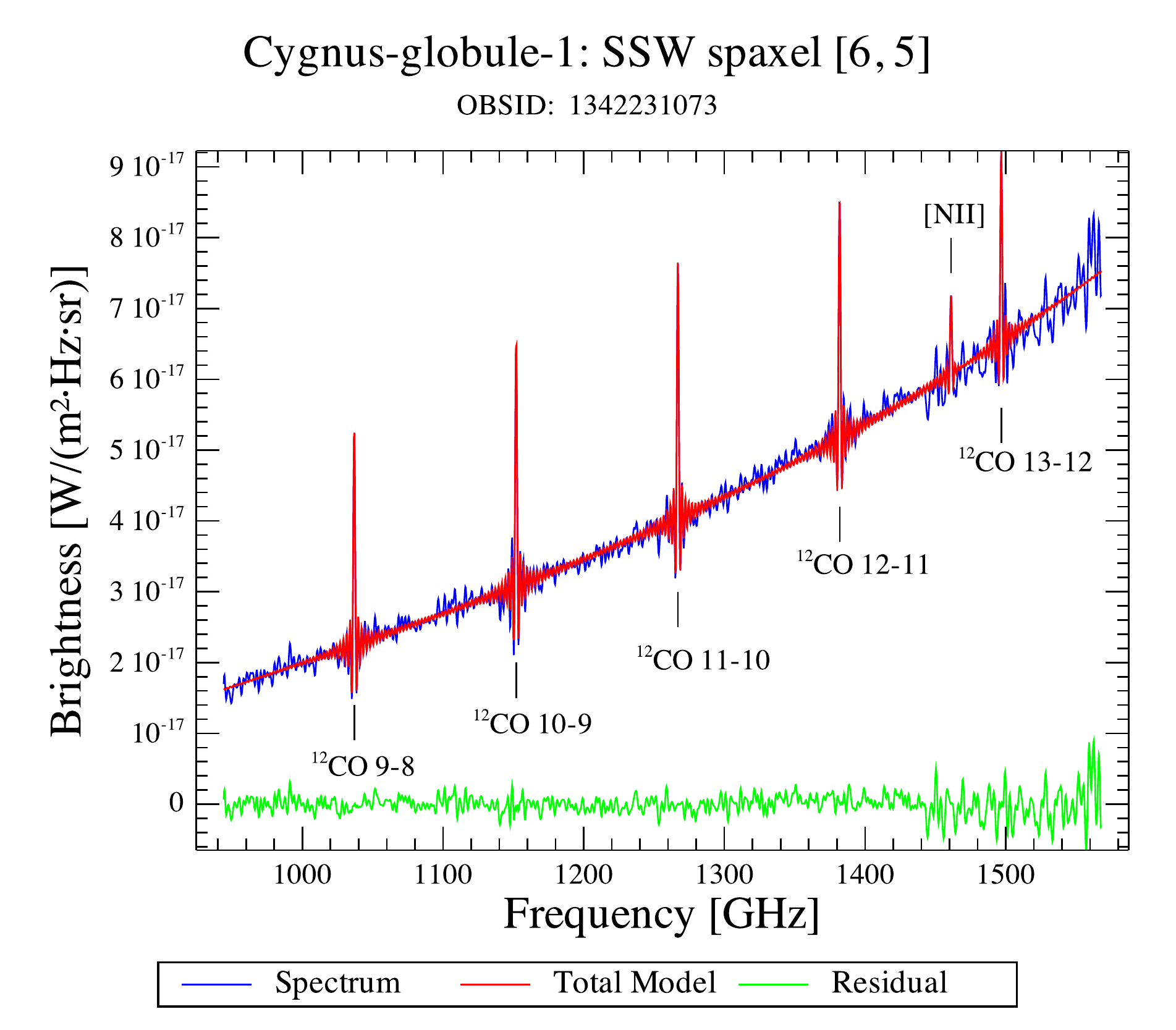}
\includegraphics[angle=0,width=8cm]{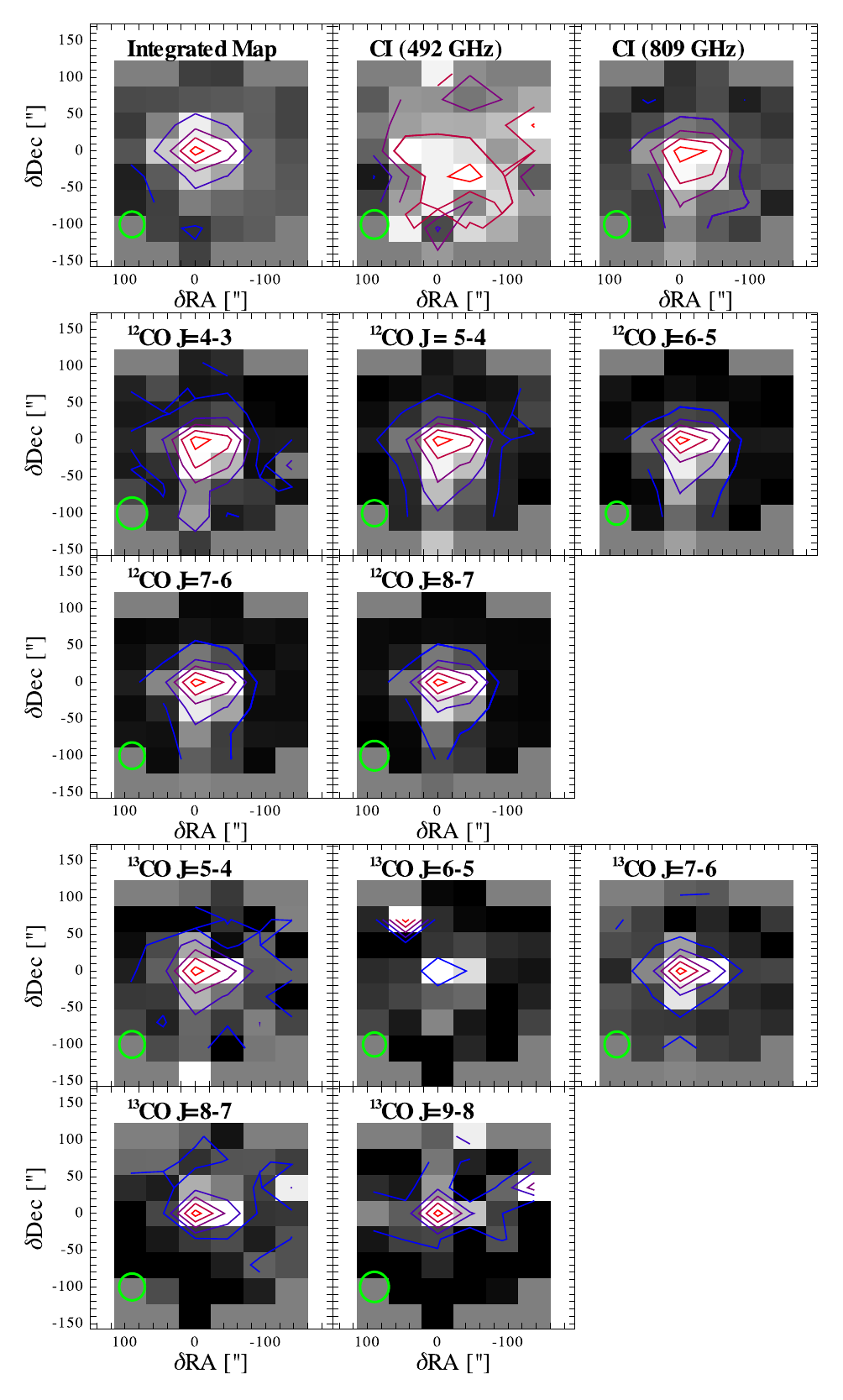}
\includegraphics[angle=0,width=8cm]{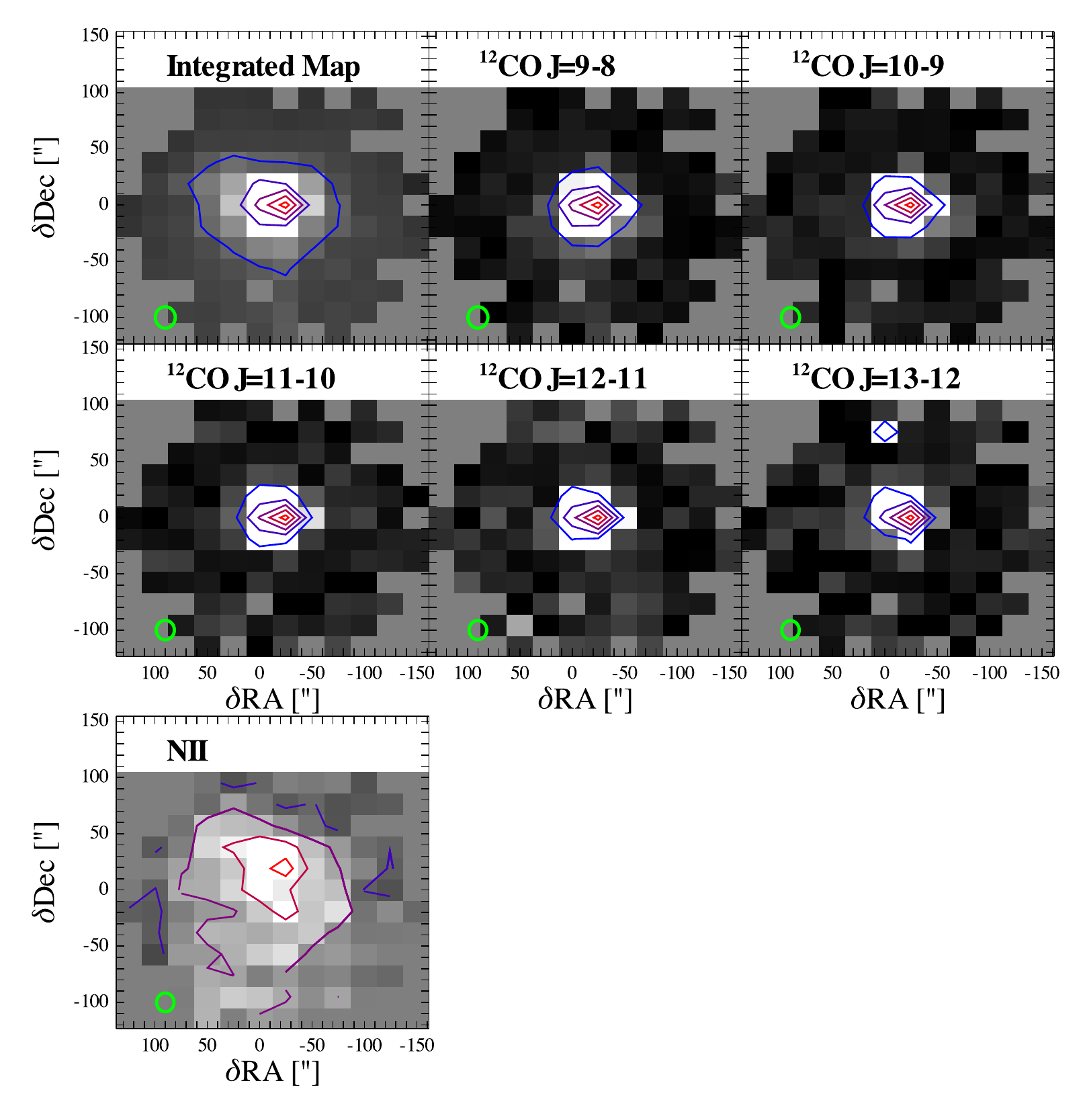}
\label{spire1}
\end{center}  
\caption [] {SPIRE spectroscopy results for the globule head. Top: FTS
  spectrum (blue) and line fit (red) for one spaxel of the SPIRE
  spectrometer for SLW (left) and SSW (right). The positions of the
  spectral lines included in the fit are indicated. Bottom: Full SPIRE
  spectral maps showing the CO-ladder and the \CI\ and
  \NII\ lines. The intensity scale has been set relative to the peak
  brightness in each map with contour levels at 0.1, 0.3, 0.5, 0.7 and
  0.9 of the peak (from blue to red). The SPIRE beam size varies
  between 31-43$''$ for SLW and 16-20”$''$ for SSW 
  \citep{swinyard2014}.}
\end{figure*}

\begin{figure*}[ht] 
\begin{center}  
  \includegraphics[angle=0,width=8cm]{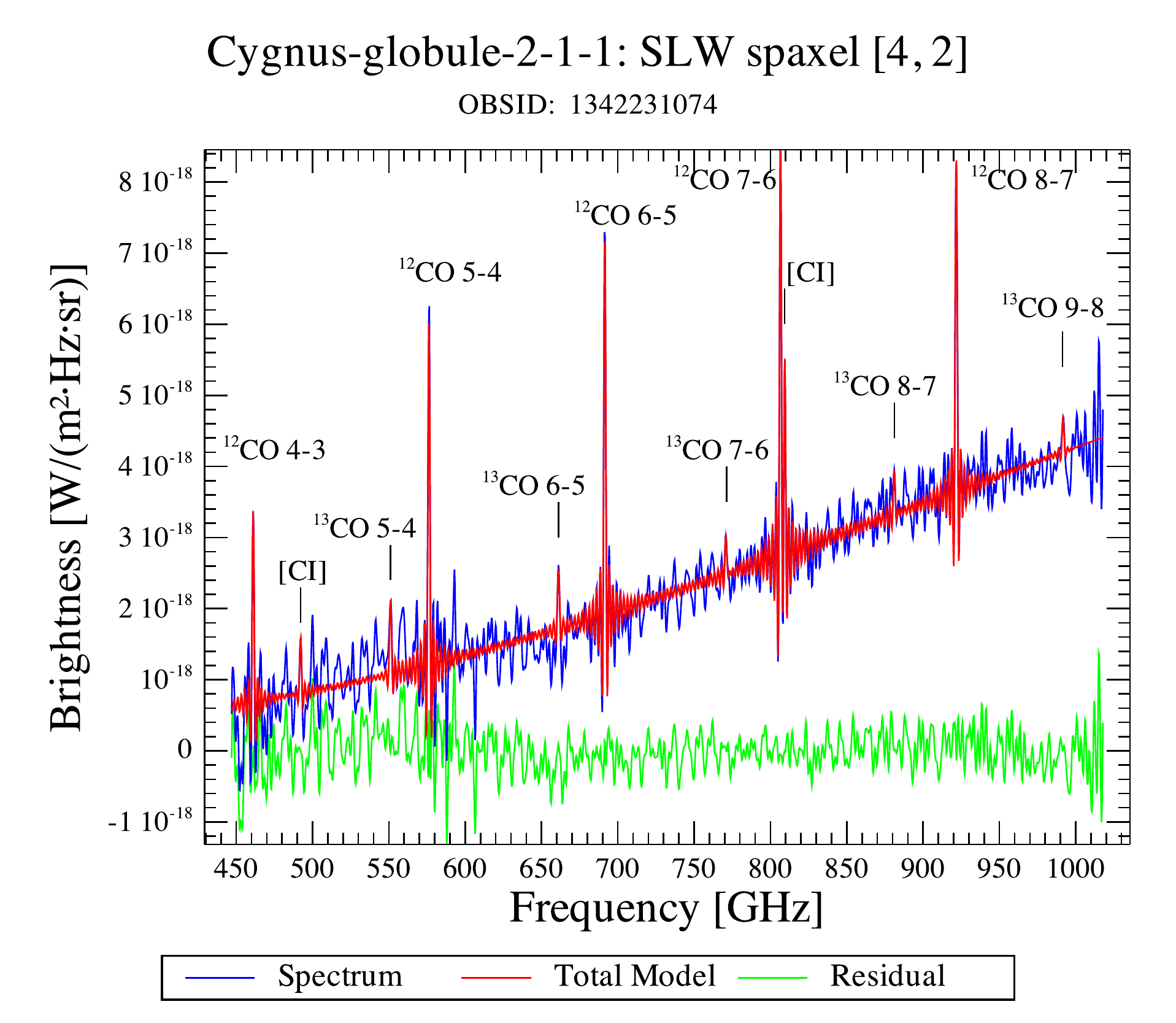}
  \includegraphics[angle=0,width=8cm]{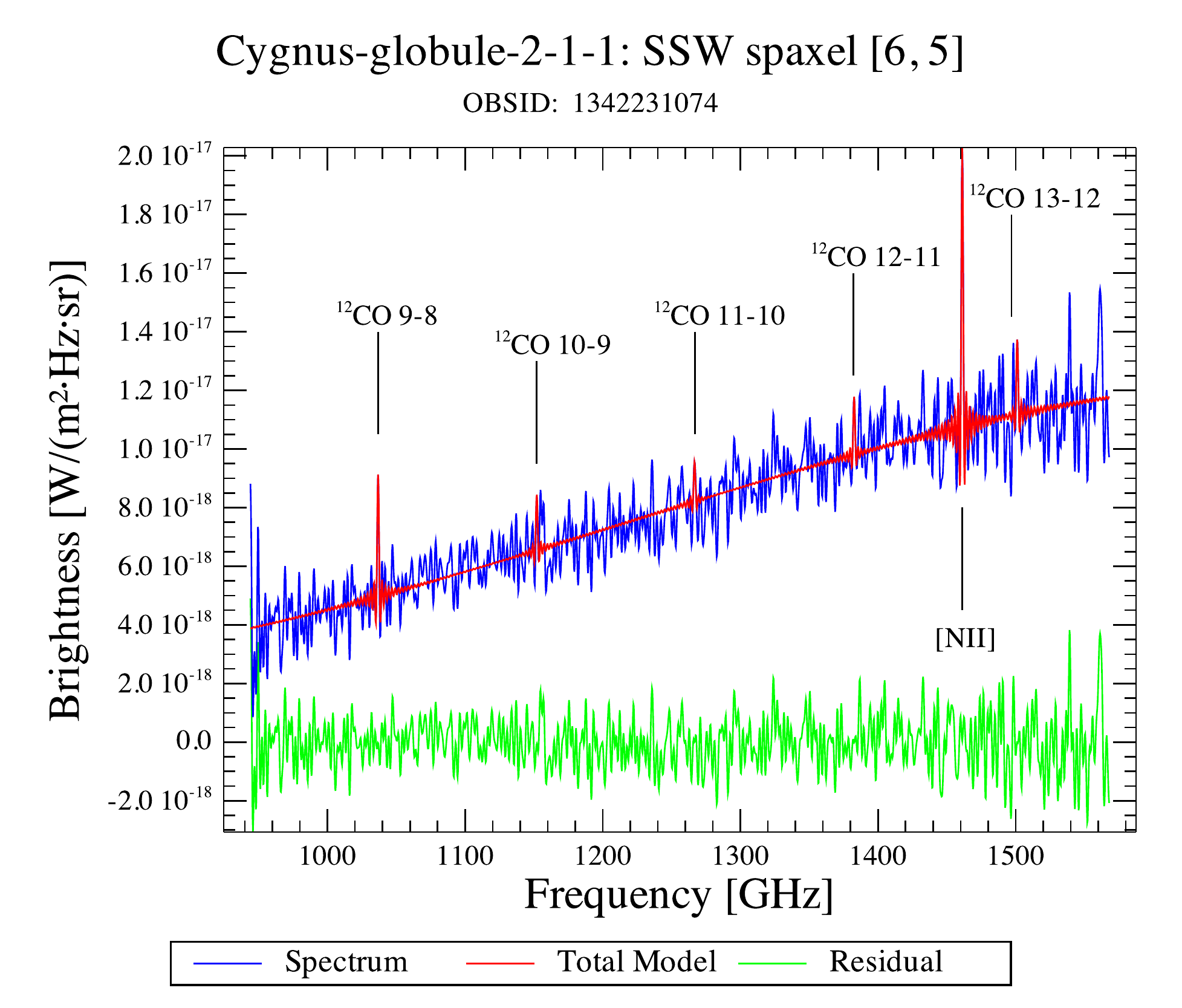}
  \includegraphics[angle=0,width=8cm]{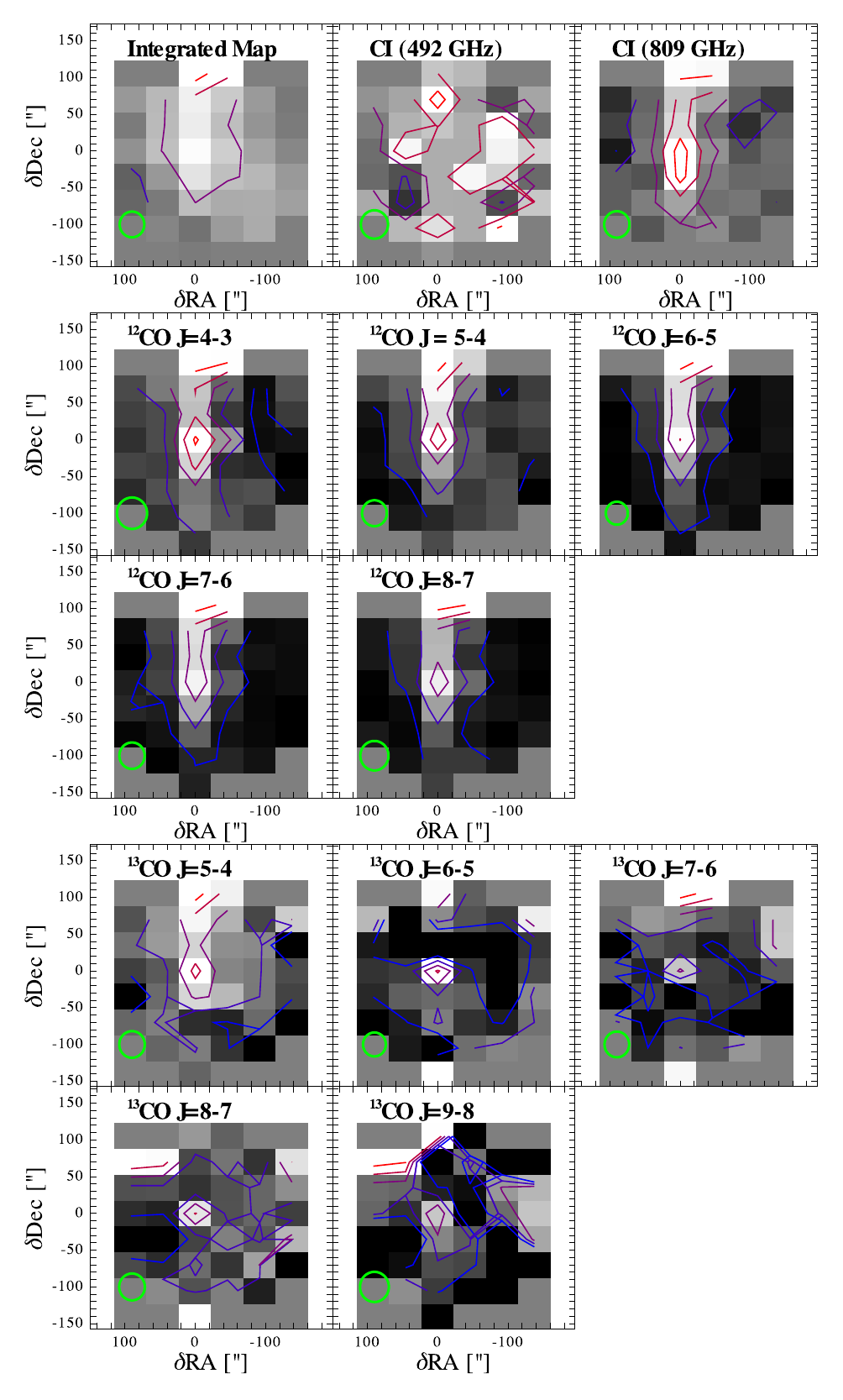}
  \includegraphics[angle=0,width=8cm]{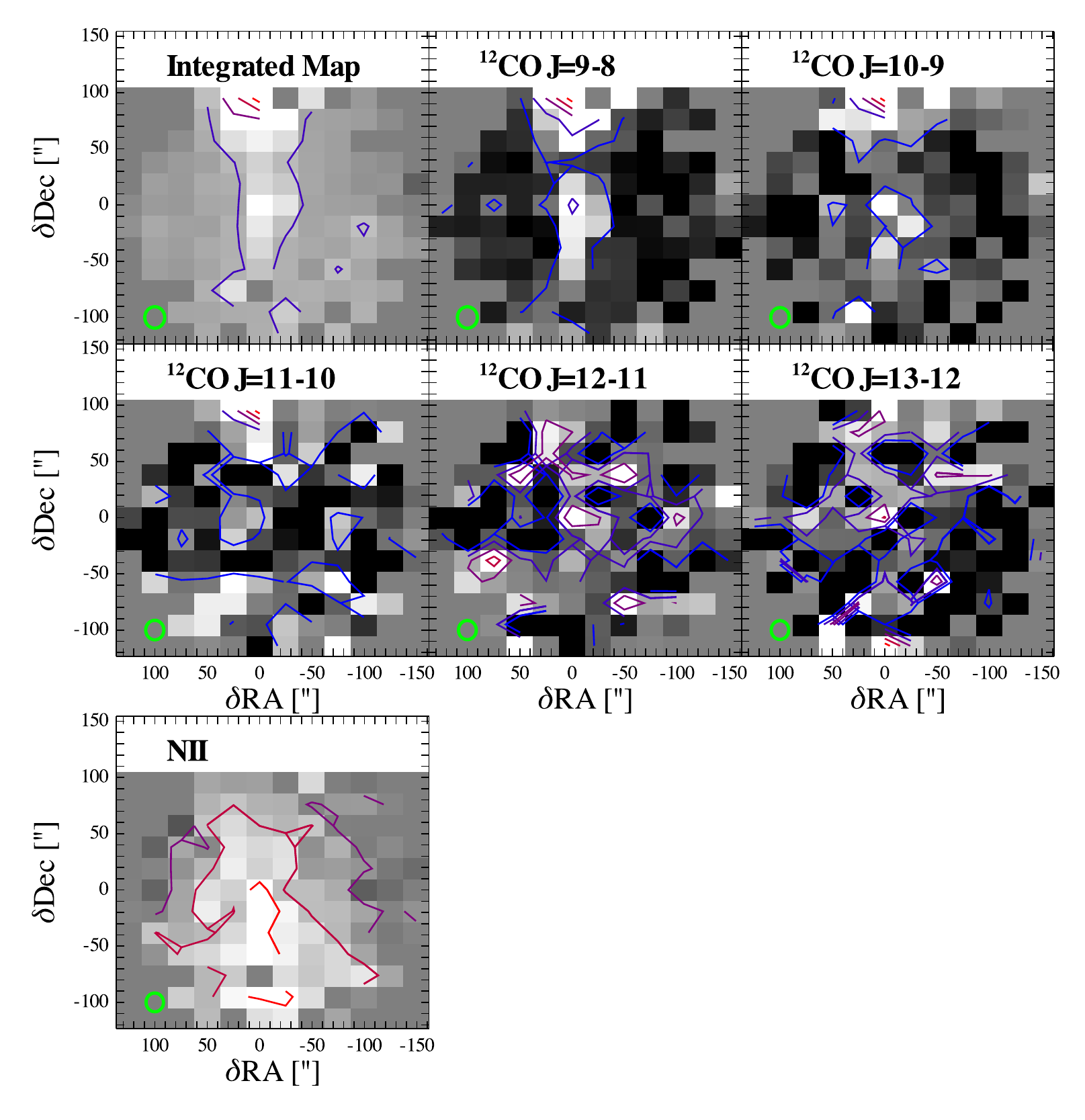}
 \label{spire2}
\end{center}  
\caption [] {SPIRE spectroscopy results for the globule tail. Details are the same as Fig.~A.2.}
\end{figure*}

\section{PDR model input parameters}

\begin{table}[htb]
\begin{center}
\caption{Overview of the most important model parameters (see also \citet{labsch2017}). All abundances
are given with respect to the total H abundance. }\label{parameter}
\begin{tabular}{lll}
\hline\hline
\multicolumn{3}{c}{\rule[-3mm]{0mm}{8mm}\bf Model Input Parameters}\\ \hline
\rule[2mm]{0mm}{2mm}He/H&0.0851&(1)\\
O/H            & 4.47 10$^{-4}$  & (2)\\
C/H            & 2.34 10$^{-4}$  & (2)\\
$^{13}$C/H      & 3.52 10$^{6}$   & (3)\tablefootmark{a}\\
$^{18}$O/H      & 8.93 10$^{-7}$  & (4)\tablefootmark{b}\\
N/H            & 8.32 10$^{-5}$  & (2)\\
S/H            & 7.41 10$^{-6}$  &(2)\\
F/H            & 6.68 10$^{-9}$  &(2)\\
$Z$            & 1              &solar metallicity\\
$\zeta_{CR}$    & 2 10$^{-16}$ s$^{-1}$ & CR ionisation rate (5)\\
$R_\mathrm{V}$  &5.5             & visual extinction/reddening (7,8) \\
$\sigma_\mathrm{D}$& 8.41 $^{-22}$~cm$^2$& UV dust cross section per H (8)\\
$\langle A(\lambda)/A_\mathrm{V} \rangle$&$2.40$&mean FUV extinction\\
$\tau_\mathrm{UV}$&$2.2 A_V$&FUV dust attenuation\\
$v_b$ & 1~km~s$^{-1}$&Doppler width\\
$n_0$&$10^{3,\ldots,7}$~cm$^{-3}$&total surface gas density\\
$M$&$10^{-3\ldots,3}$~M$_\odot$&cloud mass\\
$\chi$ & $10^{0\ldots,6}$&FUV intensity w.r.t. (6)\tablefootmark{c}\\
$\alpha$&1.5&density power law index\\
$R_\mathrm{core}$&$0.2 R_\mathrm{tot}$&size of constant density core\\
$N_\mathrm{tot}/A_\mathrm{V}$& 1.89 10$^{21}$~cm$^{-2}$&(8)\\
\hline\\
\end{tabular}
\end{center}
\vspace{-1cm}
\tablebib{
(1) \citet{asplund2005}; (2) \citet{simon-Diaz2011}; (3) \citet{langer1990}; (4) \citet{polehampton2005}; (5) \citet{hollenbach2012}; 
(6) \citet{draine1978}; (7) \citet{roellig2013}; (8) \citet{weingartner2001a}.}
\tablefoot{
\tablefoottext{a}{based on a $^{12}$C/$^{13}$C ratio of 67}
\tablefoottext{b}{based on a $^{16}$O/$^{18}$O ratio of 500}
\tablefoottext{c}{$\chi =1.71~G_0$ where G$_0$ is the mean ISRF from \citep{draine1978}.}
}
\end{table}

\end{appendix}

\end{document}